\documentclass[reprint,prl,twocolumn,twoside,aps,superscriptaddress]{revtex4-2}

\usepackage{amsmath,amssymb,amsfonts}
\usepackage{graphicx}
\usepackage{siunitx}
\usepackage{natbib}
\usepackage{braket}
\usepackage{multirow}
\usepackage[normalem]{ulem}
\usepackage{xspace}
\usepackage{url}
\usepackage[breaklinks]{hyperref}
\usepackage{braket}
\usepackage{lipsum}
\usepackage{xcolor}
\hypersetup{
    unicode=true,
    colorlinks=true,
    linkcolor=blue,
    citecolor=blue,
    urlcolor=blue
  }
\usepackage{breakurl}

\DeclareMathOperator{\sinc}{sinc}

\begin{document} 

\title{A coherent quantum interface between a neutral atom and a polar molecule}

\newcommand{\physics}{Department of Physics, Durham University, South Road, Durham, DH1 3LE, United Kingdom}
\newcommand{\jqc}{Joint Quantum Centre Durham-Newcastle, Durham University, South Road, Durham, DH1 3LE, United Kingdom}
\newcommand{\granada}{Departamento de F\'{i}sica At\'{o}mica, Molecular y Nuclear, Universidad de Granada, 18071 Granada, Spain}
\newcommand{\ici}{Instituto Carlos I de F\'{i}sica Te\'{o}rica y Computacional, Universidad de Granada, 18071 Granada, Spain}

\author{Daniel~K.~Ruttley}
\email{daniel.k.ruttley@durham.ac.uk}
\affiliation{\physics}
\author{Tom~R.~Hepworth}
\affiliation{\physics}
\author{Juan~M.~Garc\'ia-Garrido}
\affiliation{\granada}
\author{Caleb~J.~H.~Rich}
\affiliation{\physics}
\author{Rosario~Gonz\'{a}lez-F\'{e}rez}
\affiliation{\granada}
\affiliation{\ici}
\author{Alexander~Guttridge}
\affiliation{\physics}
\author{Simon~L.~Cornish}
\email{s.l.cornish@durham.ac.uk}
\affiliation{\physics}

\begin{abstract}
Arrays of trapped neutral atoms and polar molecules have separately emerged as powerful and complementary platforms for quantum science. Neutral atoms enable fast, programmable interactions through excitation to Rydberg states, whereas polar molecules possess long-lived rotational states that are attractive for quantum memories and qudits. Combining these platforms would create new possibilities, but requires a coherent interface between individual atoms and molecules, which so far has not been realised. Here, we establish such an interface between a single neutral atom and a single polar molecule. By tuning atomic and molecular transitions into resonance, we realise strong state-dependent dipolar interactions at micrometre separations in an optical tweezer platform with single-particle control. We exploit this interface to perform quantum operations between the two particles, demonstrating atom-mediated readout of a molecular qubit, coherent spin exchange, and entanglement using a blockade-based controlled-NOT operation. These results establish a new hybrid quantum platform at the intersection of atomic and molecular physics, in which long-lived molecular quantum information can be rapidly mapped onto an atom for readout or onward coherent transfer. The platform provides a route to hybrid quantum processors utilising atom-mediated readout and entanglement of molecular qubits, mixed-species quantum simulators of dipolar systems, and macroscopic entangled molecular states for quantum-enhanced metrology and precision sensing.
\end{abstract}

\date{\today}
\maketitle

Arrays of individually trapped neutral atoms and polar molecules have separately emerged as leading platforms for quantum science and technology.
Both support long-range interactions and can be controlled using optical tweezer arrays. 
Neutral atoms support rapid, programmable interactions through excitation to Rydberg states, enabling high-fidelity quantum gates and scalable quantum processors~\cite{Saffman2010,Browaeys2020,Wu2021,Defenu2023}. 
In contrast, polar molecules possess rich internal structures with long-lived states that support second-scale coherences~\cite{Cornish2024,Ruttley2025}, enabling dense quantum memories~\cite{Sawant2020}, large synthetic dimensions~\cite{Raghuram2026}, and new opportunities for precision metrology~\cite{DeMille2024}.
Despite these complementary strengths, these architectures have thus far developed independently due to the lack of a coherent quantum interface connecting individual atoms and molecules.

An atom--molecule interface would combine the fast, programmable interactions of atoms with the long-lived rotational states of molecules, providing a route towards hybrid quantum processors and new forms of quantum simulation. 
For example, the dense encoding of information available in molecules could reduce the number of physical qubits required for quantum information processing~\cite{Sawant2020}, addressing a key challenge in scaling quantum technologies.
More broadly, hybrid atom--molecule systems have been proposed theoretically for quantum computation~\cite{Kuznetsova2011,Zhang2022,Wang2022,Bai2026,Zhang2026}, quantum simulation~\cite{Kuznetsova2018,Dobrzyniecki2023}, and non-destructive molecular detection~\cite{Kuznetsova2016,Zeppenfeld2017,Jarisch2018,Gawlas2020,Patsch2022,Young2026}.
Beyond quantum information processing, hybrid atom--molecule systems would provide a platform for studying fundamentally new phenomena, including Rydberg-mediated molecular cooling~\cite{Zhao2012,Huber2012,Zhang2024cooling} and exotic polyatomic bound states~\cite{Rittenhouse2010,Rittenhouse2011,GonzalezFerez2020}.
Realising these opportunities requires strong, controllable, and coherent interactions between individual atoms and molecules.

Although dipolar interactions between atoms and molecules have previously been explored in gas-phase~\cite{Petitjean1986,Zhu2025} and beam~\cite{Gawlas2020,Zou2022} experiments, these systems do not provide the individual particle control required for quantum operations. In particular, the lack of control over particle separations precludes control of the interactions and, hence, coherent dynamics. Consequently, capabilities such as coherent quantum-state transfer and entanglement between individual atoms and molecules have remained out of reach.

Here, we establish a coherent quantum interface between an individual neutral atom and an individual polar molecule. 
By confining the particles in species-specific optical tweezers and tuning atomic and molecular transitions into resonance, we engineer MHz-scale, state-dependent dipolar interactions at micrometre separations. 
We exploit this interface to map molecular quantum states onto an atom for readout, observe coherent atom--molecule spin exchange, and generate entanglement between an atom and a molecule.
These results establish a new hybrid quantum platform that combines the complementary capabilities of neutral atoms and polar molecules at the level of individual particles, opening new opportunities for quantum science and technology.

\subsection*{Engineering an atom--molecule interface}
\begin{figure*}
\includegraphics[width=\hsize]{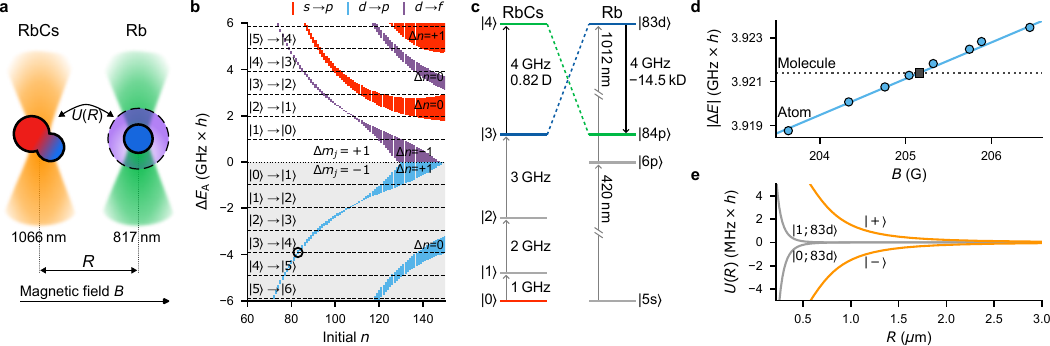}
\caption{\textbf{Engineering resonant interactions between a molecule and an atom.}
\textbf{a}, Schematic of the experiment showing a RbCs molecule and a Rb Rydberg atom which are prepared in species-specific optical tweezers. 
\textbf{b}, Energies of transitions, $\Delta E_\mathrm{A}$, between the Rydberg levels of Rb as a function of the initial principal quantum number, $n$. The height of the lines show the energy range accessible by varying a magnetic field over $100\,$G. The dashed lines show when the atomic transitions become resonant with molecular transitions labelled as $\ket{N}\to\ket{N'}$. The highlighted point shows the transitions that we use in this work.
\textbf{c}, Energy levels used in this work, as described in the text. We tune the pair state $\ket{3;83\mathrm{d}}$ (blue) to be resonant with $\ket{4;84\mathrm{p}}$ (green). 
\textbf{d}, Microwave spectroscopy near the resonance condition. We show the frequencies $|\Delta E_\mathrm{A}|/h$ (blue) and $|\Delta E_\mathrm{M}|/h$ (black) of the atomic and molecular transitions as a function of magnetic field, $B$. Resonance is achieved at $B_\mathrm{res}=205.16(2)\,$G. The error bars show the $1\sigma$ confidence intervals and, on average, there are 1214 experimental shots per data point.
\textbf{e}, Potential-energy curves $U(R)$ of as a function of the interparticle separation, $R$, for different pair states at $B = B_\mathrm{res}$.
\label{fig:overview}}
\end{figure*}

We establish a coherent atom--molecule interface by engineering resonant dipolar interactions between a single Rb atom and a single RbCs molecule. 
The particles are confined in species-specific optical tweezers, providing precise control of their relative position (see Methods). 
We prepare the atom in the electronic ground state $\ket{5\mathrm{s}}$ and the molecule in the absolute ground state $\ket{0}$, labelled by the atomic and molecular quantum numbers $nl$ and $N$, respectively (see Methods for the full state labels). 
An applied magnetic field of $B\approx200\,\mathrm{G}$ defines the quantisation axis, which is aligned with the interparticle vector (see Fig.~\ref{fig:overview}a).

The key to creating a strong atom--molecule interaction is to transform the negligible interaction between the ground-state particles into a resonant exchange interaction between an atomic Rydberg transition and a molecular rotational transition. 
This produces a first-order dipole--dipole interaction.
To achieve this, we tune an electric-dipole-allowed transition in the Rydberg manifold into resonance with a rotational transition of the molecule, satisfying the condition $\delta \equiv (\Delta E_\mathrm{A}+\Delta E_\mathrm{M})/h=0$,
where $\Delta E_\mathrm{A}$ and $\Delta E_\mathrm{M}$ are the atomic and molecular transition energies. 

In Fig.~\ref{fig:overview}b, we show the energies $\Delta E_\mathrm{A}$ of atomic transitions from states experimentally accessible with two-photon excitation from $\ket{5\mathrm{s}}$ (see Methods). 
We coarsely tune $\delta$ by choice of atomic and molecular states, and then use the magnetic field to bring the transitions into exact resonance. 
The vertical extent of each transition in Fig.~\ref{fig:overview}b indicates the range of tunability of $\Delta E_\mathrm{A}$ over a $100\,\mathrm{G}$ magnetic-field range (see Methods). 
Resonance occurs where these atomic transitions intersect the allowed molecular transitions ($-\Delta E_\mathrm{M}$, dashed lines).

In this work, we engineer resonance between the pair states $\ket{3;83\mathrm{d}}$ and $\ket{4;84\mathrm{p}}$, labelled as $\ket{N;nl}$. 
Fig.~\ref{fig:overview}c shows the corresponding atomic and molecular energy levels, with the resonant pair states highlighted in blue and green. 
The molecule is rotationally excited using microwave pulses~\cite{Ruttley2024} and the excited states are long-lived~\cite{Ruttley2025}.
The atom is optically excited from $\ket{5\mathrm{s}}$ to $\ket{83\mathrm{d}}$ using a two-photon transition (see Methods).
These pair states balance the number of microwave pulses required for molecular state preparation against the practical challenges of atomic excitation to high-$n$ Rydberg states~\cite{Adams2020}.

We determine the resonance condition using microwave spectroscopy of the single-particle transitions. 
Fig.~\ref{fig:overview}d shows the measured transition frequencies $|\Delta E_\mathrm{A}|/h$ (blue) and $|\Delta E_\mathrm{M}|/h$ (black) as a function of magnetic field. 
The atomic transition frequency varies strongly with magnetic field. In contrast, the molecular transition frequency is effectively field independent because the two rotational states have identical nuclear-spin projections and negligible rotational Zeeman shifts. 
The two transitions cross at a resonance field of $B_\mathrm{res}=205.16(2)\,\mathrm{G}$, where $\Delta E_\mathrm{M}=-\Delta E_\mathrm{A}\approx3.9\,\mathrm{GHz}\times h$.

Figure~\ref{fig:overview}e shows the calculated potential-energy curves $U(R)$ for different atom--molecule pair states. These energies are obtained using the charge--dipole Hamiltonian, which approaches the dipole--dipole interaction at long range (see Methods). For non-resonant pair states such as $\ket{0;83\mathrm{d}}$ and $\ket{1;83\mathrm{d}}$ (grey lines), the interaction is at the kHz scale for micrometre separations, occurring only through second-order processes at these distances.

By tuning the atomic and molecular transitions into resonance ($\delta\rightarrow0$), we access the first-order dipolar coupling. 
The resonant pair states $\ket{3;83\mathrm{d}}$ and $\ket{4;84\mathrm{p}}$ then strongly hybridise to produce MHz-scale interaction energies at micrometre separations (orange lines). 
The resulting potentials are well approximated by $U(R)=\pm C_3/R^3$, where $C_3=-d_\mathrm{A}d_\mathrm{M}/(4\pi\epsilon_0)$ for our geometry (see Methods). Here, $d_\mathrm{A}=-14.5\,\mathrm{kD}$ and $d_\mathrm{M}=0.82\,\mathrm{D}$ are the atomic and molecular transition dipole moments, respectively, giving $C_3/h=1.79\,\mathrm{MHz}\,\mu\mathrm{m}^3$. 
The eigenstates are coherent superpositions of the atomic and molecular pair states, $\ket{\pm}=\frac{1}{\sqrt{2}}(\ket{3;83\mathrm{d}}\pm\ket{4;84\mathrm{p}})$, which provide the basis for coherent exchange between the two particles.
These strong and controllable interactions provide the foundation for the atom--molecule quantum operations demonstrated below.

\subsection*{State-selective atom--molecule interactions}
First, we demonstrate the strong state dependence of the atom--molecule interaction underlying the coherent interface. We characterise this state selectivity through measurement of the blockade of the atomic transition $\ket{5\mathrm{s}}\rightarrow\ket{83\mathrm{d}}$ arising from resonant atom--molecule interactions. 
We hold the particles at a separation of $R=0.9(1)\,\mu$m and drive the atomic transition using a resonant $\pi$ pulse with two-photon Rabi frequency $\Omega_\mathrm{Ryd}/2\pi=931(5)\,$kHz.
The optical tweezers are antitrapping for Rydberg atoms, meaning population of the Rydberg state results in atomic loss. We detect this loss via fluorescence imaging at the end of the experimental sequence.
For all measurements presented in this work, we postselect experimental runs in which an atom--molecule pair is successfully prepared and the molecule is detected at the end of the sequence (see Methods).

\begin{figure*}
\includegraphics[width=\textwidth]{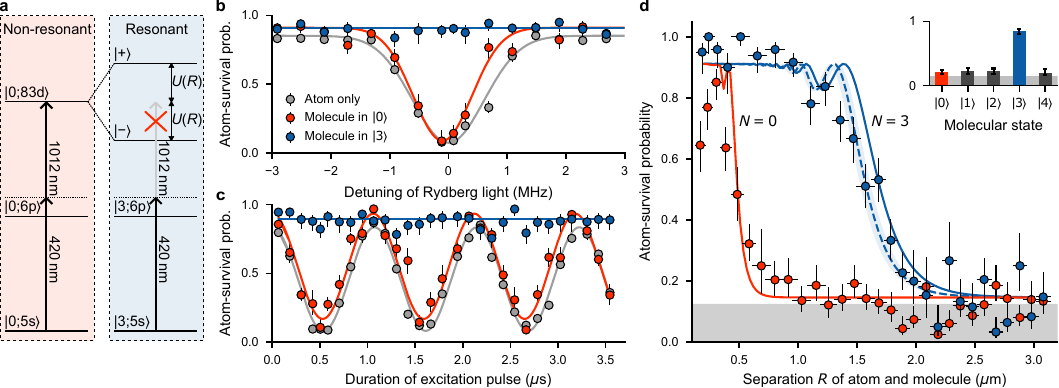}
\caption{\textbf{Rydberg blockade caused by resonant atom--molecule interactions.}
\textbf{a}, Energy levels of the atom--molecule system for the non-resonant (red) and resonant (blue) cases at $R\approx1\,\mu$m, as described in the text.
\textbf{b}, Atomic survival probability as a function of the two-photon detuning of the Rydberg excitation light. Red and blue points show data for molecules prepared in the states $\ket{0}$ and $\ket{3}$, respectively, at $R=0.9(1)\,\mu$m. Grey points show the spectrum measured in the absence of a molecule. When the molecule is prepared in $\ket{3}$, excitation of the atom is strongly suppressed.
\textbf{c}, Atomic survival probability as a function of the duration of the Rydberg-excitation pulse. Colours are as in b.
\textbf{d}, Characterisation of the distance dependence of the interactions. The atomic-survival probability is shown as a function of the separation $R$ of the atom and molecule. Blue (red) points show data for molecules prepared in $\ket{3}$ ($\ket{0}$). Solid lines show the predicted behaviour, as described in the text. The dashed line is a fit to the blue data assuming a purely dipolar interaction; the blue shaded region shows the $1\sigma$ confidence interval.
The grey region indicates the atom-only excitation contrast. Inset: atom-survival probability for different molecular rotational states at $R=0.9(1)\,\mu$m.
In all panels, the error bars show the $1\sigma$ confidence intervals and, on average, there are 91 experimental shots per data point.
\label{fig:blockade}}
\end{figure*}

In the absence of resonant interactions, such as when the molecule is in the state $\ket{0}$, the atom and molecule are effectively decoupled (Fig.~\ref{fig:blockade}a, red). 
By varying the two-photon detuning of the excitation light, we measure an excitation spectrum (Fig.~\ref{fig:blockade}b, red) that approximately agrees with a reference measurement performed without a molecule (Fig.~\ref{fig:blockade}b, grey). Crucially, the atom-loss probability when the light is on resonance, which provides a measure of the likelihood of Rydberg excitation, is consistent between the two cases (0.91(5) for the molecule in state $\ket{0}$ and 0.93(2) for the atom-only case).

By contrast, when the molecule is prepared in the state $\ket{3}$, atomic excitation to $\ket{83\mathrm{d}}$ is strongly suppressed. In this regime, the resonant atom--molecule interaction hybridises the pair states into eigenstates $\ket{\pm}$  (Fig.~\ref{fig:blockade}a, blue). At $R=0.9(1)\,\mu$m, the interaction strength $|U(R)|/h \approx 2\,\mathrm{MHz}$ exceeds the power-broadened linewidth of the transition and blockades excitation (i.e.\ $|U(R)| \gtrsim \hbar\Omega_\mathrm{Ryd}$). Consistent with this, the blue data in Fig.~\ref{fig:blockade}b show a clear suppression of atomic excitation.
We do not resolve excitation to the interaction-shifted states $\ket{\pm}$, which we attribute to fluctuations in the interaction strength and coupling to undetected hyperfine states in the molecule (see Methods).

In Fig.~\ref{fig:blockade}c, we show the effect of varying the duration of the Rydberg-excitation pulse on resonance with the bare-atom transition. 
In the absence of blockade (grey: no molecule, red: molecule in the state $\ket{0}$), we observe clear Rabi oscillations. The lower bound of the atom-recovery probability is limited by infidelity in atomic-state preparation and Rydberg excitation (see Methods).
Conversely, when we prepare the molecule in the state $\ket{3}$, the Rabi oscillation is strongly suppressed. 
The blockade is robust because moderate fluctuations in $\delta$ and $R$ do not compromise the condition that the interaction shift exceeds the transition linewidth (see Methods).

We use the blockade as a probe of the engineered atom--molecule interaction, characterising its distance dependence as shown in Fig.~\ref{fig:blockade}d. We perform this measurement with a two-photon Rabi frequency $\Omega_\mathrm{Ryd}/2\pi=318(3)$\,kHz and record the atomic survival after driving $\pi$ pulse on the transition to the Rydberg state as a function of interparticle separation $R$.
As we vary $R$, we compensate for light shifts of the Rydberg transition caused by varying the tweezer overlap (see Methods). 
The atom-only excitation contrast is indicated by the grey shaded region.

The blue data in Fig.~\ref{fig:blockade}d show the atomic recovery when the molecule is prepared in the state $\ket{3}$. 
We compare these data to the predicted survival probability (blue line) from a model in which we consider point-like particles with an interaction strength following the state $\ket{+}$ shown in Fig.~\ref{fig:overview}e (see Methods).
The oscillatory features observed for $R \lesssim 1.4\,\mu$m arise from the use of a square excitation pulse: at the minima of these features, the $\sinc^2$ sidelobes in the Fourier spectrum of the pulse overlap with the interaction-shifted transition (see Methods).

To quantify the interaction strength, we fit these data assuming perfect resonance, point-like particles, and a pure dipole--dipole interaction (see Methods). From this fit (blue dashed line), we obtain $C_3/h = 1.36(15)\,\mathrm{MHz}\,\mu\mathrm{m}^3$, with the corresponding shaded region indicating the $1\sigma$ confidence interval.
We expect that this is slightly weaker than the theoretical estimate presented earlier for point-like particles due to the finite spatial extent of the particle wavefunctions~\cite{Ruttley2025}. 

The red data and corresponding model line show the behaviour when the molecule is prepared in the state $\ket{0}$. In this case, we observe no significant blockade until $R \lesssim 0.5\,\mu$m, where the molecule is within the Rydberg-electron wavefunction. 
Furthermore, this blockade is less robust, as small fluctuations in $R$ are now sufficient to break the blockade condition (see Methods).

The inset of Fig.~\ref{fig:blockade}d demonstrates the state-selectivity of the interaction for the states $\ket{N}$ with $N\in[0,4]$ at $R=0.9(1)\,\mu$m.
Consistent with the resonance condition, we observe significant blockade only when the molecule is in the state $\ket{3}$.

The strong state-selectivity of the interaction at distances $R\approx1\,\mu$m provides the control required for a coherent atom--molecule quantum interface. 
In the following sections, we exploit this interface to demonstrate three key advances.
First, we demonstrate non-destructive readout of the molecular state, in which the state of the molecule is inferred without requiring a change in its internal state.
Second, we show how this interaction can be used to realise coherent spin-exchange.
Finally, we generate entanglement between the atom and the molecule using a controlled-NOT operation.

\subsection*{Molecular-state readout}
We use the atom--molecule interface to realise atom-mediated readout of the molecular state. 
This is highly advantageous because state-resolved molecular detection is generally destructive, relying either on dissociation of the molecule~\cite{Ruttley2024,Picard2024SiteSelective} or photon scattering that can alter the internal state~\cite{Anderegg2019,Holland2023Imaging,Vilas2024}. By mapping specific molecular populations to the resonant state, we selectively blockade the Rydberg excitation of the atom. Measuring the atomic state therefore infers the molecular state without direct interrogation, leaving the molecule intact for subsequent quantum operations.

Here, we focus on a strong, projective measurement of the molecular state, which represents an essential capability for hybrid quantum systems. 
Such measurements are required for quantum error correction through the extraction of error syndromes~\cite{Bluvstein2024,Miles2026} and for probing complex many-body observables, including out-of-time-order correlators~\cite{mi2021}. 
More broadly, they enable heralded state preparation~\cite{Wallquist2009} and measurement-based quantum information protocols~\cite{Wei2021}.

\begin{figure}
\centering
\includegraphics{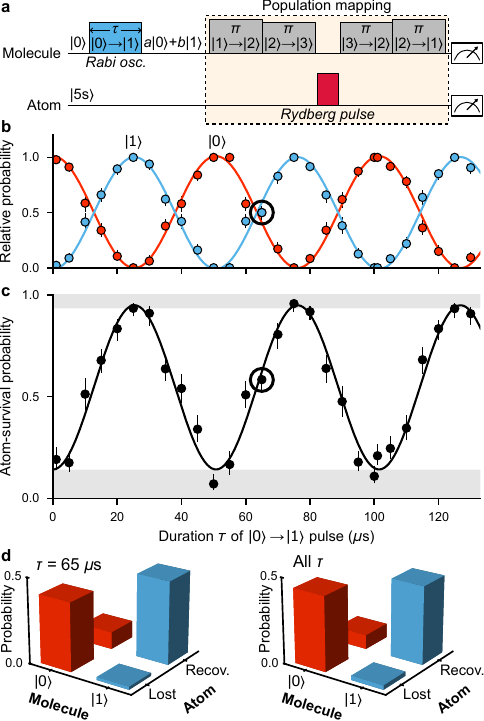}
\caption{\textbf{Molecular readout with a Rydberg atom.}
\textbf{a}, Schematic of the experimental sequence used to map the state of a molecule onto an atom. First, we drive a Rabi oscillation on the molecular transition $\ket{0}\to\ket{1}$ to prepare it in a superposition. Then, we map the population of one of the component states (here, $\ket{1}$) to the resonant state $\ket{3}$ with a series of microwave $\pi$ pulses. Next, we attempt to resonantly excite the atom to the Rydberg state. This is blockaded state selectively and any Rydberg atoms are ejected from the tweezer. Finally, the molecule is returned to its original state and we readout both particles.
\textbf{b}, Rabi oscillations on the molecular transition $\ket{0}\to\ket{1}$. We show the relative occupations of the two states.
\textbf{c}, Atom-survival probability after this detection scheme. The shaded regions indicate the contrast of the Rydberg excitation when we prepare no molecules.
\textbf{d}, The correlations between the molecular population and atom recovery and loss. The left panel shows data from $\tau=65\,\mu$s (highlighted point), the right panel shows data from the whole Rabi oscillation. The correlations are as expected, with the largest infidelity being recovery of the atom when the molecule is in $\ket{0}$.
In all panels, the error bars show the $1\sigma$ confidence intervals and, on average, there are 51 experimental shots per data point.
\label{fig:molecule_readout}}
\end{figure}

We demonstrate this readout technique by tracking a Rabi oscillation on the molecular transition $\ket{0}\rightarrow\ket{1}$. The experimental sequence is shown in Fig.~\ref{fig:molecule_readout}a.
First, the molecular population is driven using a microwave pulse of variable duration $\tau$.
The resulting state populations are shown in Fig.~\ref{fig:molecule_readout}b and are extracted by mapping the molecular state onto distinct configurations of the constituent atoms following dissociation of the molecule at the end of the sequence~\cite{Ruttley2024} (see Methods).

To perform the atom-mediated readout, we map the population of one molecular state onto the state which causes Rydberg blockade.
This readout sequence is shown in the highlighted region of Fig.~\ref{fig:molecule_readout}a.
We transfer molecular population in the state $\ket{1}$ to the state $\ket{3}$ using two additional microwave pulses (see Methods). 
Then, we apply a $\pi$ pulse on the bare-atom transition $\ket{5\mathrm{s}}\rightarrow\ket{83\mathrm{d}}$ at $R=1.1(1)\,\mu$m. 
As in Fig.~\ref{fig:blockade}, this excitation is blockaded when the molecule occupies $\ket{3}$, but proceeds freely when it remains in $\ket{0}$. 
Following this, we transfer population in $\ket{3}$ back to $\ket{1}$. 
This protocol implements the conditional mapping $\ket{0;5\mathrm{s}}\rightarrow\ket{0;83\mathrm{d}}$ and $\ket{1;5\mathrm{s}}\rightarrow\ket{1;5\mathrm{s}}$. 
Ejection of the Rydberg atom maps these outcomes to atomic loss and survival, respectively. 

Figure~\ref{fig:molecule_readout}c shows the resulting atom-survival probability. 
The signal closely follows the population of the molecular state $\ket{1}$, demonstrating successful mapping of the molecular state onto the atom.
This is highlighted in Fig.~\ref{fig:molecule_readout}d which shows the correlations between the measured molecular state and the atomic outcome for $\tau=65\,\mu$s (left) and all $\tau$ (right).
From the fit in Fig.~\ref{fig:molecule_readout}c, we extract the fidelity of atom-mediated molecular readout as $F_\mathrm{meas}=\frac{1}{2}(F_{0|0}+F_{1|1})=0.91(1)$, where $F_{N|N}$ is the probability of inferring from the atomic measurement that the molecule is in state $\ket{N}$, given that it was prepared in $\ket{N}$. Specifically, we obtain $F_{0|0}=0.86(2)$ and $F_{1|1}=0.95(2)$.
These fidelities are primarily limited by imperfections in atomic control with the dominant error corresponding to atom survival when the molecule occupies $\ket{0}$, as shown in Fig.~\ref{fig:molecule_readout}d. 
This is consistent with the imperfect Rydberg excitation that we observe in the absence of a molecule, indicated by the grey shaded regions in Fig.~\ref{fig:molecule_readout}c.
These technical imperfections can be mitigated, and the detection fidelities substantially improved, by adopting the high-fidelity optical control routinely demonstrated in modern neutral-atom processors.
Additionally, from the fit in Fig.~\ref{fig:molecule_readout}b, we determine that the probability of the atomic measurement inducing a molecular bit flip is consistent with zero (see Methods).
The Rydberg-excitation light induces molecular loss with a probability of $9(6)\%$. 
This constitutes an erasure error that is removed via postselection and could be eliminated in future implementations (see Methods).

\subsection*{Spin exchange and entanglement}

\begin{figure*}
\includegraphics[width=\textwidth]{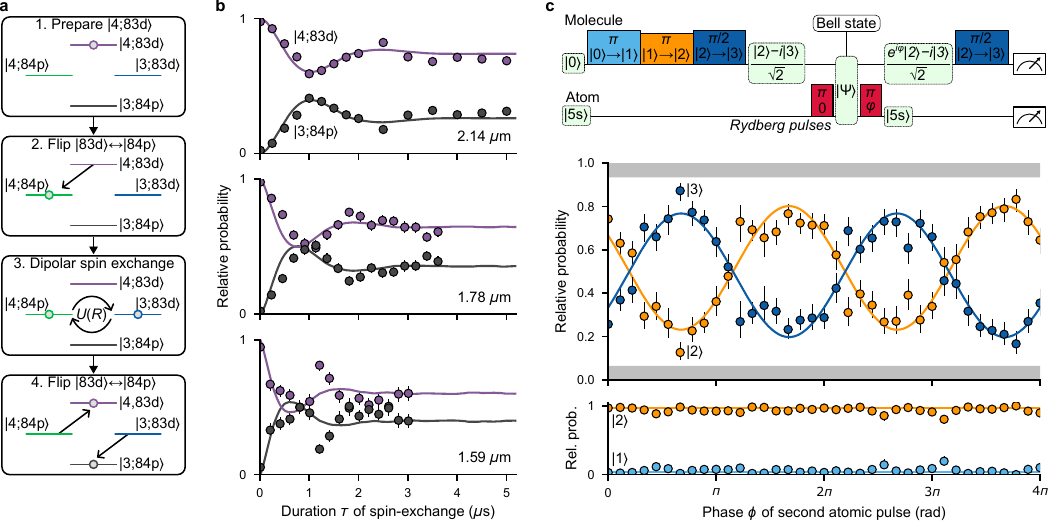}
\caption{\textbf{Dipolar spin-exchange and entanglement of an atom and molecule.} \textbf{a}, Experimental sequence (see text) that we use to observe atom--molecule spin exchange. \textbf{b}, Relative populations of the states $\ket{4}\ket{83\mathrm{d}}$ (purple) and $\ket{3}\ket{84\mathrm{p}}$ (black) with varying spin-exchange duration $\tau$. The different panels show different interparticle separations. We perform a global fit to extract the parameters given in the text. 
\textbf{c}, Entanglement of the atom and molecule using state-selective dipolar interactions. Upper: experimental sequence, as described in the text. Lower: Entanglement-mediated phase transfer between particles. We prepare the maximally entangled Bell state $\ket{\Psi}$ by performing a \texttt{CNOT} gate on the atom with the excitation light while the molecule is in the state $(\ket{2}-i\ket{3})/\sqrt{2}$. We verify this entanglement by disentangling the particles with an atomic $\pi$ pulse of varying phase $\varphi$. When the particles are entangled, this phase is coherently mapped to the molecular state (large panel). In contrast, when the molecule are prepared in the state $(\ket{1}-i\ket{2})/\sqrt{2}$, there are no resonant interactions. Therefore, we do not entangle the particles and the molecule does not acquire the phase of the atomic pulse (small panel).
In both cases, the atom and molecule are prepared at $R=1.1\,\mu$m.
In all panels, the error bars show the $1\sigma$ confidence intervals and, on average, there are 107 experimental shots per data point.
\label{fig:entanglement}}
\end{figure*}

Next, we demonstrate coherent dipolar spin exchange between an atom and a molecule.
For this, we implement the sequence shown in Fig.~\ref{fig:entanglement}a. 
First, we prepare the system in the pair state $\ket{4;83\mathrm{d}}$ (green) by preparing the molecule in $\ket{4}$ and driving a $\pi$ pulse on the Rydberg transition.
This pair state is non-resonant, so the atomic excitation can proceed freely.
Spin-exchange dynamics are initiated by driving a microwave $\pi$ pulse to the pair state $\ket{4;84\mathrm{p}}$ with a Rabi frequency $\Omega_\mathrm{MW}$ that is much greater than the dipolar interaction with the resonantly coupled $\ket{3;83\mathrm{d}}$ (i.e.\ $\hbar\Omega_\mathrm{MW}\gg |U(R)|$). 
In terms of the interacting eigenstates, this initialises the system in $\ket{4;84\mathrm{p}} = \frac{1}{\sqrt{2}}(\ket{+}-\ket{-})$.
The relative phase of this superposition then evolves with time due to the different eigenenergies of $\ket{+}$ and $\ket{-}$. 
We map this onto an oscillation in the populations by applying a second atomic microwave $\pi$ pulse after time $\tau$ that also freezes the dynamics prior to readout.

With this protocol,
we ideally prepare the state
\begin{equation*}
\ket{\psi(\tau)} =
\cos\left(\frac{\tau\Delta(R)}{\hbar}\right)\ket{4;83\mathrm{d}}
- i\sin\left(\frac{\tau\Delta(R)}{\hbar}\right)\ket{3;84\mathrm{p}}.
\end{equation*}
Here, $\Delta(R)$ is the energy splitting between $\ket{+}$ and $\ket{-}$. It follows that the maximally entangled state, $\frac{1}{\sqrt{2}}(\ket{4;83\mathrm{d}}
- i\ket{3;84\mathrm{p}})$, is prepared when $\tau=h/(4\Delta(R))$.

Fig.~\ref{fig:entanglement}b shows the resulting dynamics for several values of $R$. 
We plot the relative populations of the states $\ket{4;83\mathrm{d}}$ and $\ket{3;84\mathrm{p}}$ and ignore other pair states: this corrects for technical errors like imperfect state preparation (see Methods).
Clear oscillations are observed with a rate that depends on the distance $R$, demonstrating coherent spin exchange between the atom and molecule.

We fit the data using a global Monte Carlo model that accounts for shot-to-shot fluctuations in interparticle separation and detuning from resonance. In the model, we sample the separation $R$ from a Gaussian distribution with mean $\bar R$ and width $\sigma_R$, and the detuning $\delta$ from a Gaussian distribution with mean $\bar\delta$ and width $\sigma_\delta$.
We independently measure $\bar R$ (see Methods) and extract parameters $\sigma_R = 0.36(2)\,\mu$m, $\bar\delta = 0.29(2)\,\mathrm{MHz}$, and $\sigma_\delta = 0.12(4)\,\mathrm{MHz}$. 
We attribute the separation fluctuations to the ejection of the Rydberg atom and finite wavefunction spreads of the particles.
The detuning fluctuations are attributed to magnetic-field noise (see Methods).

Whilst the spin-exchange sequence can, in principle, generate maximally entangled states of the atom and molecule, in practice its performance is limited by noise on the interaction strength. 
Proposals using quantum optimal control to mitigate these effects have been developed for dipolar systems~\cite{Goerz2011,Muller2011,Hughes2020,Bergonzoni2025}, but they are challenging to implement here because we cannot currently perform single-particle operations on the molecule while the atom occupies a Rydberg state.
This is because the Rydberg atom has a much larger electric dipole moment than the molecule, leading to strong off-resonant driving of the atom when attempting to address the molecule.

A more robust route to entanglement is provided by the blockade mechanism, which requires only that $|U(R)| \gtrsim \hbar\Omega_\mathrm{Ryd}$. The trade-off is the entangling operation is slower.
We implement such an approach using the protocol shown in the top panel of Fig.~\ref{fig:entanglement}c.
We first prepare the system in the state $\frac{1}{\sqrt{2}}(\ket{2;5\mathrm{s}} - i\ket{3;5\mathrm{s}})$ by applying a sequence of microwave pulses to the molecule at $R=1.1(1)\,\mu$m.
We then drive an atomic $\pi$ pulse which is resonant when the molecule occupies $\ket{2}$ but blockaded when it occupies $\ket{3}$. 
Ideally, this implements a controlled-NOT operation and prepares the Bell state
\begin{equation*}
\ket{\Psi} = \frac{1}{\sqrt{2}}\left(\ket{2;83\mathrm{d}} - i\ket{3;5\mathrm{s}}\right),
\end{equation*}
which is a maximally entangled state of the atom and molecule.

Full state tomography would require single-particle rotations that overcome the blockade condition ($\hbar\Omega_\mathrm{Ryd} \gtrsim |U(R)|$), which are not presently available with our Rydberg-excitation system (see Methods). Instead, we probe the coherence generated during Bell-state preparation through coherent phase transfer between the atom and molecule.

Starting from $\ket{\Psi}$, we apply a second atomic $\pi$ pulse with controllable phase $\phi$ relative to the first pulse (we take the phase of the first pulse to be zero).
This maps the state $\ket{\Psi}$ to $\frac{1}{\sqrt{2}}\left(e^{i\phi}\ket{2;5\mathrm{s}} - i\ket{3;5\mathrm{s}}\right)$,
transferring the phase of the atomic drive onto the molecular superposition. A molecular $\pi/2$ pulse then converts this phase into populations that are measured via state-selective molecular readout. 
Then varying the phase $\phi$ we expect to observe Ramsey-type oscillations in the molecular populations.

The results of this measurement are shown in the central panel of Fig.~\ref{fig:entanglement}c.
We postselect on the recovery of both the atom and molecule, excluding the 22\% of experimental runs in which imperfect transfer on the Rydberg transition resulted in atom loss.
The measured fringe contrast is $0.57(1)$.
Observation of these oscillations demonstrates coherent transfer of phase information between the atom and molecule.
This requires coherence between the two components of the Bell state.
Therefore, the measured contrast provides a measurement of the Bell-state coherence.
From this, we estimate a SPAM-corrected entanglement fidelity of $0.77(3)$ and an uncorrected fidelity of $0.52(2)$, most likely limited  by imperfections in the atomic and molecular excitations and decoherence during the sequence (see Methods). The grey shaded region indicates the contrast obtained when the same pulse sequence is applied with the Rydberg-excitation light detuned from resonance, and highlights the fidelity of the molecular microwave transfers.

To verify that the observed fringes originate from the state-dependent blockade, we repeat the experiment with the particles initially prepared in the superposition $\frac{1}{\sqrt{2}}(\ket{1;5\mathrm{s}} - i\ket{2;5\mathrm{s}})$, for which neither molecular component is resonantly coupled to the atom. In this case the atomic pulse sequence imparts only a global phase,
$e^{i\phi}\frac{1}{\sqrt{2}}(\ket{1;5\mathrm{s}} - i\ket{2;5\mathrm{s}})$,
which cannot affect any observable. As shown in the lower panel of Fig.~\ref{fig:entanglement}c, no phase-dependent oscillations are observed. 
The observation of fringes only in the resonant case confirms that the measured coherence originates from the interaction-mediated atom--molecule entangled state.

\subsection*{Outlook} 
We have established a coherent quantum interface between an individual neutral atom and an individual polar molecule by engineering resonant dipolar interactions in species-specific optical tweezers. 
Exploiting this interface, we have, to our knowledge, demonstrated three capabilities previously unrealised between neutral atoms and polar molecules: atom-mediated readout of a molecular state, atom--molecule spin exchange, and the generation of entanglement between the particles.

The performance of this interface can be rapidly advanced by addressing its current bottleneck: the fidelity of the atomic operations. State-of-the-art neutral atom processors routinely achieve infidelities approaching $0.1 \%$~\cite{Evered2026}, two orders of magnitude lower than those reported here. 
Adopting these established capabilities, including encoding the atomic qubit in the ground hyperfine manifold, will significantly improve the performance. 
Crucially, implementing the ground-state encoding will facilitate mid-circuit measurement protocols where the atomic qubit can be read out, reset, and reused~\cite{Finkelstein2024} for repetitive interrogation of the molecular state. 
Finally, implementing quantum optimal control promises to further enhance the robustness of the atom--molecule blockade.

Ultimately, the atom--molecule interface demonstrated here establishes the experimental foundation for a new class of hybrid quantum architectures.
It provides a route towards hybrid quantum processors in which atoms perform fast quantum logic while molecules provide long-lived quantum memories with dense storage capabilities~\cite{Cornish2024,Sawant2020}. 
The atom-mediated readout can be extended to multiplexed measurements of molecular states, enabling scalable probing of molecular quantum registers and quantum simulators~\cite{Baranov2012}. Furthermore, the strong atom--molecule interactions central to the interface enable fast molecule-molecule quantum gates mediated by Rydberg atoms~\cite{Kuznetsova2011,Zhang2022,Wang2022}. Such gates would lay the foundation for macroscopic molecular entanglement~\cite{Zhang2026}, opening new opportunities for quantum-enhanced metrology and precision sensing~\cite{DeMille2024,Zhang2023DecoherenceFree,Ruttley2025}.
Together, these capabilities open new avenues for quantum science, where the complementary strengths of atomic and molecular systems can be combined to create quantum architectures beyond the reach of either platform alone.

% \bibliography{ref}

\section*{Acknowledgments}
We thank K.~Wadenpfuhl, C.~S.~Adams, and J.~D.~Pritchard for helpful discussions on experimental matters, M.~Bergonzoni and G.~Pupillo for discussions on molecular gates mediated by atoms, and H.~R.~Sadeghpour and A.~M.~Rey for theoretical discussions.
\subsection*{Funding}
We acknowledge support from the UK Engineering and Physical Sciences Research Council (EPSRC) Grants EP/P01058X/1, EP/W00299X/1, EP/W016141/1, EP/Y01510X/1 and UKRI2226 funded through the Programme Grant Scheme, UK Research and Innovation (UKRI) Frontier Research Grant EP/X023354/1, the Royal Society, and Durham University.
R.G.F. and J.M.G.G. gratefully acknowledge financial support by the Spanish project PID2023-147039NB-I00. 
J.M.G.G. acknowledges financial
support from the grant PRE2021-099603 funded by MICIU/
AEI/10.13039/501100011033 and by ‘‘ESF+’’.
\section*{Author contributions}
DKR, TRH, and CJHR performed the experiments. JMGC, RGF, and TRH performed the theoretical calculations. DKR, AG, and SLC conceptualised the experiments. All authors contributed to the analysis of the results. AG and SLC acquired funding for the experimental apparatus. DKR wrote the initial draft of the manuscript and all authors reviewed and edited it.  RGF supervised the theoretical work. SLC supervised the overall project.
\section*{Competing interests}
There are no competing interests to declare.
\section*{Data, code and materials availability}
The data that support the findings of this study are available at [link to be inserted].

\clearpage

\renewcommand{\thefigure}{S\arabic{figure}}
\renewcommand{\thetable}{S\arabic{table}}
\renewcommand{\theequation}{S\arabic{equation}}
\renewcommand{\thepage}{S\arabic{page}}
\setcounter{figure}{0}
\setcounter{table}{0}
\setcounter{equation}{0}
\setcounter{page}{1}

\onecolumngrid
\begin{center}
    {\large\bfseries
    Supplementary Materials for\\A coherent quantum interface between a neutral atom and a polar molecule\par}
\end{center}
\twocolumngrid

\section*{Materials and Methods}
\subsection*{Atomic and molecular states}
Throughout this work, we use shorthand notation for the atomic and molecular states of interest. Here, we define the full state labels used for both species.

\subsubsection*{Atomic states} The Rb atoms are initially prepared in the hyperfine state $\ket{5\mathrm{s}} \equiv \ket{5\mathrm{s}_{1/2},f=2,m_f=2}$.
Here, the electronic manifold of the valence electron is labelled by the quantum numbers $\ket{n l_j}$, where $n$ is the principal quantum number, $l$ is the orbital angular momentum, and $j$ is the total angular momentum.  
This state is populated by optical pumping at low magnetic field ($4.78\,\mathrm{G}$) (Ref.~\cite{Spence22}) after the ground-state molecule has been formed.

We excite the atom to the Rydberg state $\ket{83\mathrm{d}} \equiv \ket{83\mathrm{d}_{5/2},m_j=5/2}$ by driving two $\sigma^+$ transitions from $\ket{5\mathrm{s}}$.
At the high magnetic fields that we use in this work, the Rydberg states are in the diamagnetic regime, with GHz-scale Zeeman shifts (see below).
We do not resolve their hyperfine structure because the relevant energy scale ($\sim100\,\mathrm{kHz}\times h$, see Ref.~\cite{Tauschinsky2013}) is smaller than the excitation linewidth.
The Rydberg excitation occurs via a virtual level which is blue-detuned from the intermediate state $\ket{6\mathrm{p}} \equiv \ket{6\mathrm{p}_{3/2},m_j=3/2}$ (see Ref.~\cite{Guttridge2023} and below).

To engineer resonant atom--molecule interactions, we use the atomic transition $\ket{83\mathrm{d}} \rightarrow \ket{84\mathrm{p}}$, where the state $\ket{84\mathrm{p}} \equiv \ket{84\mathrm{p}_{3/2},m_j=3/2}$.

\subsubsection*{Molecular states} We prepare RbCs molecules in the rovibrational and hyperfine ground state $\ket{0} \equiv \ket{N=0,M_N=0,m_\mathrm{Rb}=3/2,m_\mathrm{Cs}=7/2}$.
Here, $N$ is the rotational quantum number and $M_N$ is its projection, while $m_a$ denotes the nuclear-spin projection of atom $a$.

To engineer atom--molecule resonance, we rotationally excite the molecule with resonant microwave fields~\cite{Ruttley2024,Ruttley2025,Hepworth2025}.
Electric-dipole selection rules permit transitions with $\Delta N=\pm1$ and $\Delta M_N=0,\pm1$, while the nuclear-spin projections remain unchanged.
Throughout this work, we drive only $\sigma^+$ transitions by using 10-kHz-scale Rabi frequencies which are small enough to prevent significant off-resonant excitation to other hyperfine states.
Therefore, the molecules remain within the stretched-state manifold $\ket{N} \equiv \ket{N,M_N=N,m_\mathrm{Rb}=3/2,m_\mathrm{Cs}=7/2}$.
The energies of the rotational levels are $E_N\approx B_v N(N+1)$ with $B_v/h=490\,$MHz~\cite{Gregory2016}.
We make use of the excited rotational states $\ket{1}$, $\ket{2}$, $\ket{3}$, and $\ket{4}$.

\subsection*{Theoretical calculations}
The hybrid Rydberg atom--molecule system is described by the Hamiltonian~\cite{Rittenhouse2010,gonzalez15}
\begin{equation}
    H = H_\mathrm{A} + H_\mathrm{M} + H_\mathrm{cd}.
    \label{eq:H-full}
\end{equation}
Here, $H_\mathrm{A}$ and $H_\mathrm{M}$ describe the internal structure of the atom and molecule, including their response to external electric and magnetic fields. 
$H_\mathrm{cd}$ accounts for the charge--dipole interaction between the molecular electric dipole moment $\boldsymbol{d}_\mathrm{M}$ and the electric field produced by the Rb ion core and the valence electron. It has the form
\begin{equation}
    H_\mathrm{cd}= -\frac{e}{4\pi\epsilon_0} \left(\boldsymbol{d}_{\mathrm{M}}\cdot \left[
\frac{\boldsymbol{R}}{|\boldsymbol{R}|^3}-\frac{\boldsymbol{R}-\boldsymbol{r}}{|\boldsymbol{R}-\boldsymbol{r}|^3}\right]\right),
\end{equation}
where $\boldsymbol r$ and $\boldsymbol R$ are measured from the Rb ionic core and denote, respectively, the Rydberg-electron position operator and the molecule position.
Within the Born--Oppenheimer approximation, the atom--molecule separation $\boldsymbol R$ is treated as a fixed parameter, and by solving the time-independent Schrödinger equation associated with Hamiltonian~\eqref{eq:H-full} we obtain the Born--Oppenheimer potentials (BOP) $U(R)$. For each pair state, we reference each 
potential curve to the corresponding uncoupled pair-state energy at $R\rightarrow\infty$.

\begin{figure}
\includegraphics[width=\hsize]{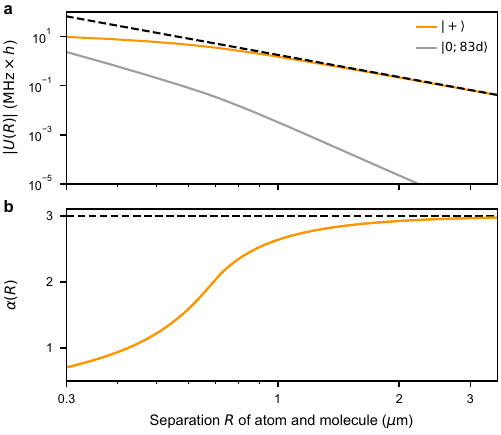}
\caption{\textbf{Comparison between charge--dipole and dipole--dipole interactions.}
\textbf{a}, Magnitudes $|U(R)|$ of the interaction-induced energy shifts for the states $\ket{+}$ (orange) and $\ket{0;83\mathrm{d}}$ (grey), calculated using the full charge--dipole Hamiltonian.
\textbf{b}, Local power-law exponent $\alpha(R)$ for the state $\ket{+}$, extracted pointwise from the curve in \textbf{a}.
In both panels, the black dashed line shows the behaviour expected from the point-dipole model with $C_3/h = 1.79\,\mathrm{MHz}\,\mu\mathrm{m}^3$. At short distances, the dipole--dipole approximation breaks down due to the finite spatial extent of the Rydberg electron, requiring the full charge--dipole description.
\label{fig:theory_methods}}
\end{figure}

The charge-resolved nature of the interaction Hamiltonian is particularly relevant
when the atomic core--molecule separation $R$ becomes comparable to the spatial extent of the Rydberg-electron wavefunction.
 In this regime, the full charge--dipole interaction is required to accurately describe this hybrid system as illustrated in Fig.~\ref{fig:theory_methods}.
Panel a shows the BOP of the state $\ket{+}$ (orange line), as in Fig.~\ref{fig:overview}e, together with the BOP predicted by a point-dipole model with dipolar coefficient $C_3/h = 1.79\,\mathrm{MHz}\,\mu\mathrm{m}^3$ (black dashed line). Deviations between the two curves become apparent at short range, {$R\lesssim1\,\mu \mathrm{m}$}, where the {point-dipole description breaks down as the molecule probes the extended Rydberg-electron charge distribution}.
To characterise this crossover quantitatively, we extract a local effective power-law exponent and coefficient from the calculated BOPs with
 \begin{equation}
\alpha(R) = -\frac{\mathrm{d}\log |U(R)|}{\mathrm{d}\log R},
\qquad
C_\alpha(R)=U(R)R^{\alpha(R)}.
\end{equation}
In panel b, we show the behaviour of $\alpha(R)$. It deviates from the ideal point dipole--dipole interaction value of 3 as $R$ decreases and the molecule probes the extended Rydberg-electron charge distribution.
The coefficient $C_\alpha(R)/h$ varies from $1.51\,\mathrm{MHz}\,\mu\mathrm{m}^{\alpha(R)}$ to $2.59\,\mathrm{MHz}\,\mu\mathrm{m}^{\alpha(R)}$ over the experimentally-relevant range $R>0.5\,\mu\mathrm{m}$, and tends to dipole--dipole limit of $1.79\,\mathrm{MHz}\,\mu\mathrm{m}^3$ as $R\to\infty$.

For comparison to the non-resonant case, in Fig.~\ref{fig:theory_methods}a, we also plot $|U(R)|$ for the state $\ket{0;83\mathrm{d}}$ (grey line).
Over the range shown here, the interaction between the molecule and atom is much weaker, and, as a consequence, the energy shift of this potential compared to its asymptotic limit is significantly smaller.

For the resonant states {at} experimental atom--molecule separations {$R\gtrsim1\,\mu \mathrm{m}$}, 
the dipole--dipole interaction provides a good
approximation to the charge-dipole one.
When $R$ is larger than the characteristic spatial extent of the Rydberg-electron wavefunction, the instantaneous electric field generated by the Rydberg atom at $\boldsymbol{R}$ is well approximated by that of a point dipole, and the interaction can be expressed in terms of effective transition dipole moments.
In the dipole--dipole limit, the interaction between the atom and molecule is~\cite{Wall2015}
\begin{equation}
    V_\mathrm{dd}(\boldsymbol{R}) =
    \frac{1}{4\pi\epsilon_0 R^3}
    \left[
    \boldsymbol{d}_\mathrm{A}\cdot\boldsymbol{d}_\mathrm{M}
    - 3(\boldsymbol{d}_\mathrm{A}\cdot{\boldsymbol{e}_R})(\boldsymbol{d}_\mathrm{M}\cdot{\boldsymbol{e}_R})
    \right],
\end{equation}
where $\boldsymbol{d}_\mathrm{A}$ and $\boldsymbol{d}_\mathrm{M}$ are the dipole operators of the atom and molecule, and $\boldsymbol{e}_R\equiv \boldsymbol{R}/|\boldsymbol{R}|$ is the interparticle unit vector.

In spherical coordinates, with the z-axis aligned along the quantisation axis, this can be written as~\cite{Ravets2015,Wadenpfuhl2025}
\begin{equation}
    V_\mathrm{dd}(R,\theta,\varphi) =
    \frac{1}{4\pi\epsilon_0 R^3}
    \left(\mathcal{M}_0+\mathcal{M}_1+\mathcal{M}_2\right),
\end{equation}
where 
$\theta$ and $\varphi$ are the polar and azimuthal angles
of $\boldsymbol{R}$ and 
\begin{align}
\mathcal{M}_0 &= (1-3\cos^2\theta)\,
\left[d_\mathrm{M}^z d_\mathrm{A}^z + \tfrac{1}{2}(d_\mathrm{M}^+d_\mathrm{A}^- + d_\mathrm{M}^- d_\mathrm{A}^+) \right], \\
\mathcal{M}_1 &= -\frac{3}{\sqrt{2}}\sin\theta \cos\theta
\begin{aligned}[t]
&\Big[ e^{i\varphi}(d_\mathrm{M}^z d_\mathrm{A}^- + d_\mathrm{M}^- d_\mathrm{A}^z) \\
&\quad - e^{-i\varphi}(d_\mathrm{M}^z d_\mathrm{A}^+ + d_\mathrm{M}^+ d_\mathrm{A}^z)\Big],
\end{aligned} \\
\mathcal{M}_2 &= -\frac{3}{2}\sin^2\theta
\left[e^{2i\varphi} d_\mathrm{M}^- d_\mathrm{A}^- + e^{-2i\varphi} d_\mathrm{M}^+ d_\mathrm{A}^+\right],
\end{align}
where $d^\pm \equiv \mp (d^x \pm id^y)/\sqrt{2}$ are the spherical components of the dipole operator, and $d^x$, $d^y$, and $d^z$ are its Cartesian components.
The operators  $\mathcal{M}_0$, $\mathcal{M}_1$, and $\mathcal{M}_2$ couple states with  total angular-momentum projections $M_F+m_j$ differing by $0$, $\pm1$, and $\pm2$, respectively. 
Here, $M_F \equiv M_N + m_\mathrm{Rb} + m_\mathrm{Cs}$ is the projection of the total molecular angular momentum. 

In this work, the interparticle axis is aligned with the quantisation axis (set by the magnetic field), as shown in Fig.~\ref{fig:overview}a, so that $\theta=\varphi=0$. 
In this geometry, $\mathcal{M}_1$ and $\mathcal{M}_2$ vanish, and the interaction conserves the total projection $M_F+m_j$.
The stretched states which are used in the experiment are
$\ket{3;83\mathrm{d}}$ (with $j=m_j=5/2$ and $M_N=3$) and $\ket{4;84\mathrm{p}}$ (with
$j=m_j=3/2$ and $M_N=4$). 
These are coupled through the term
$d_\mathrm{M}^+ d_\mathrm{A}^-$ in $\mathcal{M}_0$, and the matrix element of the 
dipole--dipole interaction reads
\begin{align}
    V_\mathrm{dd}(R)
    &\equiv
    \braket{4;84\mathrm{p}|V_\mathrm{dd}(R,0,0)|3;83\mathrm{d}} \nonumber\\
    &=
    -\frac{\braket{4|d_\mathrm{M}^+|3}
    \braket{84\mathrm{p}|d_\mathrm{A}^-|83\mathrm{d}}}{4\pi\epsilon_0 R^3},
\end{align}
which we write as $C_3/R^3 \equiv -d_\mathrm{A}d_\mathrm{M}/(4\pi\epsilon_0R^3)$ in the main text (likewise for $\braket{3;83\mathrm{d}|V_\mathrm{dd}(R,0,0)|4;84\mathrm{p}}$).
Then, the interaction between the resonant states can be described by the dipole--dipole Hamiltonian \cite{Wu2021}
\begin{equation}
    H_\mathrm{dd} = \begin{pmatrix}
    0 & C_3/R^3 \\
    C_3/R^3 & h\delta
    \end{pmatrix},
\end{equation}
in the basis $\{\ket{3;83\mathrm{d}},\ket{4;84\mathrm{p}}\}$ (all other states are effectively decoupled).
At resonance ($\delta\to0$), the eigenstates of this Hamiltonian are $\ket{\pm} = \frac{1}{\sqrt{2}}(\ket{3;83\mathrm{d}} \pm \ket{4;84\mathrm{p}})$ with energies $U(R) = \pm  C_3/R^3$.

If the interparticle axis were not exactly aligned with the quantisation axis,
the matrix elements of $\mathcal{M}_1$ and $\mathcal{M}_2$ would no longer vanish 
and 
$M_F+m_j$ would not be conserved. 
In this regime, pair states involving different molecular hyperfine states could couple to our chosen pair states~\cite{garciagarrido2026Unpublished}.
These pair states are numerous and closely spaced in energy, with detunings on the kHz scale, and their admixture of $\ket{83\mathrm{d}}$ character makes them optically accessible with our laser system.
 
We do not measure population in these states experimentally as we readout only the populations of the stretched molecular states, and so transfer to the non-stretched states appears as a lack of molecule recovery which is removed in our postselection routine (see below).
We attribute the absence of resolvable interaction-shifted features to coupling to these states, which can be caused by technical imperfections in alignment of the interparticle axis ($\theta \neq 0$), shot-to-shot fluctuations in the relative geometry, and the finite wavefunction spread of the particles.
We note, however, that these couplings do not constitute a fundamental limitation for blockade-based gates, since these rely only on the presence of an interaction shift exceeding the excitation linewidth ($|U(R)
|\gtrsim \hbar\Omega_\mathrm{Ryd}$), rather than isolation of individual hyperfine channels.

\subsection*{Experimental apparatus}
Our experimental apparatus has been described in previous publications~\cite{Ruttley2024}. 
The platform enables the preparation, control, and detection of single atoms and molecules trapped in an ultrahigh-vacuum (UHV) environment with species-specific optical tweezers.

The experiments presented here begin with two Rb atoms and one Cs atom, each trapped in species-specific optical tweezers~\cite{Brooks21}. 
The optical tweezers are formed by a high numerical-aperture objective lens mounted below a UHV glass cell.
The atoms are cooled to the motional ground state via Raman sideband cooling~\cite{Spence22} and rearranged using tweezer transport to form a co-trapped Rb--Cs atom pair and a single Rb atom.
The atom pair is then magnetoassociated into a weakly bound molecule~\cite{Ruttley2023}, which is subsequently transferred to the rovibrational ground state $\ket{0}$ using a two-photon STIRAP sequence~\cite{Guttridge2023}. 
After molecular preparation, we reconfigure the optical tweezers to set a controlled interparticle separation $R$ between the RbCs molecule and the Rb atom.
The Rb atom is then excited to the Rydberg manifold using a two-photon excitation scheme (see Ref.~\cite{Guttridge2023} and below). 
At the end of each experimental sequence, we dissociate the molecule and map its internal states onto distinct atomic configurations~\cite{Ruttley2024}. 
Final detection is performed via fluorescence imaging of the tweezer occupations.

Below, we provide further details of techniques and experimental upgrades which are critical for the measurements presented in this work.

\subsubsection*{Control of the interparticle distance $R$}
To precisely control the distance between the particles, we use species-specific optical tweezers.
The molecule is primarily trapped by a tweezer using light at a wavelength of 1066\,nm (the ``molecule tweezer'') and the atom is primarily trapped by a tweezer using light at a wavelength of 817\,nm (the ``atom tweezer'')~\cite{Guttridge2023}. The positions of the tweezers are controlled by optical elements in their beam path prior to the objective lens which forms the traps and images the atoms.
Specifically, the position of the molecule tweezer is set using a spatial light modulator and remains fixed throughout experimental sequences.
The position of the atom tweezer (and the other 817\nobreakdash-nm tweezers used in experiments) is controlled by an acousto-optic deflector (AOD).
By changing the frequencies of the radio-frequency (RF) tones applied to the AOD, we can dynamically and precisely translate the 817\nobreakdash-nm tweezers.
We calibrate the relationship between RF frequency and tweezer position using fluorescence images of individual Rb atoms recorded at different AOD frequencies, and have independently verified this calibration by measuring the interaction shift between two Rydberg atoms, as described in Ref.~\cite{Guttridge2023}.

\begin{figure}
\includegraphics[width=\hsize]{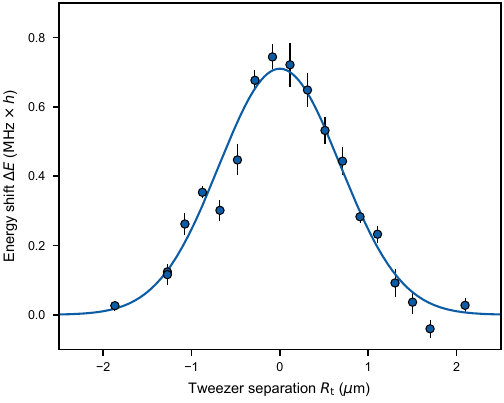}
\caption{\textbf{Calibration of the relative tweezer positions.} We show the measured frequency of the atomic transition $\ket{5\mathrm{s}}\rightarrow\ket{83\mathrm{d}}$ as a function of the separation $R_\mathrm{t}$ between the centres of the atom and molecule tweezers. The molecule tweezer induces a differential ac Stark shift of the transition, with a magnitude that depends on the local intensity experienced by the atom. The transition frequency is maximised when the tweezers are spatially overlapped ($R_\mathrm{t}=0$).
The solid line is a Gaussian fit used to determine the relative radial alignment of the tweezers. 
The error bars show the $1\sigma$ confidence intervals and, on average, there are 432 experimental shots per data point.
\label{fig:trap-overlap}}
\end{figure}

We calibrate the relative position of the atom tweezer and the molecule tweezer by measuring the differential ac Stark shift $\Delta E$ of the atomic transition $\ket{5\mathrm{s}}\rightarrow\ket{83\mathrm{d}}$ induced by the molecule tweezer. 
An example measurement is shown in Fig.~\ref{fig:trap-overlap}. 
A Rb atom is initially prepared in the atom tweezer, which has an intensity of $63\,\mathrm{kW/cm^2}$. 
The atom tweezer is then translated towards the molecule tweezer (intensity $18\,\mathrm{kW/cm^2}$), held at a separation $R_\mathrm{t}$, and the atomic transition frequency is measured.
When the tweezers are overlapped ($R_\mathrm{t}=0$), the transition frequency is maximised because the atom experiences the largest differential ac Stark shift from the molecule tweezer. 
This method is sensitive to the radial overlap of the tweezers but relatively insensitive to overlap along the propagation direction because the tweezers have micron-scale Rayleigh ranges. 
Along this axis, we optimise the overlap by maximising the contrast of the atom--molecule spin-exchange signal. 
With our postselection scheme, this contrast is maximised when the interparticle vector is aligned with the quantisation axis, because couplings to other molecular hyperfine states (which are not detected) are minimised~\cite{garciagarrido2026Unpublished}.

As in Ref.~\cite{Guttridge2023}, we determine the interparticle separation $R$ by modelling the three-dimensional trapping potential experienced by each particle. For the atom, we use the polarisabilities $\alpha^\mathrm{Rb}_{1066} = 687\times4\pi\varepsilon_0 a_0^3$~\cite{UDportal} and $\alpha^\mathrm{Rb}_{817} = 4307\times4\pi\varepsilon_0 a_0^3$~\cite{UDportal}. For the molecule, the polarisability of the stretched state $\ket{N}$ is
\begin{equation}
    \alpha^\mathrm{RbCs}_\lambda(N) = a^{(0)}_\lambda - \frac{N}{2N+3} a^{(2)}_\lambda,
\end{equation}
where $a^{(0)}_\lambda$ and $a^{(2)}_\lambda$ are the isotropic and anisotropic polarisabilities at wavelength $\lambda$. 
For the molecule tweezer, we take $a^{(0)}_{1066} = 2.0\times10^3\times4\pi\varepsilon_0 a_0^3$~\cite{Ruttley2024} and $a^{(2)}_{1066}=1980\times4\pi\varepsilon_0 a_0^3$~\cite{Ruttley2024}. For the atom tweezer, $a^{(0)}_{817} = 4.0\times10^2\times4\pi\varepsilon_0 a_0^3$~\cite{Guttridge2023} and $a^{(2)}_{817}=-2814\times4\pi\varepsilon_0 a_0^3$~\cite{Ruttley2024}.
At large separations, the interparticle separation $R$ is approximately equal to the separation $R_\mathrm{t}$ between the tweezer centres. 
However, at small $R_\mathrm{t}$, the particles experience attractive forces from both tweezers.
This means that they move closer together, such that $R<R_\mathrm{t}$~\cite{Guttridge2023}.

\subsubsection*{Molecule detection}
Direct fluorescence imaging of RbCs molecules is not  possible because there are no closed optical cycling transitions suitable for repeated photon scattering.
Instead, we detect molecules by dissociating them into their constituent atoms and subsequently imaging the resulting atoms.

The dissociation process is state-selective and only molecules in the rotational ground state $\ket{0}$ are dissociated. This enables state-resolved molecular detection.
To measure the population in a rotational state $\ket{N}$, we first transfer molecules in $\ket{N}$ to $\ket{0}$ and then apply the dissociation sequence.
For example, to distinguish populations in $\ket{2}$ and $\ket{3}$, molecules in $\ket{2}$ are first transferred to $\ket{0}$, dissociated, and the resulting atoms moved to a designated storage location within the tweezer array.
Molecules in $\ket{3}$ are then transferred to $\ket{0}$ and detected in an analogous manner.
The resulting atoms from each detection step are rearranged into distinct locations within the tweezer array.
In this way, the rotational state of the molecule is mapped onto the final positions of the dissociated atoms, allowing the populations of multiple molecular states to be determined from a single fluorescence image.
A detailed description of the detection protocol is given in Ref.~\cite{Ruttley2024}.

In this work, we use this technique to determine both the presence of a molecule and its rotational state.
First, we identify experimental shots in which a molecule was not formed ($\approx 50\%$) by storing the Rb atom from the atom pair in a dedicated ``detection tweezer''~\cite{Guttridge2023}. 
Subsequently, the Cs atom is removed using resonant light, ensuring that the molecule tweezer is empty.
This procedure provides a reference data set corresponding to experimental runs in which no molecule was present, such as the grey data shown in Fig.~\ref{fig:blockade}.
For experimental shots in which a molecule is successfully formed, this detection step does not affect the molecule.
We then apply postselection to discard runs in which no atoms are recovered following the molecular readout sequence.
Such events correspond primarily to molecule loss during the experimental sequence.
These loss processes are approximately independent of the molecular rotational state~\cite{Ruttley2024,Ruttley2025} and therefore do not significantly affect the relative state populations that are the focus of this work.

\subsubsection*{Rydberg excitation}
We excite the atoms to the Rydberg state $\ket{83\mathrm{d}}$ using the inverted two-photon scheme described in Ref.~\cite{Guttridge2023}. As illustrated in Fig.~\ref{fig:overview}c, we drive the $\sigma^{+}$ transition from the state $\ket{\mathrm{5s}}$ to a virtual level detuned by approximately $1.6\,$GHz from $\ket{6\mathrm{p}}$ using light at a wavelength of 420\,nm. The 420-nm beam propagates perpendicular to the quantisation axis and is linearly polarised orthogonal to it. We couple this virtual level to $\ket{83\mathrm{d}}$ using light at a wavelength of 1012\,nm from a beam that propagates approximately parallel to the quantisation axis and is approximately circularly polarised (see below). The beam waists are $52(6)\,\mu$m and $35(2)\,\mu$m for the 420-nm and 1012\nobreakdash-nm beams, respectively.

This two-photon excitation scheme allows us to access Rydberg states with orbital angular momentum $l=0$ ($\mathrm{s}$) and $l=2$ ($\mathrm{d}$). When identifying resonances, we therefore consider electric-dipole-allowed transitions originating from these states. The corresponding transition energies are shown in Fig.~\ref{fig:overview}b, with $\mathrm{s}\rightarrow\mathrm{p}$ transitions shown in red, $\mathrm{d}\rightarrow\mathrm{p}$ transitions in blue, and $\mathrm{d}\rightarrow\mathrm{f}$ transitions in purple. Although electric-dipole selection rules place no restriction on $\Delta n$, only a narrow range of $\Delta n$ values yields atomic transition frequencies comparable to the molecular transition frequencies relevant here.

Our polarisation configuration could in principle also address Rydberg states with $m_j=1/2$. However, the transitions to these states are significantly detuned from the transition $\ket{\mathrm{5s}}\rightarrow\ket{83\mathrm{d}}$ at the magnetic fields used in this work (see below) and are therefore not appreciably driven.

To drive the Rydberg transition, we first switch on the 1012\nobreakdash-nm light and then apply a pulse of 420-nm light using an acousto-optic modulator (AOM). 
The 1012\nobreakdash-nm light is switched off after the pulse, such that the duration of the 420-nm pulse defines the duration with which we drive the Rydberg transition.

\subsubsection*{Molecule loss caused by Rydberg-excitation light}
We find that the 1012\nobreakdash-nm light can strongly reduce the probability of recovering a molecule when its intensity is too high. We attribute this effect to light-induced avoided crossings between hyperfine states within the molecular rotational manifolds. The applied light produces differential ac Stark shifts between hyperfine states and, when its polarisation is not purely linear along the quantisation axis, can couple states with different values of $M_F \equiv M_N + m_\mathrm{Rb} + m_\mathrm{Cs}$~\cite{Blackmore2020}.

\begin{figure}
\includegraphics[width=\hsize]{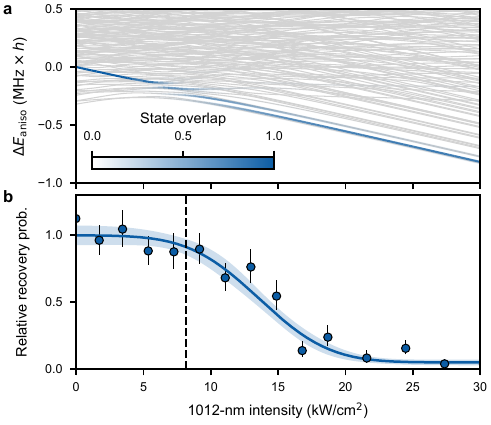}
\caption{\textbf{Molecular loss induced by the 1012-nm Rydberg-excitation light.}
\textbf{a}, Calculated energy shifts $\Delta E_\mathrm{aniso}$ due to the anisotropic part of the molecular polarisability in the $N=3$ rotational manifold as a function of the 1012-nm intensity. Colours indicate the overlap with the state $\ket{3}$ at zero intensity. A dense network of avoided crossings emerges near $8\,\mathrm{kW/cm^{2}}$.
\textbf{b}, Measured relative probability of recovering a molecule prepared in $\ket{3}$ as a function of the 1012-nm intensity. The reduction in recovery probability coincides with the onset of the avoided crossings shown in \textbf{a}, consistent with population transfer into hyperfine states that are not detected by the imaging sequence.
The solid line is a fit to the error function; the shaded region shows the $1\sigma$ confidence interval of the fit.
The dashed line corresponds to the 1012-nm power that we typically use. 
The error bars show the $1\sigma$ confidence intervals and, on average, there are 138 experimental shots per data point.
\label{fig:1012_loss}}
\end{figure}

To model this effect, we use the Python package \texttt{diatomic-py}~\cite{blackmore2023}, extended to include optical fields with arbitrary polarisation. The polarisation of the 1012\nobreakdash-nm beam is measured before the vacuum chamber using a polarimeter (Thorlabs PAX1000IR1/M). The beam propagates at an angle of $7^\circ$ to the quantisation axis and is approximately left-handed circularly polarised, with azimuthal angle $28^\circ$ and ellipticity $-39^\circ$. Fig.~\ref{fig:1012_loss}a shows the calculated energy shifts $\Delta E_\mathrm{aniso}$ of the hyperfine states in the $N=3$ manifold due to the anisotropic part of the molecular polarisability as a function of the 1012\nobreakdash-nm intensity.
The isotropic shift common to all states has been removed. The calculations assume $a^{(2)}_{1012}=3.3\times10^3\times4\pi\varepsilon_0 a_0^3$~\cite{Vexiau2017}$,$ and a trapping intensity of $3.5\,\mathrm{kW}/\mathrm{cm}^{2}$ from the molecular tweezer. The colour scale indicates the overlap of each eigenstate with the state $\ket{3}$ in the absence of the 1012\nobreakdash-nm light. A dense network of avoided crossings appears when the 1012\nobreakdash-nm intensity reaches approximately $8\,\mathrm{kW}/\mathrm{cm}^{2}$. At this intensity, the light shift of the state $\ket{3}$ caused by the anisotropic polarisability is approximately $-200\,\mathrm{kHz}$. 
The behaviour of the corresponding states in the $N=4$ manifold is similar, resulting in only a small differential shift of the frequency of the transition $\ket{3}\to\ket{4}$.

Fig.~\ref{fig:1012_loss}b shows the relative probability of recovering a molecule initially prepared in $\ket{3}$ after the 1012\nobreakdash-nm beam is pulsed for $532\,\mu$s at varying intensities. 
The plotted intensities are inferred from the beam power measured outside the vacuum chamber assuming perfect transmission and alignment; the actual intensity at the molecules may therefore be somewhat lower. 
We observe a sharp reduction in the recovery probability once the intensity reaches the approximate region where the avoided crossings occur. 
We attribute this behaviour to population transfer into other hyperfine states as the crossings are traversed. Because these states are not detected by our state-sensitive imaging sequence, the transfer appears experimentally as a loss of molecules.

Experimentally, these avoided crossings and the associated loss features constrain the maximum usable intensity of the 1012\nobreakdash-nm light during Rydberg excitation while maintaining molecular survival.
Typically, we use $100\,\mathrm{mW}$ of 1012\nobreakdash-nm power, corresponding to an intensity of approximately $8\,\mathrm{kW/cm^{2}}$ under ideal alignment (Fig.~\ref{fig:1012_loss}b, dashed line). 
The onset of loss can be shifted to higher intensities by increasing the depth of the molecular tweezer, which increases the tweezer-induced anisotropic light shifts and thereby separates the relevant hyperfine states further.
To estimate the probability of molecule loss induced by the 1012\nobreakdash-nm light, we fit the data in Fig.~\ref{fig:1012_loss}b with the error function (solid line). 
The shaded region indicates the $1\sigma$ confidence interval of the fit.
From this, we extract a loss probability of $9(6)\%$ at our typical operating intensity. 
This corresponds to an erasure error on the molecular qubit that is removed by postselecting on the recovery of the molecule at the end of the experimental sequence.

In future, these effects may be mitigated either by employing 1012\nobreakdash-nm light linearly polarised along the quantisation axis, which would require selecting alternative Rydberg states, or by delivering the 1012\nobreakdash-nm beam through the objective lens so that it addresses only the atoms.

\subsubsection*{Control of optical and microwave pulses}
The optical and microwave pulses used for atomic and molecular excitation are generated using a direct-digital-synthesis platform (AMD RFSoC 4x2, controlled via the \texttt{QICK} firmware~\cite{Stefanazzi2022})~\cite{Raghuram2026}. 
This system provides precise control over pulse duration, amplitude, frequency, and phase.

One output drives an external dipole antenna via an RF amplifier (Minicircuits ZVA-183-S+) to generate the microwave fields used for molecular excitation~\cite{Ruttley2024}, while the second output controls the AOM that switches the 420-nm light used for atomic Rydberg excitation~\cite{Guttridge2023}. 
This enables a well-defined relative phase between the atomic and molecular control channels, which is essential for the phase mapping presented in Fig.~\ref{fig:entanglement}c which we use to characterise the atom--molecule entanglement.

\subsection*{Rydberg spectroscopy}
\subsubsection*{Ground-to-Rydberg Zeeman spectroscopy}
\begin{figure}
\includegraphics[width=\hsize]{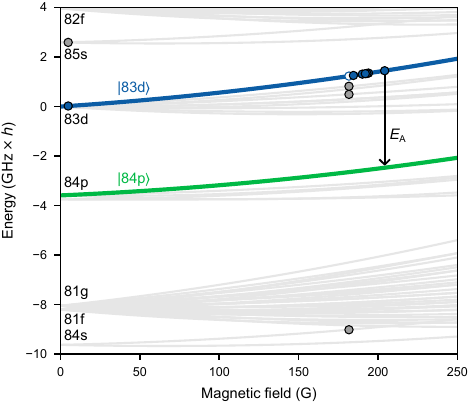}
\caption{\textbf{Effect of a magnetic field on Rydberg states with $l < 6$ around the state $\ket{83\mathrm{d}}$.} Lines show the calculated state energies and we highlight the states used in this work.
Data points show measured energies which are referenced to the blue empty point, which is set to lie on the calculated curve.
The error bars corresponding to the $1\sigma$ confidence intervals are smaller than the markers and, on average, there are 971 experimental shots per data point.
\label{fig:zeeman_diagram}}
\end{figure}

Fig.~\ref{fig:zeeman_diagram} shows the energies of Rydberg states near $\ket{83\mathrm{d}}$ as a function of magnetic field. The lines show state energies calculated using the Python package \texttt{PairInteraction}~\cite{Weber2017,Mogerle2026}. We include states with $l\leq6$ and neglect hyperfine structure. Energies are referenced to the state $\ket{83\mathrm{d}}$ at zero magnetic field and are calculated assuming zero electric field.

The interaction of the atom with an external magnetic field $\boldsymbol{B}$ is described by~\cite{Weber2017,Mogerle2026}
\begin{equation}
    H_B =
    -\boldsymbol{\mu}\cdot\boldsymbol{B}
    +
    \frac{1}{8m_\mathrm{e}}
    |\boldsymbol{d}\times\boldsymbol{B}|^2,
    \label{eq:diamagnetic}
\end{equation}
where $\boldsymbol{\mu}$ is the magnetic dipole operator, $\boldsymbol{d}$ is the electric dipole operator, and $m_\mathrm{e}$ is the electron mass.
The first term gives rise to the linear Zeeman shift, while the second describes the diamagnetic interaction.

For the Rydberg states used in this work ($n\approx83$), diamagnetic effects become significant at magnetic fields above approximately $10\,$G and therefore strongly influence the Rydberg spectrum over the magnetic-field range relevant to our experiments. It is thus essential to include this contribution when calculating $\Delta E_\mathrm{A}$ in order to identify and tune resonances.
It gives rise to a wide variation in the tuning of $\Delta E_\mathrm{A}$ that is possible for transitions with different initial $n$.
This can be seen in Fig.~\ref{fig:overview}b, where we show how $\Delta E_\mathrm{A}$ varies in the field range $150\,$G to $250\,$G.
The diamagnetic interaction can mix states with the same value of $m_j$ within a given $nl$ manifold. This further motivates our use of stretched states, which possess a unique value of $m_j$ for each $n$ and $l$. As a result, their state composition remains unchanged over the magnetic-field range that we use.

The data points in Fig.~\ref{fig:zeeman_diagram} show measured energies of Rydberg states accessible with our two-photon excitation scheme. 
Error bars are smaller than the symbol size; typical statistical uncertainties are approximately $100\,\mathrm{kHz}\times h$, obtained from fits to spectroscopic features such as those shown in Fig.~\ref{fig:blockade}b. 
When comparing measurements taken at different magnetic fields, we correct for the linear Zeeman shift of the atomic ground state~\cite{Steck_Rb}.
Additional systematic uncertainties arise from differential light shifts induced by the excitation lasers and from stray electric fields, but we estimate these to be on the order of approximately $1\,\mathrm{MHz}\times h$ which is much smaller than the energy range shown here.

To facilitate comparison with theory, all measured energies are referenced to the empty point in Fig.~\ref{fig:zeeman_diagram}, which is constrained to lie on the corresponding calculated curve. We determine energy differences rather than absolute transition frequencies because the quoted accuracy of our wavemeter (Bristol Instruments 671A-NIR) is $\pm0.2$ parts per million, corresponding to a systematic uncertainty of approximately $200\,\mathrm{MHz}$ on the two-photon transition frequency.

\subsubsection*{Ground-to-Rydberg dc Stark spectroscopy}
Our apparatus incorporates an array of electrodes~\cite{Ruttley2024} that provides control of the electric field both parallel to the magnetic field defining the quantisation axis and along one perpendicular direction. These electrodes are used to compensate stray electric fields in the corresponding directions.

\begin{figure}
\includegraphics[width=\hsize]{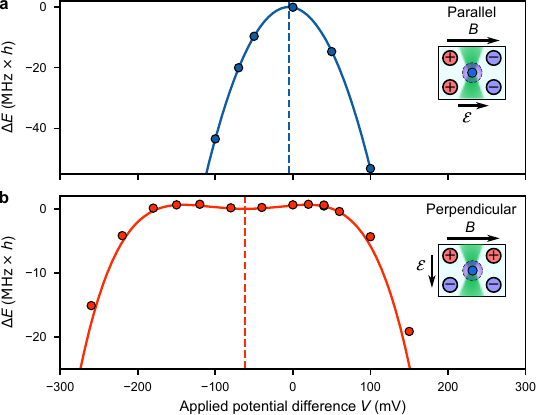}
\caption{\textbf{dc Stark spectroscopy of the transition $\ket{5\mathrm{s}}\rightarrow\ket{83\mathrm{d}}$.}
Measured Stark shifts $\Delta E$ as a function of applied electrode voltage for electric fields parallel (\textbf{a}) and perpendicular (\textbf{b}) to the quantisation axis. Solid lines show calculations using \texttt{PairInteraction}~\cite{Weber2017,Mogerle2026}. Insets show the corresponding electrode configurations. Fits to the spectra are used to calibrate the applied fields and determine the residual stray fields. Dashed lines indicate the voltages used to compensate the stray electric fields. 
The error bars corresponding to the $1\sigma$ confidence intervals are smaller than the markers and, on average, there are 951 experimental shots per data point.
\label{fig:dc-Stark}}
\end{figure}

To determine the stray electric fields, we perform dc Stark spectroscopy of the transition $\ket{5\mathrm{s}}\rightarrow\ket{83\mathrm{d}}$. We generate an electric field $\mathcal{E}$ by applying a potential difference $V$ across the electrode array, with the electrode voltages set to $\pm V/2$. The measured spectra are shown in Fig.~\ref{fig:dc-Stark} for electric fields parallel (a) and perpendicular (b) to the quantisation axis. The insets show the corresponding electrode configurations.
The solid lines show calculations performed using \texttt{PairInteraction}~\cite{Weber2017,Mogerle2026} at $B=181.7\,$G, the magnetic field used for these measurements. By fitting the measured transition frequencies to these calculations, we obtain calibration factors of $\mathcal{E}_\parallel/V_\parallel = 1.065(2)\,(\mathrm{mV}/\mathrm{cm})/\mathrm{mV}$ and $\mathcal{E}_\perp/V_\perp = 0.957(4)\,(\mathrm{mV}/\mathrm{cm})/\mathrm{mV}$. From the fitted offsets, we infer stray fields of $\mathcal{E}_\parallel=-5.4(1)\,\mathrm{mV}/\mathrm{cm}$ and $\mathcal{E}_\perp=-58.9(8)\,\mathrm{mV}/\mathrm{cm}$ when no voltages are applied to the electrodes.
For all measurements presented in the main text, we compensate these stray fields by applying the corresponding offset voltages.

We do not have independent control over the second electric-field component perpendicular to the quantisation axis. From microwave spectroscopy of the transition $\ket{83\mathrm{d}}\rightarrow\ket{84\mathrm{p}}$, we estimate the magnitude of this residual field to be approximately $12\,\mathrm{mV}/\mathrm{cm}$ (see below).

\subsubsection*{Rydberg-to-Rydberg spectroscopy}
We perform spectroscopy of the transition $\ket{83\mathrm{d}}\rightarrow\ket{84\mathrm{p}}$ to determine the atomic transition energy $\Delta E_\mathrm{A}$ and thereby identify the atom--molecule resonance condition.

The spectroscopy sequence is based on a resonant $2\pi$ pulse on the transition $\ket{5\mathrm{s}}\rightarrow\ket{83\mathrm{d}}$. 
In the absence of additional driving, this pulse returns the atom to the trapped state $\ket{5\mathrm{s}}$. 
During the middle of the pulse, however, we apply a short microwave $\pi$ pulse near resonance with the transition $\ket{83\mathrm{d}}\rightarrow\ket{84\mathrm{p}}$ using the antenna normally employed for molecular control. 
If the microwave pulse transfers population to $\ket{84\mathrm{p}}$, the atom is no longer returned to $\ket{5\mathrm{s}}$ and is subsequently lost from the tweezer. 
We therefore detect the Rydberg-to-Rydberg transition through atomic loss.

This sequence allows us to perform the Rydberg-to-Rydberg spectroscopy in the presence of the excitation light. 
This ensures that any light shifts caused by the excitation beams are included in the measured transition frequency and therefore correspond directly to the conditions under which interaction-induced blockade is observed.

\begin{figure}
\includegraphics[width=\hsize]{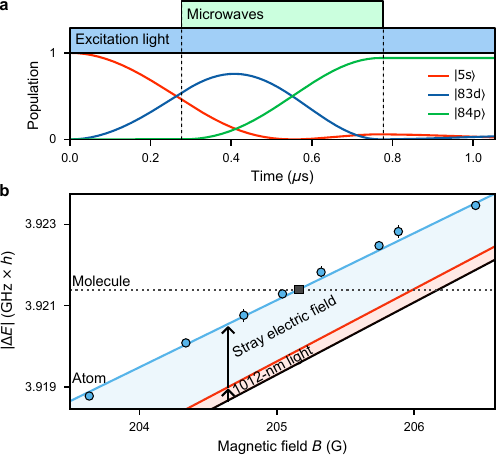}
\caption{\textbf{Spectroscopy of the transition $\ket{83\mathrm{d}}\rightarrow\ket{84\mathrm{p}}$.}
\textbf{a}, Pulse sequence used to probe the $\ket{83\mathrm{d}}\rightarrow\ket{84\mathrm{p}}$ transition, as described in the text. The panel also shows a simulation of the state evolution under resonant driving of both transitions, performed using \texttt{QuTiP}~\cite{Johansson2012,Johansson2013,Lambert2026}. \textbf{b}, Measured transition frequencies as a function of magnetic field compared with calculations from \texttt{PairInteraction}~\cite{Weber2017,Mogerle2026} (black line). The calculated resonance lies approximately $1.7\,\mathrm{MHz}$ below the observed transition. The red shaded region indicates the contribution from the light shift of $\ket{83\mathrm{d}}$ induced by the 1012-nm excitation beam. The remaining shift (blue) is attributed to a residual electric field of approximately $12\,\mathrm{mV}/\mathrm{cm}$ along the direction which we cannot control. The black point and dashed line show the energy of the molecular transition $\ket{3}\to\ket{4}$.
The error bars show the $1\sigma$ confidence intervals and, on average, there are 1214 experimental shots per data point.
\label{fig:83d-84p-spec}}
\end{figure}

For the measurements shown in Fig.~\ref{fig:overview}d, we drive the microwave transition $\ket{83\mathrm{d}}\rightarrow\ket{84\mathrm{p}}$ with a Rabi frequency of approximately $1\,\mathrm{MHz}$, while the transition $\ket{5\mathrm{s}}\rightarrow\ket{83\mathrm{d}}$ is driven with a Rabi frequency of $0.95(6)\,\mathrm{MHz}$.
In Fig.~\ref{fig:83d-84p-spec}a, we show this pulse sequence together with a simulation of the state populations under resonant driving of both transitions, calculated with the Python package \texttt{QuTiP}~\cite{Johansson2012,Johansson2013,Lambert2026}. The simulation shows near-complete transfer from $\ket{5\mathrm{s}}$ to $\ket{84\mathrm{p}}$ when both drives are on resonance.

In Fig.~\ref{fig:83d-84p-spec}b, we reproduce the data shown in Fig.~\ref{fig:overview}d and compare them with calculations from \texttt{PairInteraction}~\cite{Weber2017,Mogerle2026}. 
The theoretical calculations assume zero electric field and are shown by the black line.
The calculated transition frequencies are approximately 1.7\,MHz below the observed frequencies.
Of this offset, $0.33(2)\,$MHz (red shaded region) comes from the shift of the state $\ket{83\mathrm{d}}$ from the 1012\nobreakdash-nm light, which we have independently measured.
We attribute the remaining difference (blue shaded region) to a stray electric field in the direction that we cannot control with our electrode configuration; from calculations with \texttt{PairInteraction}~\cite{Weber2017,Mogerle2026} we estimate this field is approximately $12\,\mathrm{mV}/\mathrm{cm}$.

For comparison, the black square in Fig.~\ref{fig:overview}d (and Fig.~\ref{fig:83d-84p-spec}b) shows the energy of the transition $\ket{3}\to\ket{4}$ measured under the conditions used to engineer resonance (i.e.\ with the Rydberg-excitation light on), but with no atom present.
We determine this transition frequency using microwave spectroscopy of the molecule, as in Ref.~\cite{Ruttley2024}.
The differential Zeeman shift of the transition is negligible: using \texttt{diatomic-py}~\cite{blackmore2023}, we calculate the magnetic moments of the molecular states $\ket{3}$ and $\ket{4}$ to be $5.335\mu_\mathrm{N}$ and $5.329\mu_\mathrm{N}$, respectively.
The resulting differential magnetic moment of the transition is approximately $6\times10^{-3}\mu_\mathrm{N}=5\,\mathrm{Hz/G}\times h$, meaning that the transition frequency is effectively constant over the range of magnetic fields explored here (dashed line).

\subsection*{The $R$ dependence of blockade}
We model the dependence of the Rydberg blockade on $R$, as shown in Fig.~\ref{fig:blockade}d, using a minimal three-level model. 
For the resonant case, we use the basis $\{\ket{3;5\mathrm{s}}, \ket{3;83\mathrm{d}}, \ket{4;84\mathrm{p}}\}$ and assume that the resonance condition is exactly satisfied, such that $E_\mathrm{A}=-E_\mathrm{M}$. Under the rotating-wave approximation, the Hamiltonian describing the system is
\begin{equation}
H =
    \begin{pmatrix}
        0 & \frac{1}{2}\hbar\Omega_\mathrm{Ryd} & 0 \\
        \frac{1}{2}\hbar\Omega_\mathrm{Ryd} & 0 & V(R) \\
        0 & V(R) & 0
    \end{pmatrix},
\end{equation}
where $\Omega_\mathrm{Ryd}/2\pi=318(3)\,\mathrm{kHz}$ is the Rabi frequency with which we resonantly drive the atomic transition $\ket{5\mathrm{s}}\rightarrow\ket{83\mathrm{d}}$ and $V(R)$ is the coupling between the pair states $\ket{3;83\mathrm{d}}$ and $\ket{4;84\mathrm{p}}$.
In the dipole-dipole regime, $V(R)=V_\mathrm{dd}(R)=C_3/R^3$ (as in the main text). However, this Hamiltonian remains valid in the charge-dipole regime because the relevant pair states remain strongly hybridised~\cite{garciagarrido2026Unpublished}.
In the absence of the Rydberg-excitation light, the energies of the hybridised pair states $\ket{\pm}$ are $U(R)=\pm V(R)$, as described in the main text and shown by the orange lines in Fig.~\ref{fig:overview}e.

We initialise the system in the state $\ket{3;5\mathrm{s}}$ and allow it to evolve for a duration $\tau = \pi/\Omega_\mathrm{Ryd}$, corresponding to a $\pi$ pulse on the bare atomic transition. The state after this pulse is $e^{-iH\tau/\hbar}\ket{3;5\mathrm{s}}$.
Experimentally, Rydberg atoms are ejected from the optical tweezer before detection, such that the measured atom-survival probability is proportional to the population remaining in the state $\ket{3;5\mathrm{s}}$. 
Ideally, the population in $\ket{3;5\mathrm{s}}$ after the pulse is
\begin{equation}
\label{eq:state-pop-blockade-vs-distance}
\left|
\braket{3;5\mathrm{s}|e^{-iH\tau/\hbar}|3;5\mathrm{s}}
\right|^2
=
\frac{
\left[
1+\kappa^2
\cos\!\left(
\frac{\pi}{2\kappa}\sqrt{1+\kappa^2}
\right)
\right]^2
}
{(1+\kappa^2)^2},
\end{equation}
where $\kappa \equiv \hbar\Omega_\mathrm{Ryd}/[2V(R)]$.
We use Eq.~\eqref{eq:state-pop-blockade-vs-distance} both to generate the predicted survival probability from the theoretically calculated interaction strength $V(R)$ (Fig.~\ref{fig:blockade}d, solid blue line) and to extract an effective dipolar coefficient $C_3/h=1.36(15)\,\mathrm{MHz}\,\mu\mathrm{m}^3$ by fitting the experimental data under the assumption $V(R) = V_\mathrm{dd}(R)=C_3/R^3$ (Fig.~\ref{fig:blockade}d, dashed blue line).
The only other free parameters are scaling parameters that account for the experimental contrast.

For the experimental parameters used here, Eq.~\eqref{eq:state-pop-blockade-vs-distance} predicts oscillations in the atom-survival probability for $R \lesssim 1.4\,\mu\mathrm{m}$. 
These arise from the square excitation pulse, whose finite duration produces a $\mathrm{sinc}^2$ spectral envelope. At sufficiently small $R$, the resulting Fourier sidelobes can become resonant with the interaction-shifted transition, leading to the oscillatory features. 
Such effects could be suppressed in future experiments using shaped excitation pulses.

For the non-resonant case, we use the basis $\{\ket{0;5\mathrm{s}}, \ket{0;83\mathrm{d}}\}$ and the Hamiltonian describing the system is
\begin{equation}
H' =
    \begin{pmatrix}
        0 & \frac{1}{2}\hbar\Omega_\mathrm{Ryd}  \\
        \frac{1}{2}\hbar\Omega_\mathrm{Ryd} & U(R)
    \end{pmatrix}.
\end{equation}
Here, we include $U(R)$ directly in the Hamiltonian as an energy shift of the pair state $\ket{0;83\mathrm{d}}$. Unlike the resonant case, the off-resonant interactions do not lead to strong coupling with a single nearby pair state, and their effect is therefore well described by the energy shift given by the BOP shown in Fig.~\ref{fig:overview}e (grey line).
After the Rydberg-excitation pulse, we expect the population of the state $\ket{0;83\mathrm{d}}$ to be
\begin{equation}
\label{eq:state-pop-blockade-vs-distance-non-res}
\left|
\braket{0;5\mathrm{s}|e^{-iH'\tau/\hbar}|0;5\mathrm{s}}
\right|^2 = 
\frac{
1 + \xi^2 \cos^2\!\left(
\frac{\pi}{2\xi}\sqrt{1+\xi^2}
\right)
}{
1 + \xi^2
}
\end{equation}
where $\xi\equiv \hbar\Omega_\mathrm{Ryd}/U(R)$.
We use Eq.~\eqref{eq:state-pop-blockade-vs-distance-non-res} to plot the solid red line in Fig.~\ref{fig:blockade}d, using the experimental contrast that we fit to the blue data.

\subsection*{Fidelities in the hybrid system}
\subsubsection*{Atom-mediated readout of the molecule}
To determine fidelities associated with the atom-mediated readout of the molecule, we fit the data shown in Fig.~\ref{fig:molecule_readout}. The fidelity of successfully determining the state of the molecule by measuring the atom is $F_\mathrm{meas}=\frac{1}{2}(F_{0|0}+F_{1|1})=0.91(1)$. Here, $F_{N|N}$ is the probability of inferring from the atomic measurement that the molecule is in state $\ket{N}$, given that it was prepared in $\ket{N}$. The values of $F_{0|0}=0.86(2)$ and $F_{1|1}=0.95(2)$ are obtained from a least-squares fit to the data in Fig.~\ref{fig:molecule_readout}c. As discussed in the main text, these fidelities are limited by the fidelity of atomic control in our experiment.

To quantify the probability of a measurement-induced bit flip in the molecule, we fit the data in Fig.~\ref{fig:molecule_readout}b. 
The contrast of this measurement corresponds to the probability that the molecular populations are unchanged by the atomic readout protocol. 
A least-squares fit to the data yields cosine oscillations with a peak-to-peak contrast of $1.014(9)$ (Fig.~\ref{fig:molecule_readout}b, solid lines). 
The fitted contrast from this unconstrained fit is slightly larger than the physical maximum of unity due to statistical fluctuations. 
It corresponds to a lower bound on the probability of avoiding a measurement-induced bit flip of $0.996$ at the 95\% confidence level.

The analysis above considers only bit-flip errors, since all fidelities are conditioned on successful recovery of the molecule at the end of the sequence. 
As throughout this work, experimental runs in which the molecule is lost (that is, erasure errors) are discarded through postselection. 
As discussed above, these losses arise primarily from coupling to other hyperfine states of the molecule caused by the 1012-nm light when its intensity is too high.

\subsubsection*{Atom--molecule entanglement}
We quantify the fidelity of atom--molecule entanglement using the data shown in Fig.~\ref{fig:entanglement}d.
As described in the main text, the particles are entangled using a $\pi$ pulse on the Rydberg-excitation transition.
The coherence of the entangled state is then measured by applying a second $\pi$ pulse with variable phase and mapping this phase onto the molecular populations using a Ramsey-style measurement.
The phase of the atomic Rydberg drive is transferred to the molecule when coherence exists between the two components of the Bell state $\ket{\Psi}$.
Therefore, the observed fringe contrast directly measures the entanglement coherence, giving $\mathcal{C}=0.57(1)$.
The width of the region between the grey bands corresponds to the fidelity of molecular-state preparation which is $0.87(4)$.
Correcting for the imperfect molecular-state preparation gives a coherence of $\mathcal{C}=0.66(3)$ for the entangled state.

As in experiments where entanglement is generated between identical molecules~\cite{Holland2023Entanglement,Bao2023Entanglement,Picard2025,Ruttley2025}, we calculate the Bell-state fidelity $F=\frac{1}{2}(P_{\uparrow\downarrow}+P_{\downarrow\uparrow}+\mathcal{C})$. Here, $P_{\uparrow\downarrow}$ and $P_{\downarrow\uparrow}$ are the populations of the Bell-state components $\ket{3;5\mathrm{s}}$ and $\ket{2;83\mathrm{d}}$.
From the data in Fig.~\ref{fig:molecule_readout}d, we estimate $P_{\uparrow\downarrow}+P_{\downarrow\uparrow} = 0.88(4)$ when correcting for the state preparation of the molecule.
Combining this with the measured coherence $\mathcal{C}$ gives an entanglement fidelity of $F=0.77(3)$ when correcting for the imperfect state preparation of the molecule (this fidelity is $F=0.67(3)$ when not performing this correction).

The measured coherences and fidelities are primarily limited by imperfect driving of the atomic transition $\ket{5\mathrm{s}}\rightarrow\ket{83\mathrm{d}}$.
In particular, the measured coherence is a lower bound on the true coherence because the same imperfect Rydberg drive is used both to create and to analyse the entangled state.
We emphasise that the values above are conditioned on successful recovery of the atom at the end of the experimental sequence, corresponding to postselection that excludes the $22\%$ of runs in which the atom is lost.
Without this postselection, the corresponding entanglement fidelity is $0.60(2)$ when correcting for the molecular-state preparation, or $0.52(2)$ when not performing this correction.

\subsection*{Modelling the spin-exchange dynamics}
We model the spin-exchange oscillations shown in Fig.~\ref{fig:entanglement}b using a two-state model. 
In these experiments, we initially prepare the system in the pair state $\ket{4;83\mathrm{d}}$ by first exciting the molecule to $\ket{4}$ and subsequently exciting the atom to $\ket{83\mathrm{d}}$. The pair state $\ket{4;83\mathrm{d}}$ is non-resonant, so the atomic excitation is not blockaded (see Fig.~\ref{fig:blockade}d, inset) and the state does not evolve under any spin-exchange dynamics.

Spin-exchange dynamics are initiated by transferring the atom to $\ket{84\mathrm{p}}$ via a microwave $\pi$ pulse on the transition $\ket{83\mathrm{d}}\to\ket{84\mathrm{p}}$ with Rabi frequency $\Omega_\mathrm{MW}/2\pi=8.49(9)\,$MHz which is much greater than the coupling between the interacting pair states. This prepares the system in the state $\ket{4;84\mathrm{p}}$, which is coherently coupled to $\ket{3;83\mathrm{d}}$ via the spin-exchange interaction. The dynamics in this subspace are described by the Hamiltonian
\begin{equation}
H_\mathrm{se} =
\begin{pmatrix}
0 & V(R) \\
V(R) & h\delta
\end{pmatrix},
\end{equation}
written in the basis $\{\ket{3;83\mathrm{d}},\ket{4;84\mathrm{p}}\}$.
As above, $V(R)$ is the coupling between the states $\ket{3;83\mathrm{d}}$ and $\ket{4;84\mathrm{p}}$.
During the interaction window, dynamics are restricted to this two-state manifold, while other pair states remain uncoupled.

To read out the spin-exchange dynamics, a second $\pi$ pulse on the transition $\ket{83\mathrm{d}}\leftrightarrow\ket{84\mathrm{p}}$ maps population in $\ket{3;83\mathrm{d}}$ to $\ket{3;84\mathrm{p}}$, while population remaining in $\ket{4;84\mathrm{p}}$ is mapped back to $\ket{4;83\mathrm{d}}$. A subsequent $\pi$ pulse on the transition $\ket{83\mathrm{d}}\to\ket{5\mathrm{s}}$ transfers population in $\ket{4;83\mathrm{d}}$ to the atomic ground state, resulting in atom recovery, whereas population in $\ket{3;84\mathrm{p}}$ remains in a Rydberg state and is detected as atom loss.
Finally, we detect remaining atoms and readout the molecular state as described above.

With this protocol, after microwave transfers, the population in $\ket{3;84\mathrm{p}}$ (Fig.~\ref{fig:entanglement}b, black points) is
\begin{equation}
P_{3;84\mathrm{p}}(\tau;R,\delta)
=
\frac{V(R)^2}{\widetilde V}(R,\delta)^2
\sin^2\!\left(
2\pi\tau\frac{\widetilde V(R,\delta)}{h}
\right),
\end{equation}
where we have defined the generalised spin-exchange coupling
\begin{equation}
\widetilde{V}(R,\delta) = \sqrt{V(R)^2 + (h\delta/2)^2}\,.
\end{equation}
The remaining population remains in $\ket{4;83\mathrm{d}}$ (Fig.~\ref{fig:entanglement}b, purple points).

Imperfect state preparation results in approximately $38\%$ of experimental shots populating the two undesired subspaces: atom recovery when the molecule is in $\ket{3}$, and atom loss when the molecule is in $\ket{4}$. These events can be identified using the molecular state readout (see above) and are removed via postselection because they do not contribute to the spin-exchange dynamics of interest.
The first class of events (approximately $3\%$) arises from imperfect excitation of the molecule to $\ket{4}$ such that the molecule remains in $\ket{3}$. Then, the atomic excitation is blockaded and the atom stays in the ground state throughout the sequence. 
The second class (approximately $35\%$) corresponds to infidelity in the Rydberg excitation and could be substantially improved in the future.
Nevertheless, in both cases, no spin-exchange dynamics occur and the populations in these states are independent of the interaction time $\tau$.

To model our experimental data, we perform a Monte Carlo simulation that accounts for shot-to-shot fluctuations in both the interparticle separation $R$ and the detuning $\delta$. For each realisation, $R$ is sampled from a Gaussian distribution with mean $\bar{R}$ and standard deviation $\sigma_R$, and $\delta$ is sampled from a Gaussian distribution with mean $\bar{\delta}$ and standard deviation $\sigma_\delta$.
We take the interaction strength to be $V(R) = V_\mathrm{dd}(R) = C_3/R^3$, with $C_3/h = 1.36(15)\,\mathrm{MHz}\,\mu\mathrm{m}^3$ obtained from the fit to the blockade measurements shown in Fig.~\ref{fig:blockade}d (see above). 
For each Monte Carlo realisation, we evaluate the spin-exchange dynamics over an interaction time $\tau$ using the model described above.
The experimentally observable probabilities are obtained by averaging over $N=10^4$ iterations and are
\begin{align}
\overline{P}_{3;84\mathrm{p}}(\tau)
&=
\frac{1}{N}
\sum_{i=1}^{N}
P_{3;84\mathrm{p}}(\tau;R_i,\delta_i),\\
\overline{P}_{4;83\mathrm{d}}(\tau)
&=
1-\overline{P}_{3;84\mathrm{p}}(\tau).
\end{align}
The mean separation $\bar{R}$ is independently determined from ac Stark shift measurements described above. The remaining free parameters $\sigma_R$, $\sigma_\delta$, and $\bar{\delta}$ are extracted via least-squares fitting to the measured populations.
We attribute the fitted detuning fluctuations primarily to magnetic-field noise, while the fluctuations in $R$ arise from the finite wavefunction spread of the particles and the ejection of Rydberg atoms from the optical tweezers.

%%%%%%%%%%%%%%%% SUPPLEMENTARY FIGURES %%%%%%%%%%%%%%%


\begin{thebibliography}{76}%
\makeatletter
\providecommand \@ifxundefined [1]{%
 \@ifx{#1\undefined}
}%
\providecommand \@ifnum [1]{%
 \ifnum #1\expandafter \@firstoftwo
 \else \expandafter \@secondoftwo
 \fi
}%
\providecommand \@ifx [1]{%
 \ifx #1\expandafter \@firstoftwo
 \else \expandafter \@secondoftwo
 \fi
}%
\providecommand \natexlab [1]{#1}%
\providecommand \enquote  [1]{``#1''}%
\providecommand \bibnamefont  [1]{#1}%
\providecommand \bibfnamefont [1]{#1}%
\providecommand \citenamefont [1]{#1}%
\providecommand \href@noop [0]{\@secondoftwo}%
\providecommand \href [0]{\begingroup \@sanitize@url \@href}%
\providecommand \@href[1]{\@@startlink{#1}\@@href}%
\providecommand \@@href[1]{\endgroup#1\@@endlink}%
\providecommand \@sanitize@url [0]{\catcode `\\12\catcode `\$12\catcode `\&12\catcode `\#12\catcode `\^12\catcode `\_12\catcode `\%12\relax}%
\providecommand \@@startlink[1]{}%
\providecommand \@@endlink[0]{}%
\providecommand \url  [0]{\begingroup\@sanitize@url \@url }%
\providecommand \@url [1]{\endgroup\@href {#1}{\urlprefix }}%
\providecommand \urlprefix  [0]{URL }%
\providecommand \Eprint [0]{\href }%
\providecommand \doibase [0]{https://doi.org/}%
\providecommand \selectlanguage [0]{\@gobble}%
\providecommand \bibinfo  [0]{\@secondoftwo}%
\providecommand \bibfield  [0]{\@secondoftwo}%
\providecommand \translation [1]{[#1]}%
\providecommand \BibitemOpen [0]{}%
\providecommand \bibitemStop [0]{}%
\providecommand \bibitemNoStop [0]{.\EOS\space}%
\providecommand \EOS [0]{\spacefactor3000\relax}%
\providecommand \BibitemShut  [1]{\csname bibitem#1\endcsname}%
\let\auto@bib@innerbib\@empty
%</preamble>
\bibitem [{\citenamefont {Saffman}\ \emph {et~al.}(2010)\citenamefont {Saffman}, \citenamefont {Walker},\ and\ \citenamefont {M\o{}lmer}}]{Saffman2010}%
  \BibitemOpen
  \bibfield  {author} {\bibinfo {author} {\bibfnamefont {M.}~\bibnamefont {Saffman}}, \bibinfo {author} {\bibfnamefont {T.~G.}\ \bibnamefont {Walker}},\ and\ \bibinfo {author} {\bibfnamefont {K.}~\bibnamefont {M\o{}lmer}},\ }\bibfield  {title} {\bibinfo {title} {Quantum information with {R}ydberg atoms},\ }\href {https://doi.org/10.1103/RevModPhys.82.2313} {\bibfield  {journal} {\bibinfo  {journal} {Rev. Mod. Phys.}\ }\textbf {\bibinfo {volume} {82}},\ \bibinfo {pages} {2313} (\bibinfo {year} {2010})}\BibitemShut {NoStop}%
\bibitem [{\citenamefont {Browaeys}\ and\ \citenamefont {Lahaye}(2020)}]{Browaeys2020}%
  \BibitemOpen
  \bibfield  {author} {\bibinfo {author} {\bibfnamefont {A.}~\bibnamefont {Browaeys}}\ and\ \bibinfo {author} {\bibfnamefont {T.}~\bibnamefont {Lahaye}},\ }\bibfield  {title} {\bibinfo {title} {Many-body physics with individually controlled {R}ydberg atoms},\ }\href {https://doi.org/10.1038/s41567-019-0733-z} {\bibfield  {journal} {\bibinfo  {journal} {Nat. Phys.}\ }\textbf {\bibinfo {volume} {16}},\ \bibinfo {pages} {132} (\bibinfo {year} {2020})}\BibitemShut {NoStop}%
\bibitem [{\citenamefont {Wu}\ \emph {et~al.}(2021)\citenamefont {Wu}, \citenamefont {Liang}, \citenamefont {Tian}, \citenamefont {Yang}, \citenamefont {Chen}, \citenamefont {Liu}, \citenamefont {Tey},\ and\ \citenamefont {You}}]{Wu2021}%
  \BibitemOpen
  \bibfield  {author} {\bibinfo {author} {\bibfnamefont {X.}~\bibnamefont {Wu}}, \bibinfo {author} {\bibfnamefont {X.}~\bibnamefont {Liang}}, \bibinfo {author} {\bibfnamefont {Y.}~\bibnamefont {Tian}}, \bibinfo {author} {\bibfnamefont {F.}~\bibnamefont {Yang}}, \bibinfo {author} {\bibfnamefont {C.}~\bibnamefont {Chen}}, \bibinfo {author} {\bibfnamefont {Y.-C.}\ \bibnamefont {Liu}}, \bibinfo {author} {\bibfnamefont {M.~K.}\ \bibnamefont {Tey}},\ and\ \bibinfo {author} {\bibfnamefont {L.}~\bibnamefont {You}},\ }\bibfield  {title} {\bibinfo {title} {A concise review of {{Rydberg}} atom based quantum computation and quantum simulation*},\ }\href {https://doi.org/10.1088/1674-1056/abd76f} {\bibfield  {journal} {\bibinfo  {journal} {Chin. Phys. B}\ }\textbf {\bibinfo {volume} {30}},\ \bibinfo {pages} {020305} (\bibinfo {year} {2021})}\BibitemShut {NoStop}%
\bibitem [{\citenamefont {Defenu}\ \emph {et~al.}(2023)\citenamefont {Defenu}, \citenamefont {Donner}, \citenamefont {Macr{\`i}}, \citenamefont {Pagano}, \citenamefont {Ruffo},\ and\ \citenamefont {Trombettoni}}]{Defenu2023}%
  \BibitemOpen
  \bibfield  {author} {\bibinfo {author} {\bibfnamefont {N.}~\bibnamefont {Defenu}}, \bibinfo {author} {\bibfnamefont {T.}~\bibnamefont {Donner}}, \bibinfo {author} {\bibfnamefont {T.}~\bibnamefont {Macr{\`i}}}, \bibinfo {author} {\bibfnamefont {G.}~\bibnamefont {Pagano}}, \bibinfo {author} {\bibfnamefont {S.}~\bibnamefont {Ruffo}},\ and\ \bibinfo {author} {\bibfnamefont {A.}~\bibnamefont {Trombettoni}},\ }\bibfield  {title} {\bibinfo {title} {Long-range interacting quantum systems},\ }\href {https://doi.org/10.1103/RevModPhys.95.035002} {\bibfield  {journal} {\bibinfo  {journal} {Rev. Mod. Phys.}\ }\textbf {\bibinfo {volume} {95}},\ \bibinfo {pages} {035002} (\bibinfo {year} {2023})}\BibitemShut {NoStop}%
\bibitem [{\citenamefont {Cornish}\ \emph {et~al.}(2024)\citenamefont {Cornish}, \citenamefont {Tarbutt},\ and\ \citenamefont {Hazzard}}]{Cornish2024}%
  \BibitemOpen
  \bibfield  {author} {\bibinfo {author} {\bibfnamefont {S.~L.}\ \bibnamefont {Cornish}}, \bibinfo {author} {\bibfnamefont {M.~R.}\ \bibnamefont {Tarbutt}},\ and\ \bibinfo {author} {\bibfnamefont {K.~R.~A.}\ \bibnamefont {Hazzard}},\ }\bibfield  {title} {\bibinfo {title} {Quantum computation and quantum simulation with ultracold molecules},\ }\href {https://doi.org/10.1038/s41567-024-02453-9} {\bibfield  {journal} {\bibinfo  {journal} {Nat. Phys.}\ }\textbf {\bibinfo {volume} {20}},\ \bibinfo {pages} {730} (\bibinfo {year} {2024})}\BibitemShut {NoStop}%
\bibitem [{\citenamefont {Ruttley}\ \emph {et~al.}(2025)\citenamefont {Ruttley}, \citenamefont {Hepworth}, \citenamefont {Guttridge},\ and\ \citenamefont {Cornish}}]{Ruttley2025}%
  \BibitemOpen
  \bibfield  {author} {\bibinfo {author} {\bibfnamefont {D.~K.}\ \bibnamefont {Ruttley}}, \bibinfo {author} {\bibfnamefont {T.~R.}\ \bibnamefont {Hepworth}}, \bibinfo {author} {\bibfnamefont {A.}~\bibnamefont {Guttridge}},\ and\ \bibinfo {author} {\bibfnamefont {S.~L.}\ \bibnamefont {Cornish}},\ }\bibfield  {title} {\bibinfo {title} {Long-lived entanglement of molecules in magic-wavelength optical tweezers},\ }\href {https://doi.org/10.1038/s41586-024-08365-1} {\bibfield  {journal} {\bibinfo  {journal} {Nature}\ }\textbf {\bibinfo {volume} {637}},\ \bibinfo {pages} {827} (\bibinfo {year} {2025})}\BibitemShut {NoStop}%
\bibitem [{\citenamefont {Sawant}\ \emph {et~al.}(2020)\citenamefont {Sawant}, \citenamefont {Blackmore}, \citenamefont {Gregory}, \citenamefont {Mur-Petit}, \citenamefont {Jaksch}, \citenamefont {Aldegunde}, \citenamefont {Hutson}, \citenamefont {Tarbutt},\ and\ \citenamefont {Cornish}}]{Sawant2020}%
  \BibitemOpen
  \bibfield  {author} {\bibinfo {author} {\bibfnamefont {R.}~\bibnamefont {Sawant}}, \bibinfo {author} {\bibfnamefont {J.~A.}\ \bibnamefont {Blackmore}}, \bibinfo {author} {\bibfnamefont {P.~D.}\ \bibnamefont {Gregory}}, \bibinfo {author} {\bibfnamefont {J.}~\bibnamefont {Mur-Petit}}, \bibinfo {author} {\bibfnamefont {D.}~\bibnamefont {Jaksch}}, \bibinfo {author} {\bibfnamefont {J.}~\bibnamefont {Aldegunde}}, \bibinfo {author} {\bibfnamefont {J.~M.}\ \bibnamefont {Hutson}}, \bibinfo {author} {\bibfnamefont {M.~R.}\ \bibnamefont {Tarbutt}},\ and\ \bibinfo {author} {\bibfnamefont {S.~L.}\ \bibnamefont {Cornish}},\ }\bibfield  {title} {\bibinfo {title} {Ultracold polar molecules as qudits},\ }\href {https://doi.org/10.1088/1367-2630/ab60f4} {\bibfield  {journal} {\bibinfo  {journal} {New J. Phys.}\ }\textbf {\bibinfo {volume} {22}},\ \bibinfo {pages} {013027} (\bibinfo {year} {2020})}\BibitemShut {NoStop}%
\bibitem [{\citenamefont {Raghuram}\ \emph {et~al.}(2026)\citenamefont {Raghuram}, \citenamefont {Blondell}, \citenamefont {Mortlock}, \citenamefont {Maddox}, \citenamefont {Dasgupta}, \citenamefont {Middleton-Spencer}, \citenamefont {Hazzard}, \citenamefont {Price}, \citenamefont {Gregory},\ and\ \citenamefont {Cornish}}]{Raghuram2026}%
  \BibitemOpen
  \bibfield  {author} {\bibinfo {author} {\bibfnamefont {A.~P.}\ \bibnamefont {Raghuram}}, \bibinfo {author} {\bibfnamefont {F.~M.}\ \bibnamefont {Blondell}}, \bibinfo {author} {\bibfnamefont {J.~M.}\ \bibnamefont {Mortlock}}, \bibinfo {author} {\bibfnamefont {B.~P.}\ \bibnamefont {Maddox}}, \bibinfo {author} {\bibfnamefont {S.}~\bibnamefont {Dasgupta}}, \bibinfo {author} {\bibfnamefont {H.~A.~J.}\ \bibnamefont {Middleton-Spencer}}, \bibinfo {author} {\bibfnamefont {K.~R.~A.}\ \bibnamefont {Hazzard}}, \bibinfo {author} {\bibfnamefont {H.~M.}\ \bibnamefont {Price}}, \bibinfo {author} {\bibfnamefont {P.~D.}\ \bibnamefont {Gregory}},\ and\ \bibinfo {author} {\bibfnamefont {S.~L.}\ \bibnamefont {Cornish}},\ }\href {https://arxiv.org/abs/2604.00745} {\bibinfo {title} {Probing topological edge states in a molecular synthetic dimension}} (\bibinfo {year} {2026}),\ \Eprint {https://arxiv.org/abs/2604.00745} {arXiv:2604.00745 [physics.atom-ph]} \BibitemShut {NoStop}%
\bibitem [{\citenamefont {DeMille}\ \emph {et~al.}(2024)\citenamefont {DeMille}, \citenamefont {Hutzler}, \citenamefont {Rey},\ and\ \citenamefont {Zelevinsky}}]{DeMille2024}%
  \BibitemOpen
  \bibfield  {author} {\bibinfo {author} {\bibfnamefont {D.}~\bibnamefont {DeMille}}, \bibinfo {author} {\bibfnamefont {N.~R.}\ \bibnamefont {Hutzler}}, \bibinfo {author} {\bibfnamefont {A.~M.}\ \bibnamefont {Rey}},\ and\ \bibinfo {author} {\bibfnamefont {T.}~\bibnamefont {Zelevinsky}},\ }\bibfield  {title} {\bibinfo {title} {Quantum sensing and metrology for fundamental physics with molecules},\ }\href {https://doi.org/10.1038/s41567-024-02499-9} {\bibfield  {journal} {\bibinfo  {journal} {Nat. Phys.}\ }\textbf {\bibinfo {volume} {20}},\ \bibinfo {pages} {741} (\bibinfo {year} {2024})}\BibitemShut {NoStop}%
\bibitem [{\citenamefont {Kuznetsova}\ \emph {et~al.}(2011)\citenamefont {Kuznetsova}, \citenamefont {Rittenhouse}, \citenamefont {Sadeghpour},\ and\ \citenamefont {Yelin}}]{Kuznetsova2011}%
  \BibitemOpen
  \bibfield  {author} {\bibinfo {author} {\bibfnamefont {E.}~\bibnamefont {Kuznetsova}}, \bibinfo {author} {\bibfnamefont {S.~T.}\ \bibnamefont {Rittenhouse}}, \bibinfo {author} {\bibfnamefont {H.~R.}\ \bibnamefont {Sadeghpour}},\ and\ \bibinfo {author} {\bibfnamefont {S.~F.}\ \bibnamefont {Yelin}},\ }\bibfield  {title} {\bibinfo {title} {Rydberg atom mediated polar molecule interactions: a tool for molecular-state conditional quantum gates and individual addressability},\ }\href {https://doi.org/10.1039/C1CP21476D} {\bibfield  {journal} {\bibinfo  {journal} {Phys. Chem. Chem. Phys.}\ }\textbf {\bibinfo {volume} {13}},\ \bibinfo {pages} {17115} (\bibinfo {year} {2011})}\BibitemShut {NoStop}%
\bibitem [{\citenamefont {Zhang}\ and\ \citenamefont {Tarbutt}(2022)}]{Zhang2022}%
  \BibitemOpen
  \bibfield  {author} {\bibinfo {author} {\bibfnamefont {C.}~\bibnamefont {Zhang}}\ and\ \bibinfo {author} {\bibfnamefont {M.~R.}\ \bibnamefont {Tarbutt}},\ }\bibfield  {title} {\bibinfo {title} {Quantum computation in a hybrid array of molecules and {R}ydberg atoms},\ }\href {https://doi.org/10.1103/PRXQuantum.3.030340} {\bibfield  {journal} {\bibinfo  {journal} {PRX Quantum}\ }\textbf {\bibinfo {volume} {3}},\ \bibinfo {pages} {030340} (\bibinfo {year} {2022})}\BibitemShut {NoStop}%
\bibitem [{\citenamefont {Wang}\ \emph {et~al.}(2022)\citenamefont {Wang}, \citenamefont {Williams}, \citenamefont {Picard}, \citenamefont {Yao},\ and\ \citenamefont {Ni}}]{Wang2022}%
  \BibitemOpen
  \bibfield  {author} {\bibinfo {author} {\bibfnamefont {K.}~\bibnamefont {Wang}}, \bibinfo {author} {\bibfnamefont {C.~P.}\ \bibnamefont {Williams}}, \bibinfo {author} {\bibfnamefont {L.~R.~B.}\ \bibnamefont {Picard}}, \bibinfo {author} {\bibfnamefont {N.~Y.}\ \bibnamefont {Yao}},\ and\ \bibinfo {author} {\bibfnamefont {K.-K.}\ \bibnamefont {Ni}},\ }\bibfield  {title} {\bibinfo {title} {Enriching the quantum toolbox of ultracold molecules with {R}ydberg atoms},\ }\href {https://doi.org/10.1103/PRXQuantum.3.030339} {\bibfield  {journal} {\bibinfo  {journal} {PRX Quantum}\ }\textbf {\bibinfo {volume} {3}},\ \bibinfo {pages} {030339} (\bibinfo {year} {2022})}\BibitemShut {NoStop}%
\bibitem [{\citenamefont {Bai}\ \emph {et~al.}(2026)\citenamefont {Bai}, \citenamefont {Wei}, \citenamefont {Zhang}, \citenamefont {Li},\ and\ \citenamefont {Shao}}]{Bai2026}%
  \BibitemOpen
  \bibfield  {author} {\bibinfo {author} {\bibfnamefont {Y.-H.}\ \bibnamefont {Bai}}, \bibinfo {author} {\bibfnamefont {Y.}~\bibnamefont {Wei}}, \bibinfo {author} {\bibfnamefont {C.}~\bibnamefont {Zhang}}, \bibinfo {author} {\bibfnamefont {W.}~\bibnamefont {Li}},\ and\ \bibinfo {author} {\bibfnamefont {X.-Q.}\ \bibnamefont {Shao}},\ }\href {https://doi.org/10.48550/arXiv.2603.29349} {\bibinfo {title} {Multipartite controlled-{{NOT}} gates using molecules and {{Rydberg}} atoms}} (\bibinfo {year} {2026}),\ \Eprint {https://arxiv.org/abs/2603.29349} {arXiv:2603.29349 [quant-ph]} \BibitemShut {NoStop}%
\bibitem [{\citenamefont {Zhang}\ \emph {et~al.}(2026)\citenamefont {Zhang}, \citenamefont {Murciano}, \citenamefont {Tantivasadakarn},\ and\ \citenamefont {Finkelstein}}]{Zhang2026}%
  \BibitemOpen
  \bibfield  {author} {\bibinfo {author} {\bibfnamefont {C.}~\bibnamefont {Zhang}}, \bibinfo {author} {\bibfnamefont {S.}~\bibnamefont {Murciano}}, \bibinfo {author} {\bibfnamefont {N.}~\bibnamefont {Tantivasadakarn}},\ and\ \bibinfo {author} {\bibfnamefont {R.}~\bibnamefont {Finkelstein}},\ }\href {https://doi.org/10.48550/arXiv.2602.12909} {\bibinfo {title} {Quantum logic control and entanglement in hybrid atom-molecule arrays}} (\bibinfo {year} {2026}),\ \Eprint {https://arxiv.org/abs/2602.12909} {arXiv:2602.12909 [quant-ph]} \BibitemShut {NoStop}%
\bibitem [{\citenamefont {Kuznetsova}\ \emph {et~al.}(2018)\citenamefont {Kuznetsova}, \citenamefont {Rittenhouse}, \citenamefont {Beterov}, \citenamefont {Scully}, \citenamefont {Yelin},\ and\ \citenamefont {Sadeghpour}}]{Kuznetsova2018}%
  \BibitemOpen
  \bibfield  {author} {\bibinfo {author} {\bibfnamefont {E.}~\bibnamefont {Kuznetsova}}, \bibinfo {author} {\bibfnamefont {S.~T.}\ \bibnamefont {Rittenhouse}}, \bibinfo {author} {\bibfnamefont {I.~I.}\ \bibnamefont {Beterov}}, \bibinfo {author} {\bibfnamefont {M.~O.}\ \bibnamefont {Scully}}, \bibinfo {author} {\bibfnamefont {S.~F.}\ \bibnamefont {Yelin}},\ and\ \bibinfo {author} {\bibfnamefont {H.~R.}\ \bibnamefont {Sadeghpour}},\ }\bibfield  {title} {\bibinfo {title} {Effective spin-spin interactions in bilayers of {R}ydberg atoms and polar molecules},\ }\href {https://doi.org/10.1103/PhysRevA.98.043609} {\bibfield  {journal} {\bibinfo  {journal} {Phys. Rev. A}\ }\textbf {\bibinfo {volume} {98}},\ \bibinfo {pages} {043609} (\bibinfo {year} {2018})}\BibitemShut {NoStop}%
\bibitem [{\citenamefont {Dobrzyniecki}\ and\ \citenamefont {Tomza}(2023)}]{Dobrzyniecki2023}%
  \BibitemOpen
  \bibfield  {author} {\bibinfo {author} {\bibfnamefont {J.}~\bibnamefont {Dobrzyniecki}}\ and\ \bibinfo {author} {\bibfnamefont {M.}~\bibnamefont {Tomza}},\ }\bibfield  {title} {\bibinfo {title} {Quantum simulation of the central spin model with a {{Rydberg}} atom and polar molecules in optical tweezers},\ }\href {https://doi.org/10.1103/PhysRevA.108.052618} {\bibfield  {journal} {\bibinfo  {journal} {Phys. Rev. A}\ }\textbf {\bibinfo {volume} {108}},\ \bibinfo {pages} {052618} (\bibinfo {year} {2023})}\BibitemShut {NoStop}%
\bibitem [{\citenamefont {Kuznetsova}\ \emph {et~al.}(2016)\citenamefont {Kuznetsova}, \citenamefont {Rittenhouse}, \citenamefont {Sadeghpour},\ and\ \citenamefont {Yelin}}]{Kuznetsova2016}%
  \BibitemOpen
  \bibfield  {author} {\bibinfo {author} {\bibfnamefont {E.}~\bibnamefont {Kuznetsova}}, \bibinfo {author} {\bibfnamefont {S.~T.}\ \bibnamefont {Rittenhouse}}, \bibinfo {author} {\bibfnamefont {H.~R.}\ \bibnamefont {Sadeghpour}},\ and\ \bibinfo {author} {\bibfnamefont {S.~F.}\ \bibnamefont {Yelin}},\ }\bibfield  {title} {\bibinfo {title} {Rydberg-atom-mediated nondestructive readout of collective rotational states in polar-molecule arrays},\ }\href {https://doi.org/10.1103/PhysRevA.94.032325} {\bibfield  {journal} {\bibinfo  {journal} {Phys. Rev. A}\ }\textbf {\bibinfo {volume} {94}},\ \bibinfo {pages} {032325} (\bibinfo {year} {2016})}\BibitemShut {NoStop}%
\bibitem [{\citenamefont {Zeppenfeld}(2017)}]{Zeppenfeld2017}%
  \BibitemOpen
  \bibfield  {author} {\bibinfo {author} {\bibfnamefont {M.}~\bibnamefont {Zeppenfeld}},\ }\bibfield  {title} {\bibinfo {title} {Nondestructive detection of polar molecules via {R}ydberg atoms},\ }\href {https://doi.org/10.1209/0295-5075/118/13002} {\bibfield  {journal} {\bibinfo  {journal} {EPL}\ }\textbf {\bibinfo {volume} {118}},\ \bibinfo {pages} {13002} (\bibinfo {year} {2017})}\BibitemShut {NoStop}%
\bibitem [{\citenamefont {Jarisch}\ and\ \citenamefont {Zeppenfeld}(2018)}]{Jarisch2018}%
  \BibitemOpen
  \bibfield  {author} {\bibinfo {author} {\bibfnamefont {F.}~\bibnamefont {Jarisch}}\ and\ \bibinfo {author} {\bibfnamefont {M.}~\bibnamefont {Zeppenfeld}},\ }\bibfield  {title} {\bibinfo {title} {State resolved investigation of {F}örster resonant energy transfer in collisions between polar molecules and {R}ydberg atoms},\ }\href {http://dx.doi.org/10.1088/1367-2630/aaf02e} {\bibfield  {journal} {\bibinfo  {journal} {New J. Phys.}\ }\textbf {\bibinfo {volume} {20}},\ \bibinfo {pages} {113044} (\bibinfo {year} {2018})}\BibitemShut {NoStop}%
\bibitem [{\citenamefont {Gawlas}\ and\ \citenamefont {Hogan}(2020)}]{Gawlas2020}%
  \BibitemOpen
  \bibfield  {author} {\bibinfo {author} {\bibfnamefont {K.}~\bibnamefont {Gawlas}}\ and\ \bibinfo {author} {\bibfnamefont {S.~D.}\ \bibnamefont {Hogan}},\ }\bibfield  {title} {\bibinfo {title} {Rydberg-state-resolved resonant energy transfer in cold electric-field-controlled intrabeam collisions of {NH$_{3}$} with {R}ydberg {He} atoms},\ }\href {https://doi.org/10.1021/acs.jpclett.9b03290} {\bibfield  {journal} {\bibinfo  {journal} {J. Phys. Chem. Lett.}\ }\textbf {\bibinfo {volume} {11}},\ \bibinfo {pages} {83} (\bibinfo {year} {2020})}\BibitemShut {NoStop}%
\bibitem [{\citenamefont {Patsch}\ \emph {et~al.}(2022)\citenamefont {Patsch}, \citenamefont {Zeppenfeld},\ and\ \citenamefont {Koch}}]{Patsch2022}%
  \BibitemOpen
  \bibfield  {author} {\bibinfo {author} {\bibfnamefont {S.}~\bibnamefont {Patsch}}, \bibinfo {author} {\bibfnamefont {M.}~\bibnamefont {Zeppenfeld}},\ and\ \bibinfo {author} {\bibfnamefont {C.~P.}\ \bibnamefont {Koch}},\ }\bibfield  {title} {\bibinfo {title} {Rydberg atom-enabled spectroscopy of polar molecules via {F}{\"o}rster resonance energy transfer},\ }\href {https://doi.org/10.1021/acs.jpclett.2c02521} {\bibfield  {journal} {\bibinfo  {journal} {J. Phys. Chem. Lett.}\ }\textbf {\bibinfo {volume} {13}},\ \bibinfo {pages} {10728} (\bibinfo {year} {2022})}\BibitemShut {NoStop}%
\bibitem [{\citenamefont {Young}\ \emph {et~al.}(2026)\citenamefont {Young}, \citenamefont {Ni},\ and\ \citenamefont {Gorshkov}}]{Young2026}%
  \BibitemOpen
  \bibfield  {author} {\bibinfo {author} {\bibfnamefont {J.~T.}\ \bibnamefont {Young}}, \bibinfo {author} {\bibfnamefont {K.-K.}\ \bibnamefont {Ni}},\ and\ \bibinfo {author} {\bibfnamefont {A.~V.}\ \bibnamefont {Gorshkov}},\ }\href {https://arxiv.org/abs/2601.08921} {\bibinfo {title} {Simultaneous nondestructive measurement of many polar molecules using {R}ydberg atoms}} (\bibinfo {year} {2026}),\ \Eprint {https://arxiv.org/abs/2601.08921} {arXiv:2601.08921 [quant-ph]} \BibitemShut {NoStop}%
\bibitem [{\citenamefont {Zhao}\ \emph {et~al.}(2012)\citenamefont {Zhao}, \citenamefont {Glaetzle}, \citenamefont {Pupillo},\ and\ \citenamefont {Zoller}}]{Zhao2012}%
  \BibitemOpen
  \bibfield  {author} {\bibinfo {author} {\bibfnamefont {B.}~\bibnamefont {Zhao}}, \bibinfo {author} {\bibfnamefont {A.~W.}\ \bibnamefont {Glaetzle}}, \bibinfo {author} {\bibfnamefont {G.}~\bibnamefont {Pupillo}},\ and\ \bibinfo {author} {\bibfnamefont {P.}~\bibnamefont {Zoller}},\ }\bibfield  {title} {\bibinfo {title} {Atomic {Rydberg} reservoirs for polar molecules},\ }\href {https://link.aps.org/doi/10.1103/PhysRevLett.108.193007} {\bibfield  {journal} {\bibinfo  {journal} {Phys. Rev. Lett.}\ }\textbf {\bibinfo {volume} {108}},\ \bibinfo {pages} {193007} (\bibinfo {year} {2012})}\BibitemShut {NoStop}%
\bibitem [{\citenamefont {Huber}\ and\ \citenamefont {B\"{u}chler}(2012)}]{Huber2012}%
  \BibitemOpen
  \bibfield  {author} {\bibinfo {author} {\bibfnamefont {S.~D.}\ \bibnamefont {Huber}}\ and\ \bibinfo {author} {\bibfnamefont {H.~P.}\ \bibnamefont {B\"{u}chler}},\ }\bibfield  {title} {\bibinfo {title} {Dipole-interaction-mediated laser cooling of polar molecules to ultracold temperatures},\ }\href {https://doi.org/10.1103/PhysRevLett.108.193006} {\bibfield  {journal} {\bibinfo  {journal} {Phys. Rev. Lett.}\ }\textbf {\bibinfo {volume} {108}},\ \bibinfo {pages} {193006} (\bibinfo {year} {2012})}\BibitemShut {NoStop}%
\bibitem [{\citenamefont {Zhang}\ \emph {et~al.}(2024)\citenamefont {Zhang}, \citenamefont {Rittenhouse}, \citenamefont {Tscherbul}, \citenamefont {Sadeghpour},\ and\ \citenamefont {Hutzler}}]{Zhang2024cooling}%
  \BibitemOpen
  \bibfield  {author} {\bibinfo {author} {\bibfnamefont {C.}~\bibnamefont {Zhang}}, \bibinfo {author} {\bibfnamefont {S.~T.}\ \bibnamefont {Rittenhouse}}, \bibinfo {author} {\bibfnamefont {T.~V.}\ \bibnamefont {Tscherbul}}, \bibinfo {author} {\bibfnamefont {H.~R.}\ \bibnamefont {Sadeghpour}},\ and\ \bibinfo {author} {\bibfnamefont {N.~R.}\ \bibnamefont {Hutzler}},\ }\bibfield  {title} {\bibinfo {title} {Sympathetic cooling and slowing of molecules with {R}ydberg atoms},\ }\href {https://doi.org/10.1103/PhysRevLett.132.033001} {\bibfield  {journal} {\bibinfo  {journal} {Phys. Rev. Lett.}\ }\textbf {\bibinfo {volume} {132}},\ \bibinfo {pages} {033001} (\bibinfo {year} {2024})}\BibitemShut {NoStop}%
\bibitem [{\citenamefont {Rittenhouse}\ and\ \citenamefont {Sadeghpour}(2010)}]{Rittenhouse2010}%
  \BibitemOpen
  \bibfield  {author} {\bibinfo {author} {\bibfnamefont {S.~T.}\ \bibnamefont {Rittenhouse}}\ and\ \bibinfo {author} {\bibfnamefont {H.~R.}\ \bibnamefont {Sadeghpour}},\ }\bibfield  {title} {\bibinfo {title} {Ultracold giant polyatomic {R}ydberg molecules: Coherent control of molecular orientation},\ }\href {https://doi.org/10.1103/PhysRevLett.104.243002} {\bibfield  {journal} {\bibinfo  {journal} {Phys. Rev. Lett.}\ }\textbf {\bibinfo {volume} {104}},\ \bibinfo {pages} {243002} (\bibinfo {year} {2010})}\BibitemShut {NoStop}%
\bibitem [{\citenamefont {Rittenhouse}\ \emph {et~al.}(2011)\citenamefont {Rittenhouse}, \citenamefont {Mayle}, \citenamefont {Schmelcher},\ and\ \citenamefont {Sadeghpour}}]{Rittenhouse2011}%
  \BibitemOpen
  \bibfield  {author} {\bibinfo {author} {\bibfnamefont {S.~T.}\ \bibnamefont {Rittenhouse}}, \bibinfo {author} {\bibfnamefont {M.}~\bibnamefont {Mayle}}, \bibinfo {author} {\bibfnamefont {P.}~\bibnamefont {Schmelcher}},\ and\ \bibinfo {author} {\bibfnamefont {H.~R.}\ \bibnamefont {Sadeghpour}},\ }\bibfield  {title} {\bibinfo {title} {Ultralong-range polyatomic {Rydberg} molecules formed by a polar perturber},\ }\href {https://doi.org/10.1088/0953-4075/44/18/184005} {\bibfield  {journal} {\bibinfo  {journal} {J. Phys. B}\ }\textbf {\bibinfo {volume} {44}},\ \bibinfo {pages} {184005} (\bibinfo {year} {2011})}\BibitemShut {NoStop}%
\bibitem [{\citenamefont {González-Férez}\ \emph {et~al.}(2020)\citenamefont {González-Férez}, \citenamefont {Rittenhouse}, \citenamefont {Schmelcher},\ and\ \citenamefont {Sadeghpour}}]{GonzalezFerez2020}%
  \BibitemOpen
  \bibfield  {author} {\bibinfo {author} {\bibfnamefont {R.}~\bibnamefont {González-Férez}}, \bibinfo {author} {\bibfnamefont {S.~T.}\ \bibnamefont {Rittenhouse}}, \bibinfo {author} {\bibfnamefont {P.}~\bibnamefont {Schmelcher}},\ and\ \bibinfo {author} {\bibfnamefont {H.~R.}\ \bibnamefont {Sadeghpour}},\ }\bibfield  {title} {\bibinfo {title} {A protocol to realize triatomic ultralong range {Rydberg} molecules in an ultracold {KRb} gas},\ }\href {https://doi.org/10.1088/1361-6455/ab68b8} {\bibfield  {journal} {\bibinfo  {journal} {J. Phys. B}\ }\textbf {\bibinfo {volume} {53}},\ \bibinfo {pages} {074002} (\bibinfo {year} {2020})}\BibitemShut {NoStop}%
\bibitem [{\citenamefont {Petitjean}\ \emph {et~al.}(1986)\citenamefont {Petitjean}, \citenamefont {Gounand},\ and\ \citenamefont {Fournier}}]{Petitjean1986}%
  \BibitemOpen
  \bibfield  {author} {\bibinfo {author} {\bibfnamefont {L.}~\bibnamefont {Petitjean}}, \bibinfo {author} {\bibfnamefont {F.}~\bibnamefont {Gounand}},\ and\ \bibinfo {author} {\bibfnamefont {P.~R.}\ \bibnamefont {Fournier}},\ }\bibfield  {title} {\bibinfo {title} {Collisions of rubidium {{Rydberg-state}} atoms with ammonia},\ }\href {https://doi.org/10.1103/PhysRevA.33.143} {\bibfield  {journal} {\bibinfo  {journal} {Phys. Rev. A}\ }\textbf {\bibinfo {volume} {33}},\ \bibinfo {pages} {143} (\bibinfo {year} {1986})}\BibitemShut {NoStop}%
\bibitem [{\citenamefont {Zhu}\ \emph {et~al.}(2025)\citenamefont {Zhu}, \citenamefont {Luke}, \citenamefont {Shaham}, \citenamefont {Liu},\ and\ \citenamefont {Ni}}]{Zhu2025}%
  \BibitemOpen
  \bibfield  {author} {\bibinfo {author} {\bibfnamefont {L.}~\bibnamefont {Zhu}}, \bibinfo {author} {\bibfnamefont {J.}~\bibnamefont {Luke}}, \bibinfo {author} {\bibfnamefont {R.}~\bibnamefont {Shaham}}, \bibinfo {author} {\bibfnamefont {Y.-X.}\ \bibnamefont {Liu}},\ and\ \bibinfo {author} {\bibfnamefont {K.-K.}\ \bibnamefont {Ni}},\ }\bibfield  {title} {\bibinfo {title} {Probing dipolar interactions between {R}ydberg atoms and ultracold polar molecules},\ }\href {https://doi.org/10.1103/48rk-sxfs} {\bibfield  {journal} {\bibinfo  {journal} {Phys. Rev. Lett.}\ }\textbf {\bibinfo {volume} {135}},\ \bibinfo {pages} {153001} (\bibinfo {year} {2025})}\BibitemShut {NoStop}%
\bibitem [{\citenamefont {Zou}\ and\ \citenamefont {Hogan}(2022)}]{Zou2022}%
  \BibitemOpen
  \bibfield  {author} {\bibinfo {author} {\bibfnamefont {J.}~\bibnamefont {Zou}}\ and\ \bibinfo {author} {\bibfnamefont {S.~D.}\ \bibnamefont {Hogan}},\ }\bibfield  {title} {\bibinfo {title} {Probing van der {{Waals}} interactions and detecting polar molecules by {{F}}\"orster-resonance energy transfer with {{Rydberg}} atoms at temperatures below 100 {{mK}}},\ }\href {https://doi.org/10.1103/PhysRevA.106.043111} {\bibfield  {journal} {\bibinfo  {journal} {Phys. Rev. A}\ }\textbf {\bibinfo {volume} {106}},\ \bibinfo {pages} {043111} (\bibinfo {year} {2022})}\BibitemShut {NoStop}%
\bibitem [{\citenamefont {Ruttley}\ \emph {et~al.}(2024)\citenamefont {Ruttley}, \citenamefont {Guttridge}, \citenamefont {Hepworth},\ and\ \citenamefont {Cornish}}]{Ruttley2024}%
  \BibitemOpen
  \bibfield  {author} {\bibinfo {author} {\bibfnamefont {D.~K.}\ \bibnamefont {Ruttley}}, \bibinfo {author} {\bibfnamefont {A.}~\bibnamefont {Guttridge}}, \bibinfo {author} {\bibfnamefont {T.~R.}\ \bibnamefont {Hepworth}},\ and\ \bibinfo {author} {\bibfnamefont {S.~L.}\ \bibnamefont {Cornish}},\ }\bibfield  {title} {\bibinfo {title} {Enhanced quantum control of individual ultracold molecules using optical tweezer arrays},\ }\href {https://doi.org/10.1103/PRXQuantum.5.020333} {\bibfield  {journal} {\bibinfo  {journal} {PRX Quantum}\ }\textbf {\bibinfo {volume} {5}},\ \bibinfo {pages} {020333} (\bibinfo {year} {2024})}\BibitemShut {NoStop}%
\bibitem [{\citenamefont {Adams}\ \emph {et~al.}(2019)\citenamefont {Adams}, \citenamefont {Pritchard},\ and\ \citenamefont {Shaffer}}]{Adams2020}%
  \BibitemOpen
  \bibfield  {author} {\bibinfo {author} {\bibfnamefont {C.~S.}\ \bibnamefont {Adams}}, \bibinfo {author} {\bibfnamefont {J.~D.}\ \bibnamefont {Pritchard}},\ and\ \bibinfo {author} {\bibfnamefont {J.~P.}\ \bibnamefont {Shaffer}},\ }\bibfield  {title} {\bibinfo {title} {{Rydberg} atom quantum technologies},\ }\href {https://doi.org/10.1088/1361-6455/ab52ef} {\bibfield  {journal} {\bibinfo  {journal} {J. Phys. B}\ }\textbf {\bibinfo {volume} {53}},\ \bibinfo {pages} {012002} (\bibinfo {year} {2019})}\BibitemShut {NoStop}%
\bibitem [{\citenamefont {Picard}\ \emph {et~al.}(2024)\citenamefont {Picard}, \citenamefont {Patenotte}, \citenamefont {Park}, \citenamefont {Gebretsadkan},\ and\ \citenamefont {Ni}}]{Picard2024SiteSelective}%
  \BibitemOpen
  \bibfield  {author} {\bibinfo {author} {\bibfnamefont {L.~R.~B.}\ \bibnamefont {Picard}}, \bibinfo {author} {\bibfnamefont {G.~E.}\ \bibnamefont {Patenotte}}, \bibinfo {author} {\bibfnamefont {A.~J.}\ \bibnamefont {Park}}, \bibinfo {author} {\bibfnamefont {S.~F.}\ \bibnamefont {Gebretsadkan}},\ and\ \bibinfo {author} {\bibfnamefont {K.-K.}\ \bibnamefont {Ni}},\ }\bibfield  {title} {\bibinfo {title} {Site-selective preparation and multistate readout of molecules in optical tweezers},\ }\href {https://doi.org/10.1103/PRXQuantum.5.020344} {\bibfield  {journal} {\bibinfo  {journal} {PRX Quantum}\ }\textbf {\bibinfo {volume} {5}},\ \bibinfo {pages} {020344} (\bibinfo {year} {2024})}\BibitemShut {NoStop}%
\bibitem [{\citenamefont {Anderegg}\ \emph {et~al.}(2019)\citenamefont {Anderegg}, \citenamefont {Cheuk}, \citenamefont {Bao}, \citenamefont {Burchesky}, \citenamefont {Ketterle}, \citenamefont {Ni},\ and\ \citenamefont {Doyle}}]{Anderegg2019}%
  \BibitemOpen
  \bibfield  {author} {\bibinfo {author} {\bibfnamefont {L.}~\bibnamefont {Anderegg}}, \bibinfo {author} {\bibfnamefont {L.~W.}\ \bibnamefont {Cheuk}}, \bibinfo {author} {\bibfnamefont {Y.}~\bibnamefont {Bao}}, \bibinfo {author} {\bibfnamefont {S.}~\bibnamefont {Burchesky}}, \bibinfo {author} {\bibfnamefont {W.}~\bibnamefont {Ketterle}}, \bibinfo {author} {\bibfnamefont {K.-K.}\ \bibnamefont {Ni}},\ and\ \bibinfo {author} {\bibfnamefont {J.~M.}\ \bibnamefont {Doyle}},\ }\bibfield  {title} {\bibinfo {title} {An optical tweezer array of ultracold molecules},\ }\href {https://doi.org/10.1126/science.aax1265} {\bibfield  {journal} {\bibinfo  {journal} {Science}\ }\textbf {\bibinfo {volume} {365}},\ \bibinfo {pages} {1156} (\bibinfo {year} {2019})}\BibitemShut {NoStop}%
\bibitem [{\citenamefont {Holland}\ \emph {et~al.}(2023{\natexlab{a}})\citenamefont {Holland}, \citenamefont {Lu},\ and\ \citenamefont {Cheuk}}]{Holland2023Imaging}%
  \BibitemOpen
  \bibfield  {author} {\bibinfo {author} {\bibfnamefont {C.~M.}\ \bibnamefont {Holland}}, \bibinfo {author} {\bibfnamefont {Y.}~\bibnamefont {Lu}},\ and\ \bibinfo {author} {\bibfnamefont {L.~W.}\ \bibnamefont {Cheuk}},\ }\bibfield  {title} {\bibinfo {title} {Bichromatic imaging of single molecules in an optical tweezer array},\ }\href {https://doi.org/10.1103/PhysRevLett.131.053202} {\bibfield  {journal} {\bibinfo  {journal} {Phys. Rev. Lett.}\ }\textbf {\bibinfo {volume} {131}},\ \bibinfo {pages} {053202} (\bibinfo {year} {2023}{\natexlab{a}})}\BibitemShut {NoStop}%
\bibitem [{\citenamefont {Vilas}\ \emph {et~al.}(2024)\citenamefont {Vilas}, \citenamefont {Robichaud}, \citenamefont {Hallas}, \citenamefont {Li}, \citenamefont {Anderegg},\ and\ \citenamefont {Doyle}}]{Vilas2024}%
  \BibitemOpen
  \bibfield  {author} {\bibinfo {author} {\bibfnamefont {N.~B.}\ \bibnamefont {Vilas}}, \bibinfo {author} {\bibfnamefont {P.}~\bibnamefont {Robichaud}}, \bibinfo {author} {\bibfnamefont {C.}~\bibnamefont {Hallas}}, \bibinfo {author} {\bibfnamefont {G.~K.}\ \bibnamefont {Li}}, \bibinfo {author} {\bibfnamefont {L.}~\bibnamefont {Anderegg}},\ and\ \bibinfo {author} {\bibfnamefont {J.~M.}\ \bibnamefont {Doyle}},\ }\bibfield  {title} {\bibinfo {title} {An optical tweezer array of ultracold polyatomic molecules},\ }\href {https://doi.org/10.1038/s41586-024-07199-1} {\bibfield  {journal} {\bibinfo  {journal} {Nature}\ }\textbf {\bibinfo {volume} {628}},\ \bibinfo {pages} {282} (\bibinfo {year} {2024})}\BibitemShut {NoStop}%
\bibitem [{\citenamefont {Bluvstein}\ \emph {et~al.}(2024)\citenamefont {Bluvstein}, \citenamefont {Evered}, \citenamefont {Geim}, \citenamefont {Li}, \citenamefont {Zhou}, \citenamefont {Manovitz}, \citenamefont {Ebadi}, \citenamefont {Cain}, \citenamefont {Kalinowski}, \citenamefont {Hangleiter}, \citenamefont {Bonilla~Ataides}, \citenamefont {Maskara}, \citenamefont {Cong}, \citenamefont {Gao}, \citenamefont {Sales~Rodriguez}, \citenamefont {Karolyshyn}, \citenamefont {Semeghini}, \citenamefont {Gullans}, \citenamefont {Greiner}, \citenamefont {Vuleti{\'{c}}},\ and\ \citenamefont {Lukin}}]{Bluvstein2024}%
  \BibitemOpen
  \bibfield  {author} {\bibinfo {author} {\bibfnamefont {D.}~\bibnamefont {Bluvstein}}, \bibinfo {author} {\bibfnamefont {S.~J.}\ \bibnamefont {Evered}}, \bibinfo {author} {\bibfnamefont {A.~A.}\ \bibnamefont {Geim}}, \bibinfo {author} {\bibfnamefont {S.~H.}\ \bibnamefont {Li}}, \bibinfo {author} {\bibfnamefont {H.}~\bibnamefont {Zhou}}, \bibinfo {author} {\bibfnamefont {T.}~\bibnamefont {Manovitz}}, \bibinfo {author} {\bibfnamefont {S.}~\bibnamefont {Ebadi}}, \bibinfo {author} {\bibfnamefont {M.}~\bibnamefont {Cain}}, \bibinfo {author} {\bibfnamefont {M.}~\bibnamefont {Kalinowski}}, \bibinfo {author} {\bibfnamefont {D.}~\bibnamefont {Hangleiter}}, \bibinfo {author} {\bibfnamefont {J.~P.}\ \bibnamefont {Bonilla~Ataides}}, \bibinfo {author} {\bibfnamefont {N.}~\bibnamefont {Maskara}}, \bibinfo {author} {\bibfnamefont {I.}~\bibnamefont {Cong}}, \bibinfo {author} {\bibfnamefont {X.}~\bibnamefont {Gao}}, \bibinfo {author} {\bibfnamefont {P.}~\bibnamefont {Sales~Rodriguez}}, \bibinfo {author} {\bibfnamefont
  {T.}~\bibnamefont {Karolyshyn}}, \bibinfo {author} {\bibfnamefont {G.}~\bibnamefont {Semeghini}}, \bibinfo {author} {\bibfnamefont {M.~J.}\ \bibnamefont {Gullans}}, \bibinfo {author} {\bibfnamefont {M.}~\bibnamefont {Greiner}}, \bibinfo {author} {\bibfnamefont {V.}~\bibnamefont {Vuleti{\'{c}}}},\ and\ \bibinfo {author} {\bibfnamefont {M.~D.}\ \bibnamefont {Lukin}},\ }\bibfield  {title} {\bibinfo {title} {Logical quantum processor based on reconfigurable atom arrays},\ }\href {https://doi.org/10.1038/s41586-023-06927-3} {\bibfield  {journal} {\bibinfo  {journal} {Nature}\ }\textbf {\bibinfo {volume} {626}},\ \bibinfo {pages} {58} (\bibinfo {year} {2024})}\BibitemShut {NoStop}%
\bibitem [{\citenamefont {Miles}\ \emph {et~al.}(2026)\citenamefont {Miles}, \citenamefont {Lichtman}, \citenamefont {Scott}, \citenamefont {Scott}, \citenamefont {Norrell}, \citenamefont {Bedalov}, \citenamefont {Belknap}, \citenamefont {Cole}, \citenamefont {Eubanks}, \citenamefont {Gillette}, \citenamefont {Gokhale}, \citenamefont {Goldwin}, \citenamefont {Iliev}, \citenamefont {Jones}, \citenamefont {Kuper}, \citenamefont {Mason}, \citenamefont {Mitchell}, \citenamefont {Murphree}, \citenamefont {Neff-Mallon}, \citenamefont {Noel}, \citenamefont {Radnaev}, \citenamefont {Vinogradov},\ and\ \citenamefont {Saffman}}]{Miles2026}%
  \BibitemOpen
  \bibfield  {author} {\bibinfo {author} {\bibfnamefont {J.}~\bibnamefont {Miles}}, \bibinfo {author} {\bibfnamefont {M.~T.}\ \bibnamefont {Lichtman}}, \bibinfo {author} {\bibfnamefont {A.~M.}\ \bibnamefont {Scott}}, \bibinfo {author} {\bibfnamefont {J.}~\bibnamefont {Scott}}, \bibinfo {author} {\bibfnamefont {S.~A.}\ \bibnamefont {Norrell}}, \bibinfo {author} {\bibfnamefont {M.~J.}\ \bibnamefont {Bedalov}}, \bibinfo {author} {\bibfnamefont {D.~A.}\ \bibnamefont {Belknap}}, \bibinfo {author} {\bibfnamefont {D.~C.}\ \bibnamefont {Cole}}, \bibinfo {author} {\bibfnamefont {S.~Y.}\ \bibnamefont {Eubanks}}, \bibinfo {author} {\bibfnamefont {M.}~\bibnamefont {Gillette}}, \bibinfo {author} {\bibfnamefont {P.}~\bibnamefont {Gokhale}}, \bibinfo {author} {\bibfnamefont {J.}~\bibnamefont {Goldwin}}, \bibinfo {author} {\bibfnamefont {M.}~\bibnamefont {Iliev}}, \bibinfo {author} {\bibfnamefont {R.~A.}\ \bibnamefont {Jones}}, \bibinfo {author} {\bibfnamefont {K.~W.}\ \bibnamefont {Kuper}}, \bibinfo {author} {\bibfnamefont
  {D.}~\bibnamefont {Mason}}, \bibinfo {author} {\bibfnamefont {P.~T.}\ \bibnamefont {Mitchell}}, \bibinfo {author} {\bibfnamefont {J.~D.}\ \bibnamefont {Murphree}}, \bibinfo {author} {\bibfnamefont {N.~A.}\ \bibnamefont {Neff-Mallon}}, \bibinfo {author} {\bibfnamefont {T.~W.}\ \bibnamefont {Noel}}, \bibinfo {author} {\bibfnamefont {A.~G.}\ \bibnamefont {Radnaev}}, \bibinfo {author} {\bibfnamefont {I.~V.}\ \bibnamefont {Vinogradov}},\ and\ \bibinfo {author} {\bibfnamefont {M.}~\bibnamefont {Saffman}},\ }\href {https://arxiv.org/abs/2603.13492} {\bibinfo {title} {Qubit syndrome measurements with a high fidelity {Rb-Cs} {R}ydberg gate}} (\bibinfo {year} {2026}),\ \Eprint {https://arxiv.org/abs/2603.13492} {arXiv:2603.13492 [quant-ph]} \BibitemShut {NoStop}%
\bibitem [{\citenamefont {Mi}\ \emph {et~al.}(2021)\citenamefont {Mi}, \citenamefont {Roushan}, \citenamefont {Quintana}, \citenamefont {Mandr{\`a}}, \citenamefont {Marshall}, \citenamefont {Neill}, \citenamefont {Arute}, \citenamefont {Arya}, \citenamefont {Atalaya}, \citenamefont {Babbush}, \citenamefont {Bardin}, \citenamefont {Barends}, \citenamefont {Basso}, \citenamefont {Bengtsson}, \citenamefont {Boixo}, \citenamefont {Bourassa}, \citenamefont {Broughton}, \citenamefont {Buckley}, \citenamefont {Buell}, \citenamefont {Burkett}, \citenamefont {Bushnell}, \citenamefont {Chen}, \citenamefont {Chiaro}, \citenamefont {Collins}, \citenamefont {Courtney}, \citenamefont {Demura}, \citenamefont {Derk}, \citenamefont {Dunsworth}, \citenamefont {Eppens}, \citenamefont {Erickson}, \citenamefont {Farhi}, \citenamefont {Fowler}, \citenamefont {Foxen}, \citenamefont {Gidney}, \citenamefont {Giustina}, \citenamefont {Gross}, \citenamefont {Harrigan}, \citenamefont {Harrington}, \citenamefont {Hilton}, \citenamefont
  {Ho}, \citenamefont {Hong}, \citenamefont {Huang}, \citenamefont {Huggins}, \citenamefont {Ioffe}, \citenamefont {Isakov}, \citenamefont {Jeffrey}, \citenamefont {Jiang}, \citenamefont {Jones}, \citenamefont {Kafri}, \citenamefont {Kelly}, \citenamefont {Kim}, \citenamefont {Kitaev}, \citenamefont {Klimov}, \citenamefont {Korotkov}, \citenamefont {Kostritsa}, \citenamefont {Landhuis}, \citenamefont {Laptev}, \citenamefont {Lucero}, \citenamefont {Martin}, \citenamefont {McClean}, \citenamefont {McCourt}, \citenamefont {McEwen}, \citenamefont {Megrant}, \citenamefont {Miao}, \citenamefont {Mohseni}, \citenamefont {Montazeri}, \citenamefont {Mruczkiewicz}, \citenamefont {Mutus}, \citenamefont {Naaman}, \citenamefont {Neeley}, \citenamefont {Newman}, \citenamefont {Niu}, \citenamefont {O'Brien}, \citenamefont {Opremcak}, \citenamefont {Ostby}, \citenamefont {Pato}, \citenamefont {Petukhov}, \citenamefont {Redd}, \citenamefont {Rubin}, \citenamefont {Sank}, \citenamefont {Satzinger}, \citenamefont {Shvarts},
  \citenamefont {Strain}, \citenamefont {Szalay}, \citenamefont {Trevithick}, \citenamefont {Villalonga}, \citenamefont {White}, \citenamefont {Yao}, \citenamefont {Yeh}, \citenamefont {Zalcman}, \citenamefont {Neven}, \citenamefont {Aleiner}, \citenamefont {Kechedzhi}, \citenamefont {Smelyanskiy},\ and\ \citenamefont {Chen}}]{mi2021}%
  \BibitemOpen
  \bibfield  {author} {\bibinfo {author} {\bibfnamefont {X.}~\bibnamefont {Mi}}, \bibinfo {author} {\bibfnamefont {P.}~\bibnamefont {Roushan}}, \bibinfo {author} {\bibfnamefont {C.}~\bibnamefont {Quintana}}, \bibinfo {author} {\bibfnamefont {S.}~\bibnamefont {Mandr{\`a}}}, \bibinfo {author} {\bibfnamefont {J.}~\bibnamefont {Marshall}}, \bibinfo {author} {\bibfnamefont {C.}~\bibnamefont {Neill}}, \bibinfo {author} {\bibfnamefont {F.}~\bibnamefont {Arute}}, \bibinfo {author} {\bibfnamefont {K.}~\bibnamefont {Arya}}, \bibinfo {author} {\bibfnamefont {J.}~\bibnamefont {Atalaya}}, \bibinfo {author} {\bibfnamefont {R.}~\bibnamefont {Babbush}}, \bibinfo {author} {\bibfnamefont {J.~C.}\ \bibnamefont {Bardin}}, \bibinfo {author} {\bibfnamefont {R.}~\bibnamefont {Barends}}, \bibinfo {author} {\bibfnamefont {J.}~\bibnamefont {Basso}}, \bibinfo {author} {\bibfnamefont {A.}~\bibnamefont {Bengtsson}}, \bibinfo {author} {\bibfnamefont {S.}~\bibnamefont {Boixo}}, \bibinfo {author} {\bibfnamefont {A.}~\bibnamefont
  {Bourassa}}, \bibinfo {author} {\bibfnamefont {M.}~\bibnamefont {Broughton}}, \bibinfo {author} {\bibfnamefont {B.~B.}\ \bibnamefont {Buckley}}, \bibinfo {author} {\bibfnamefont {D.~A.}\ \bibnamefont {Buell}}, \bibinfo {author} {\bibfnamefont {B.}~\bibnamefont {Burkett}}, \bibinfo {author} {\bibfnamefont {N.}~\bibnamefont {Bushnell}}, \bibinfo {author} {\bibfnamefont {Z.}~\bibnamefont {Chen}}, \bibinfo {author} {\bibfnamefont {B.}~\bibnamefont {Chiaro}}, \bibinfo {author} {\bibfnamefont {R.}~\bibnamefont {Collins}}, \bibinfo {author} {\bibfnamefont {W.}~\bibnamefont {Courtney}}, \bibinfo {author} {\bibfnamefont {S.}~\bibnamefont {Demura}}, \bibinfo {author} {\bibfnamefont {A.~R.}\ \bibnamefont {Derk}}, \bibinfo {author} {\bibfnamefont {A.}~\bibnamefont {Dunsworth}}, \bibinfo {author} {\bibfnamefont {D.}~\bibnamefont {Eppens}}, \bibinfo {author} {\bibfnamefont {C.}~\bibnamefont {Erickson}}, \bibinfo {author} {\bibfnamefont {E.}~\bibnamefont {Farhi}}, \bibinfo {author} {\bibfnamefont {A.~G.}\ \bibnamefont
  {Fowler}}, \bibinfo {author} {\bibfnamefont {B.}~\bibnamefont {Foxen}}, \bibinfo {author} {\bibfnamefont {C.}~\bibnamefont {Gidney}}, \bibinfo {author} {\bibfnamefont {M.}~\bibnamefont {Giustina}}, \bibinfo {author} {\bibfnamefont {J.~A.}\ \bibnamefont {Gross}}, \bibinfo {author} {\bibfnamefont {M.~P.}\ \bibnamefont {Harrigan}}, \bibinfo {author} {\bibfnamefont {S.~D.}\ \bibnamefont {Harrington}}, \bibinfo {author} {\bibfnamefont {J.}~\bibnamefont {Hilton}}, \bibinfo {author} {\bibfnamefont {A.}~\bibnamefont {Ho}}, \bibinfo {author} {\bibfnamefont {S.}~\bibnamefont {Hong}}, \bibinfo {author} {\bibfnamefont {T.}~\bibnamefont {Huang}}, \bibinfo {author} {\bibfnamefont {W.~J.}\ \bibnamefont {Huggins}}, \bibinfo {author} {\bibfnamefont {L.~B.}\ \bibnamefont {Ioffe}}, \bibinfo {author} {\bibfnamefont {S.~V.}\ \bibnamefont {Isakov}}, \bibinfo {author} {\bibfnamefont {E.}~\bibnamefont {Jeffrey}}, \bibinfo {author} {\bibfnamefont {Z.}~\bibnamefont {Jiang}}, \bibinfo {author} {\bibfnamefont {C.}~\bibnamefont
  {Jones}}, \bibinfo {author} {\bibfnamefont {D.}~\bibnamefont {Kafri}}, \bibinfo {author} {\bibfnamefont {J.}~\bibnamefont {Kelly}}, \bibinfo {author} {\bibfnamefont {S.}~\bibnamefont {Kim}}, \bibinfo {author} {\bibfnamefont {A.}~\bibnamefont {Kitaev}}, \bibinfo {author} {\bibfnamefont {P.~V.}\ \bibnamefont {Klimov}}, \bibinfo {author} {\bibfnamefont {A.~N.}\ \bibnamefont {Korotkov}}, \bibinfo {author} {\bibfnamefont {F.}~\bibnamefont {Kostritsa}}, \bibinfo {author} {\bibfnamefont {D.}~\bibnamefont {Landhuis}}, \bibinfo {author} {\bibfnamefont {P.}~\bibnamefont {Laptev}}, \bibinfo {author} {\bibfnamefont {E.}~\bibnamefont {Lucero}}, \bibinfo {author} {\bibfnamefont {O.}~\bibnamefont {Martin}}, \bibinfo {author} {\bibfnamefont {J.~R.}\ \bibnamefont {McClean}}, \bibinfo {author} {\bibfnamefont {T.}~\bibnamefont {McCourt}}, \bibinfo {author} {\bibfnamefont {M.}~\bibnamefont {McEwen}}, \bibinfo {author} {\bibfnamefont {A.}~\bibnamefont {Megrant}}, \bibinfo {author} {\bibfnamefont {K.~C.}\ \bibnamefont {Miao}},
  \bibinfo {author} {\bibfnamefont {M.}~\bibnamefont {Mohseni}}, \bibinfo {author} {\bibfnamefont {S.}~\bibnamefont {Montazeri}}, \bibinfo {author} {\bibfnamefont {W.}~\bibnamefont {Mruczkiewicz}}, \bibinfo {author} {\bibfnamefont {J.}~\bibnamefont {Mutus}}, \bibinfo {author} {\bibfnamefont {O.}~\bibnamefont {Naaman}}, \bibinfo {author} {\bibfnamefont {M.}~\bibnamefont {Neeley}}, \bibinfo {author} {\bibfnamefont {M.}~\bibnamefont {Newman}}, \bibinfo {author} {\bibfnamefont {M.~Y.}\ \bibnamefont {Niu}}, \bibinfo {author} {\bibfnamefont {T.~E.}\ \bibnamefont {O'Brien}}, \bibinfo {author} {\bibfnamefont {A.}~\bibnamefont {Opremcak}}, \bibinfo {author} {\bibfnamefont {E.}~\bibnamefont {Ostby}}, \bibinfo {author} {\bibfnamefont {B.}~\bibnamefont {Pato}}, \bibinfo {author} {\bibfnamefont {A.}~\bibnamefont {Petukhov}}, \bibinfo {author} {\bibfnamefont {N.}~\bibnamefont {Redd}}, \bibinfo {author} {\bibfnamefont {N.~C.}\ \bibnamefont {Rubin}}, \bibinfo {author} {\bibfnamefont {D.}~\bibnamefont {Sank}}, \bibinfo
  {author} {\bibfnamefont {K.~J.}\ \bibnamefont {Satzinger}}, \bibinfo {author} {\bibfnamefont {V.}~\bibnamefont {Shvarts}}, \bibinfo {author} {\bibfnamefont {D.}~\bibnamefont {Strain}}, \bibinfo {author} {\bibfnamefont {M.}~\bibnamefont {Szalay}}, \bibinfo {author} {\bibfnamefont {M.~D.}\ \bibnamefont {Trevithick}}, \bibinfo {author} {\bibfnamefont {B.}~\bibnamefont {Villalonga}}, \bibinfo {author} {\bibfnamefont {T.}~\bibnamefont {White}}, \bibinfo {author} {\bibfnamefont {Z.~J.}\ \bibnamefont {Yao}}, \bibinfo {author} {\bibfnamefont {P.}~\bibnamefont {Yeh}}, \bibinfo {author} {\bibfnamefont {A.}~\bibnamefont {Zalcman}}, \bibinfo {author} {\bibfnamefont {H.}~\bibnamefont {Neven}}, \bibinfo {author} {\bibfnamefont {I.}~\bibnamefont {Aleiner}}, \bibinfo {author} {\bibfnamefont {K.}~\bibnamefont {Kechedzhi}}, \bibinfo {author} {\bibfnamefont {V.}~\bibnamefont {Smelyanskiy}},\ and\ \bibinfo {author} {\bibfnamefont {Y.}~\bibnamefont {Chen}},\ }\bibfield  {title} {\bibinfo {title} {Information scrambling in
  quantum circuits},\ }\href {https://doi.org/10.1126/science.abg5029} {\bibfield  {journal} {\bibinfo  {journal} {Science}\ }\textbf {\bibinfo {volume} {374}},\ \bibinfo {pages} {1479} (\bibinfo {year} {2021})}\BibitemShut {NoStop}%
\bibitem [{\citenamefont {Wallquist}\ \emph {et~al.}(2009)\citenamefont {Wallquist}, \citenamefont {Hammerer}, \citenamefont {Rabl}, \citenamefont {Lukin},\ and\ \citenamefont {Zoller}}]{Wallquist2009}%
  \BibitemOpen
  \bibfield  {author} {\bibinfo {author} {\bibfnamefont {M.}~\bibnamefont {Wallquist}}, \bibinfo {author} {\bibfnamefont {K.}~\bibnamefont {Hammerer}}, \bibinfo {author} {\bibfnamefont {P.}~\bibnamefont {Rabl}}, \bibinfo {author} {\bibfnamefont {M.}~\bibnamefont {Lukin}},\ and\ \bibinfo {author} {\bibfnamefont {P.}~\bibnamefont {Zoller}},\ }\bibfield  {title} {\bibinfo {title} {Hybrid quantum devices and quantum engineering},\ }\href {https://doi.org/10.1088/0031-8949/2009/T137/014001} {\bibfield  {journal} {\bibinfo  {journal} {Phys. Scr.}\ }\textbf {\bibinfo {volume} {2009}},\ \bibinfo {pages} {014001} (\bibinfo {year} {2009})}\BibitemShut {NoStop}%
\bibitem [{\citenamefont {Wei}(2021)}]{Wei2021}%
  \BibitemOpen
  \bibfield  {author} {\bibinfo {author} {\bibfnamefont {T.-C.}\ \bibnamefont {Wei}},\ }\bibfield  {title} {\bibinfo {title} {Measurement-based quantum computation},\ }in\ \href {https://doi.org/10.1093/acrefore/9780190871994.013.31} {\emph {\bibinfo {booktitle} {Oxford Research Encyclopedia of Physics}}},\ \bibinfo {editor} {edited by\ \bibinfo {editor} {\bibfnamefont {B.}~\bibnamefont {Foster}}}\ (\bibinfo  {publisher} {Oxford University Press},\ \bibinfo {address} {Oxford},\ \bibinfo {year} {2021})\BibitemShut {NoStop}%
\bibitem [{\citenamefont {Goerz}\ \emph {et~al.}(2011)\citenamefont {Goerz}, \citenamefont {Calarco},\ and\ \citenamefont {Koch}}]{Goerz2011}%
  \BibitemOpen
  \bibfield  {author} {\bibinfo {author} {\bibfnamefont {M.~H.}\ \bibnamefont {Goerz}}, \bibinfo {author} {\bibfnamefont {T.}~\bibnamefont {Calarco}},\ and\ \bibinfo {author} {\bibfnamefont {C.~P.}\ \bibnamefont {Koch}},\ }\bibfield  {title} {\bibinfo {title} {The quantum speed limit of optimal controlled phasegates for trapped neutral atoms},\ }\href {https://doi.org/10.1088/0953-4075/44/15/154011} {\bibfield  {journal} {\bibinfo  {journal} {J. Phys. B}\ }\textbf {\bibinfo {volume} {44}},\ \bibinfo {pages} {154011} (\bibinfo {year} {2011})}\BibitemShut {NoStop}%
\bibitem [{\citenamefont {M\"uller}\ \emph {et~al.}(2011)\citenamefont {M\"uller}, \citenamefont {Reich}, \citenamefont {Murphy}, \citenamefont {Yuan}, \citenamefont {Vala}, \citenamefont {Whaley}, \citenamefont {Calarco},\ and\ \citenamefont {Koch}}]{Muller2011}%
  \BibitemOpen
  \bibfield  {author} {\bibinfo {author} {\bibfnamefont {M.~M.}\ \bibnamefont {M\"uller}}, \bibinfo {author} {\bibfnamefont {D.~M.}\ \bibnamefont {Reich}}, \bibinfo {author} {\bibfnamefont {M.}~\bibnamefont {Murphy}}, \bibinfo {author} {\bibfnamefont {H.}~\bibnamefont {Yuan}}, \bibinfo {author} {\bibfnamefont {J.}~\bibnamefont {Vala}}, \bibinfo {author} {\bibfnamefont {K.~B.}\ \bibnamefont {Whaley}}, \bibinfo {author} {\bibfnamefont {T.}~\bibnamefont {Calarco}},\ and\ \bibinfo {author} {\bibfnamefont {C.~P.}\ \bibnamefont {Koch}},\ }\bibfield  {title} {\bibinfo {title} {Optimizing entangling quantum gates for physical systems},\ }\href {https://doi.org/10.1103/PhysRevA.84.042315} {\bibfield  {journal} {\bibinfo  {journal} {Phys. Rev. A}\ }\textbf {\bibinfo {volume} {84}},\ \bibinfo {pages} {042315} (\bibinfo {year} {2011})}\BibitemShut {NoStop}%
\bibitem [{\citenamefont {Hughes}\ \emph {et~al.}(2020)\citenamefont {Hughes}, \citenamefont {Frye}, \citenamefont {Sawant}, \citenamefont {Bhole}, \citenamefont {Jones}, \citenamefont {Cornish}, \citenamefont {Tarbutt}, \citenamefont {Hutson}, \citenamefont {Jaksch},\ and\ \citenamefont {Mur-Petit}}]{Hughes2020}%
  \BibitemOpen
  \bibfield  {author} {\bibinfo {author} {\bibfnamefont {M.}~\bibnamefont {Hughes}}, \bibinfo {author} {\bibfnamefont {M.~D.}\ \bibnamefont {Frye}}, \bibinfo {author} {\bibfnamefont {R.}~\bibnamefont {Sawant}}, \bibinfo {author} {\bibfnamefont {G.}~\bibnamefont {Bhole}}, \bibinfo {author} {\bibfnamefont {J.~A.}\ \bibnamefont {Jones}}, \bibinfo {author} {\bibfnamefont {S.~L.}\ \bibnamefont {Cornish}}, \bibinfo {author} {\bibfnamefont {M.~R.}\ \bibnamefont {Tarbutt}}, \bibinfo {author} {\bibfnamefont {J.~M.}\ \bibnamefont {Hutson}}, \bibinfo {author} {\bibfnamefont {D.}~\bibnamefont {Jaksch}},\ and\ \bibinfo {author} {\bibfnamefont {J.}~\bibnamefont {Mur-Petit}},\ }\bibfield  {title} {\bibinfo {title} {Robust entangling gate for polar molecules using magnetic and microwave fields},\ }\href {https://doi.org/10.1103/PhysRevA.101.062308} {\bibfield  {journal} {\bibinfo  {journal} {Phys. Rev. A}\ }\textbf {\bibinfo {volume} {101}},\ \bibinfo {pages} {062308} (\bibinfo {year} {2020})}\BibitemShut {NoStop}%
\bibitem [{\citenamefont {Bergonzoni}\ \emph {et~al.}(2025)\citenamefont {Bergonzoni}, \citenamefont {Jandura},\ and\ \citenamefont {Pupillo}}]{Bergonzoni2025}%
  \BibitemOpen
  \bibfield  {author} {\bibinfo {author} {\bibfnamefont {M.}~\bibnamefont {Bergonzoni}}, \bibinfo {author} {\bibfnamefont {S.}~\bibnamefont {Jandura}},\ and\ \bibinfo {author} {\bibfnamefont {G.}~\bibnamefont {Pupillo}},\ }\bibfield  {title} {\bibinfo {title} {{{iSWAP}} gate with polar molecules: {{Robustness}} criteria for entangling operations},\ }\href {https://doi.org/10.1103/q9sd-rfp6} {\bibfield  {journal} {\bibinfo  {journal} {Phys. Rev. A}\ }\textbf {\bibinfo {volume} {112}},\ \bibinfo {pages} {032621} (\bibinfo {year} {2025})}\BibitemShut {NoStop}%
\bibitem [{\citenamefont {Evered}\ \emph {et~al.}(2026)\citenamefont {Evered}, \citenamefont {Xu}, \citenamefont {Li}, \citenamefont {Geim}, \citenamefont {Ataides}, \citenamefont {Kalinowski}, \citenamefont {Bluvstein}, \citenamefont {Maskara}, \citenamefont {Kokail}, \citenamefont {Greiner}, \citenamefont {Vuletić},\ and\ \citenamefont {Lukin}}]{Evered2026}%
  \BibitemOpen
  \bibfield  {author} {\bibinfo {author} {\bibfnamefont {S.~J.}\ \bibnamefont {Evered}}, \bibinfo {author} {\bibfnamefont {M.}~\bibnamefont {Xu}}, \bibinfo {author} {\bibfnamefont {S.~H.}\ \bibnamefont {Li}}, \bibinfo {author} {\bibfnamefont {A.~A.}\ \bibnamefont {Geim}}, \bibinfo {author} {\bibfnamefont {J.~P.~B.}\ \bibnamefont {Ataides}}, \bibinfo {author} {\bibfnamefont {M.}~\bibnamefont {Kalinowski}}, \bibinfo {author} {\bibfnamefont {D.}~\bibnamefont {Bluvstein}}, \bibinfo {author} {\bibfnamefont {N.}~\bibnamefont {Maskara}}, \bibinfo {author} {\bibfnamefont {C.}~\bibnamefont {Kokail}}, \bibinfo {author} {\bibfnamefont {M.}~\bibnamefont {Greiner}}, \bibinfo {author} {\bibfnamefont {V.}~\bibnamefont {Vuletić}},\ and\ \bibinfo {author} {\bibfnamefont {M.~D.}\ \bibnamefont {Lukin}},\ }\href {https://arxiv.org/abs/2604.25987} {\bibinfo {title} {High-fidelity entangling gates and nonlocal circuits with neutral atoms}} (\bibinfo {year} {2026}),\ \Eprint {https://arxiv.org/abs/2604.25987} {arXiv:2604.25987
  [quant-ph]} \BibitemShut {NoStop}%
\bibitem [{\citenamefont {Finkelstein}\ \emph {et~al.}(2024)\citenamefont {Finkelstein}, \citenamefont {Tsai}, \citenamefont {Sun}, \citenamefont {Scholl}, \citenamefont {Direkci}, \citenamefont {Gefen}, \citenamefont {Choi}, \citenamefont {Shaw},\ and\ \citenamefont {Endres}}]{Finkelstein2024}%
  \BibitemOpen
  \bibfield  {author} {\bibinfo {author} {\bibfnamefont {R.}~\bibnamefont {Finkelstein}}, \bibinfo {author} {\bibfnamefont {R.~B.-S.}\ \bibnamefont {Tsai}}, \bibinfo {author} {\bibfnamefont {X.}~\bibnamefont {Sun}}, \bibinfo {author} {\bibfnamefont {P.}~\bibnamefont {Scholl}}, \bibinfo {author} {\bibfnamefont {S.}~\bibnamefont {Direkci}}, \bibinfo {author} {\bibfnamefont {T.}~\bibnamefont {Gefen}}, \bibinfo {author} {\bibfnamefont {J.}~\bibnamefont {Choi}}, \bibinfo {author} {\bibfnamefont {A.~L.}\ \bibnamefont {Shaw}},\ and\ \bibinfo {author} {\bibfnamefont {M.}~\bibnamefont {Endres}},\ }\bibfield  {title} {\bibinfo {title} {Universal quantum operations and ancilla-based read-out for tweezer clocks},\ }\href {https://doi.org/10.1038/s41586-024-08005-8} {\bibfield  {journal} {\bibinfo  {journal} {Nature}\ }\textbf {\bibinfo {volume} {634}},\ \bibinfo {pages} {321} (\bibinfo {year} {2024})}\BibitemShut {NoStop}%
\bibitem [{\citenamefont {Baranov}\ \emph {et~al.}(2012)\citenamefont {Baranov}, \citenamefont {Dalmonte}, \citenamefont {Pupillo},\ and\ \citenamefont {Zoller}}]{Baranov2012}%
  \BibitemOpen
  \bibfield  {author} {\bibinfo {author} {\bibfnamefont {M.}~\bibnamefont {Baranov}}, \bibinfo {author} {\bibfnamefont {M.}~\bibnamefont {Dalmonte}}, \bibinfo {author} {\bibfnamefont {G.}~\bibnamefont {Pupillo}},\ and\ \bibinfo {author} {\bibfnamefont {P.}~\bibnamefont {Zoller}},\ }\bibfield  {title} {\bibinfo {title} {Condensed matter theory of dipolar quantum gases},\ }\href {https://doi.org/10.1021/cr2003568} {\bibfield  {journal} {\bibinfo  {journal} {Chem. Rev.}\ }\textbf {\bibinfo {volume} {112}},\ \bibinfo {pages} {5012} (\bibinfo {year} {2012})}\BibitemShut {NoStop}%
\bibitem [{\citenamefont {Zhang}\ \emph {et~al.}(2023)\citenamefont {Zhang}, \citenamefont {Yu}, \citenamefont {Jadbabaie},\ and\ \citenamefont {Hutzler}}]{Zhang2023DecoherenceFree}%
  \BibitemOpen
  \bibfield  {author} {\bibinfo {author} {\bibfnamefont {C.}~\bibnamefont {Zhang}}, \bibinfo {author} {\bibfnamefont {P.}~\bibnamefont {Yu}}, \bibinfo {author} {\bibfnamefont {A.}~\bibnamefont {Jadbabaie}},\ and\ \bibinfo {author} {\bibfnamefont {N.~R.}\ \bibnamefont {Hutzler}},\ }\bibfield  {title} {\bibinfo {title} {Quantum-enhanced metrology for molecular symmetry violation using decoherence-free subspaces},\ }\href {https://doi.org/10.1103/PhysRevLett.131.193602} {\bibfield  {journal} {\bibinfo  {journal} {Phys. Rev. Lett.}\ }\textbf {\bibinfo {volume} {131}},\ \bibinfo {pages} {193602} (\bibinfo {year} {2023})}\BibitemShut {NoStop}%
\bibitem [{\citenamefont {Spence}\ \emph {et~al.}(2022)\citenamefont {Spence}, \citenamefont {Brooks}, \citenamefont {Ruttley}, \citenamefont {Guttridge},\ and\ \citenamefont {Cornish}}]{Spence22}%
  \BibitemOpen
  \bibfield  {author} {\bibinfo {author} {\bibfnamefont {S.}~\bibnamefont {Spence}}, \bibinfo {author} {\bibfnamefont {R.~V.}\ \bibnamefont {Brooks}}, \bibinfo {author} {\bibfnamefont {D.~K.}\ \bibnamefont {Ruttley}}, \bibinfo {author} {\bibfnamefont {A.}~\bibnamefont {Guttridge}},\ and\ \bibinfo {author} {\bibfnamefont {S.~L.}\ \bibnamefont {Cornish}},\ }\bibfield  {title} {\bibinfo {title} {Preparation of \(^{87}\){R}b and \(^{133}\){C}s in the motional ground state of a single optical tweezer},\ }\href {https://doi.org/10.1088/1367-2630/ac95b9} {\bibfield  {journal} {\bibinfo  {journal} {New J. Phys.}\ }\textbf {\bibinfo {volume} {24}},\ \bibinfo {pages} {103022} (\bibinfo {year} {2022})}\BibitemShut {NoStop}%
\bibitem [{\citenamefont {Tauschinsky}\ \emph {et~al.}(2013)\citenamefont {Tauschinsky}, \citenamefont {Newell}, \citenamefont {van Linden van~den Heuvell},\ and\ \citenamefont {Spreeuw}}]{Tauschinsky2013}%
  \BibitemOpen
  \bibfield  {author} {\bibinfo {author} {\bibfnamefont {A.}~\bibnamefont {Tauschinsky}}, \bibinfo {author} {\bibfnamefont {R.}~\bibnamefont {Newell}}, \bibinfo {author} {\bibfnamefont {H.~B.}\ \bibnamefont {van Linden van~den Heuvell}},\ and\ \bibinfo {author} {\bibfnamefont {R.~J.~C.}\ \bibnamefont {Spreeuw}},\ }\bibfield  {title} {\bibinfo {title} {Measurement of \textsuperscript{87}{Rb} {R}ydberg-state hyperfine splitting in a room-temperature vapor cell},\ }\href {https://doi.org/10.1103/PhysRevA.87.042522} {\bibfield  {journal} {\bibinfo  {journal} {Phys. Rev. A}\ }\textbf {\bibinfo {volume} {87}},\ \bibinfo {pages} {042522} (\bibinfo {year} {2013})}\BibitemShut {NoStop}%
\bibitem [{\citenamefont {Guttridge}\ \emph {et~al.}(2023)\citenamefont {Guttridge}, \citenamefont {Ruttley}, \citenamefont {Baldock}, \citenamefont {Gonz\'alez-F\'erez}, \citenamefont {Sadeghpour}, \citenamefont {Adams},\ and\ \citenamefont {Cornish}}]{Guttridge2023}%
  \BibitemOpen
  \bibfield  {author} {\bibinfo {author} {\bibfnamefont {A.}~\bibnamefont {Guttridge}}, \bibinfo {author} {\bibfnamefont {D.~K.}\ \bibnamefont {Ruttley}}, \bibinfo {author} {\bibfnamefont {A.~C.}\ \bibnamefont {Baldock}}, \bibinfo {author} {\bibfnamefont {R.}~\bibnamefont {Gonz\'alez-F\'erez}}, \bibinfo {author} {\bibfnamefont {H.~R.}\ \bibnamefont {Sadeghpour}}, \bibinfo {author} {\bibfnamefont {C.~S.}\ \bibnamefont {Adams}},\ and\ \bibinfo {author} {\bibfnamefont {S.~L.}\ \bibnamefont {Cornish}},\ }\bibfield  {title} {\bibinfo {title} {Observation of {R}ydberg blockade due to the charge-dipole interaction between an atom and a polar molecule},\ }\href {https://doi.org/10.1103/PhysRevLett.131.013401} {\bibfield  {journal} {\bibinfo  {journal} {Phys. Rev. Lett.}\ }\textbf {\bibinfo {volume} {131}},\ \bibinfo {pages} {013401} (\bibinfo {year} {2023})}\BibitemShut {NoStop}%
\bibitem [{\citenamefont {Hepworth}\ \emph {et~al.}(2025)\citenamefont {Hepworth}, \citenamefont {Ruttley}, \citenamefont {von Gierke}, \citenamefont {Gregory}, \citenamefont {Guttridge},\ and\ \citenamefont {Cornish}}]{Hepworth2025}%
  \BibitemOpen
  \bibfield  {author} {\bibinfo {author} {\bibfnamefont {T.~R.}\ \bibnamefont {Hepworth}}, \bibinfo {author} {\bibfnamefont {D.~K.}\ \bibnamefont {Ruttley}}, \bibinfo {author} {\bibfnamefont {F.}~\bibnamefont {von Gierke}}, \bibinfo {author} {\bibfnamefont {P.~D.}\ \bibnamefont {Gregory}}, \bibinfo {author} {\bibfnamefont {A.}~\bibnamefont {Guttridge}},\ and\ \bibinfo {author} {\bibfnamefont {S.~L.}\ \bibnamefont {Cornish}},\ }\bibfield  {title} {\bibinfo {title} {Long-lived multilevel coherences and spin-1 dynamics encoded in the rotational states of ultracold molecules},\ }\href {https://doi.org/10.1038/s41467-025-62275-y} {\bibfield  {journal} {\bibinfo  {journal} {Nat. Commun.}\ }\textbf {\bibinfo {volume} {16}},\ \bibinfo {pages} {7131} (\bibinfo {year} {2025})}\BibitemShut {NoStop}%
\bibitem [{\citenamefont {Gregory}\ \emph {et~al.}(2016)\citenamefont {Gregory}, \citenamefont {Aldegunde}, \citenamefont {Hutson},\ and\ \citenamefont {Cornish}}]{Gregory2016}%
  \BibitemOpen
  \bibfield  {author} {\bibinfo {author} {\bibfnamefont {P.~D.}\ \bibnamefont {Gregory}}, \bibinfo {author} {\bibfnamefont {J.}~\bibnamefont {Aldegunde}}, \bibinfo {author} {\bibfnamefont {J.~M.}\ \bibnamefont {Hutson}},\ and\ \bibinfo {author} {\bibfnamefont {S.~L.}\ \bibnamefont {Cornish}},\ }\bibfield  {title} {\bibinfo {title} {Controlling the rotational and hyperfine state of ultracold $^{87}\mathrm{Rb}^{133}\mathrm{Cs}$ molecules},\ }\href {https://doi.org/10.1103/PhysRevA.94.041403} {\bibfield  {journal} {\bibinfo  {journal} {Phys. Rev. A}\ }\textbf {\bibinfo {volume} {94}},\ \bibinfo {pages} {041403(R)} (\bibinfo {year} {2016})}\BibitemShut {NoStop}%
\bibitem [{\citenamefont {Gonz\'alez-F\'erez}\ \emph {et~al.}(2015)\citenamefont {Gonz\'alez-F\'erez}, \citenamefont {Sadeghpour},\ and\ \citenamefont {Schmelcher}}]{gonzalez15}%
  \BibitemOpen
  \bibfield  {author} {\bibinfo {author} {\bibfnamefont {R.}~\bibnamefont {Gonz\'alez-F\'erez}}, \bibinfo {author} {\bibfnamefont {H.~R.}\ \bibnamefont {Sadeghpour}},\ and\ \bibinfo {author} {\bibfnamefont {P.}~\bibnamefont {Schmelcher}},\ }\bibfield  {title} {\bibinfo {title} {Rotational hybridization, and control of alignment and orientation in triatomic ultralong-range {R}ydberg molecules},\ }\href {https://doi.org/10.1088/1367-2630/17/1/013021} {\bibfield  {journal} {\bibinfo  {journal} {New J. Phys.}\ }\textbf {\bibinfo {volume} {17}},\ \bibinfo {pages} {013021} (\bibinfo {year} {2015})}\BibitemShut {NoStop}%
\bibitem [{\citenamefont {Wall}\ \emph {et~al.}(2015)\citenamefont {Wall}, \citenamefont {Hazzard},\ and\ \citenamefont {Rey}}]{Wall2015}%
  \BibitemOpen
  \bibfield  {author} {\bibinfo {author} {\bibfnamefont {M.~L.}\ \bibnamefont {Wall}}, \bibinfo {author} {\bibfnamefont {K.~R.~A.}\ \bibnamefont {Hazzard}},\ and\ \bibinfo {author} {\bibfnamefont {A.~M.}\ \bibnamefont {Rey}},\ }\bibinfo {title} {Quantum magnetism with ultracold molecules},\ in\ \href {https://doi.org/10.1142/9789814678704_0001} {\emph {\bibinfo {booktitle} {From Atomic to Mesoscale}}}\ (\bibinfo  {publisher} {World Scientific},\ \bibinfo {address} {Singapore},\ \bibinfo {year} {2015})\ Chap.~\bibinfo {chapter} {1}, pp.\ \bibinfo {pages} {3--37}\BibitemShut {NoStop}%
\bibitem [{\citenamefont {Ravets}\ \emph {et~al.}(2015)\citenamefont {Ravets}, \citenamefont {Labuhn}, \citenamefont {Barredo}, \citenamefont {Lahaye},\ and\ \citenamefont {Browaeys}}]{Ravets2015}%
  \BibitemOpen
  \bibfield  {author} {\bibinfo {author} {\bibfnamefont {S.}~\bibnamefont {Ravets}}, \bibinfo {author} {\bibfnamefont {H.}~\bibnamefont {Labuhn}}, \bibinfo {author} {\bibfnamefont {D.}~\bibnamefont {Barredo}}, \bibinfo {author} {\bibfnamefont {T.}~\bibnamefont {Lahaye}},\ and\ \bibinfo {author} {\bibfnamefont {A.}~\bibnamefont {Browaeys}},\ }\bibfield  {title} {\bibinfo {title} {Measurement of the angular dependence of the dipole-dipole interaction between two individual {R}ydberg atoms at a {F}\"orster resonance},\ }\href {https://doi.org/10.1103/PhysRevA.92.020701} {\bibfield  {journal} {\bibinfo  {journal} {Phys. Rev. A}\ }\textbf {\bibinfo {volume} {92}},\ \bibinfo {pages} {020701(R)} (\bibinfo {year} {2015})}\BibitemShut {NoStop}%
\bibitem [{\citenamefont {Wadenpfuhl}\ and\ \citenamefont {Adams}(2025)}]{Wadenpfuhl2025}%
  \BibitemOpen
  \bibfield  {author} {\bibinfo {author} {\bibfnamefont {K.}~\bibnamefont {Wadenpfuhl}}\ and\ \bibinfo {author} {\bibfnamefont {C.~S.}\ \bibnamefont {Adams}},\ }\bibfield  {title} {\bibinfo {title} {Unraveling the structures in the van der {W}aals interactions of alkali-metal {R}ydberg atoms},\ }\href {https://doi.org/10.1103/PhysRevA.111.062803} {\bibfield  {journal} {\bibinfo  {journal} {Phys. Rev. A}\ }\textbf {\bibinfo {volume} {111}},\ \bibinfo {pages} {062803} (\bibinfo {year} {2025})}\BibitemShut {NoStop}%
\bibitem [{\citenamefont {García-Garrido}\ \emph {et~al.}(tion)\citenamefont {García-Garrido}, \citenamefont {Hepworth}, \citenamefont {Fernández-Mayo}, \citenamefont {Ruttley}, \citenamefont {Cornish},\ and\ \citenamefont {González-Férez}}]{garciagarrido2026Unpublished}%
  \BibitemOpen
  \bibfield  {author} {\bibinfo {author} {\bibfnamefont {J.~M.}\ \bibnamefont {García-Garrido}}, \bibinfo {author} {\bibfnamefont {T.~R.}\ \bibnamefont {Hepworth}}, \bibinfo {author} {\bibfnamefont {P.}~\bibnamefont {Fernández-Mayo}}, \bibinfo {author} {\bibfnamefont {D.~K.}\ \bibnamefont {Ruttley}}, \bibinfo {author} {\bibfnamefont {S.~L.}\ \bibnamefont {Cornish}},\ and\ \bibinfo {author} {\bibfnamefont {R.}~\bibnamefont {González-Férez}},\ }\bibfield  {title} {\bibinfo {title} {Resonant interaction between a {R}ydberg atom and a polar molecule: Theoretical description}} (\bibinfo {year} {in preparation})\BibitemShut {NoStop}%
\bibitem [{\citenamefont {Brooks}\ \emph {et~al.}(2021)\citenamefont {Brooks}, \citenamefont {Spence}, \citenamefont {Guttridge}, \citenamefont {Alampounti}, \citenamefont {Rakonjac}, \citenamefont {McArd}, \citenamefont {Hutson},\ and\ \citenamefont {Cornish}}]{Brooks21}%
  \BibitemOpen
  \bibfield  {author} {\bibinfo {author} {\bibfnamefont {R.~V.}\ \bibnamefont {Brooks}}, \bibinfo {author} {\bibfnamefont {S.}~\bibnamefont {Spence}}, \bibinfo {author} {\bibfnamefont {A.}~\bibnamefont {Guttridge}}, \bibinfo {author} {\bibfnamefont {A.}~\bibnamefont {Alampounti}}, \bibinfo {author} {\bibfnamefont {A.}~\bibnamefont {Rakonjac}}, \bibinfo {author} {\bibfnamefont {L.~A.}\ \bibnamefont {McArd}}, \bibinfo {author} {\bibfnamefont {J.~M.}\ \bibnamefont {Hutson}},\ and\ \bibinfo {author} {\bibfnamefont {S.~L.}\ \bibnamefont {Cornish}},\ }\bibfield  {title} {\bibinfo {title} {Preparation of one $^{87}${R}b and one $^{133}${C}s atom in a single optical tweezer},\ }\href {https://doi.org/10.1088/1367-2630/ac0000} {\bibfield  {journal} {\bibinfo  {journal} {New J. Phys.}\ }\textbf {\bibinfo {volume} {23}},\ \bibinfo {pages} {065002} (\bibinfo {year} {2021})}\BibitemShut {NoStop}%
\bibitem [{\citenamefont {Ruttley}\ \emph {et~al.}(2023)\citenamefont {Ruttley}, \citenamefont {Guttridge}, \citenamefont {Spence}, \citenamefont {Bird}, \citenamefont {Le~Sueur}, \citenamefont {Hutson},\ and\ \citenamefont {Cornish}}]{Ruttley2023}%
  \BibitemOpen
  \bibfield  {author} {\bibinfo {author} {\bibfnamefont {D.~K.}\ \bibnamefont {Ruttley}}, \bibinfo {author} {\bibfnamefont {A.}~\bibnamefont {Guttridge}}, \bibinfo {author} {\bibfnamefont {S.}~\bibnamefont {Spence}}, \bibinfo {author} {\bibfnamefont {R.~C.}\ \bibnamefont {Bird}}, \bibinfo {author} {\bibfnamefont {C.~R.}\ \bibnamefont {Le~Sueur}}, \bibinfo {author} {\bibfnamefont {J.~M.}\ \bibnamefont {Hutson}},\ and\ \bibinfo {author} {\bibfnamefont {S.~L.}\ \bibnamefont {Cornish}},\ }\bibfield  {title} {\bibinfo {title} {Formation of ultracold molecules by merging optical tweezers},\ }\href {https://doi.org/10.1103/PhysRevLett.130.223401} {\bibfield  {journal} {\bibinfo  {journal} {Phys. Rev. Lett.}\ }\textbf {\bibinfo {volume} {130}},\ \bibinfo {pages} {223401} (\bibinfo {year} {2023})}\BibitemShut {NoStop}%
\bibitem [{\citenamefont {Barakhshan}\ \emph {et~al.}(2022)\citenamefont {Barakhshan}, \citenamefont {Marrs}, \citenamefont {Bhosale}, \citenamefont {Arora}, \citenamefont {Eigenmann},\ and\ \citenamefont {Safronova}}]{UDportal}%
  \BibitemOpen
  \bibfield  {author} {\bibinfo {author} {\bibfnamefont {P.}~\bibnamefont {Barakhshan}}, \bibinfo {author} {\bibfnamefont {A.}~\bibnamefont {Marrs}}, \bibinfo {author} {\bibfnamefont {A.}~\bibnamefont {Bhosale}}, \bibinfo {author} {\bibfnamefont {B.}~\bibnamefont {Arora}}, \bibinfo {author} {\bibfnamefont {R.}~\bibnamefont {Eigenmann}},\ and\ \bibinfo {author} {\bibfnamefont {M.~S.}\ \bibnamefont {Safronova}},\ }\href {https://www.udel.edu/atom} {\bibinfo {title} {Portal for high-precision atomic data and computation (version 2.0)}},\ \bibinfo {howpublished} {[Online]} (\bibinfo {year} {2022})\BibitemShut {NoStop}%
\bibitem [{\citenamefont {Blackmore}\ \emph {et~al.}(2020)\citenamefont {Blackmore}, \citenamefont {Sawant}, \citenamefont {Gregory}, \citenamefont {Bromley}, \citenamefont {Aldegunde}, \citenamefont {Hutson},\ and\ \citenamefont {Cornish}}]{Blackmore2020}%
  \BibitemOpen
  \bibfield  {author} {\bibinfo {author} {\bibfnamefont {J.~A.}\ \bibnamefont {Blackmore}}, \bibinfo {author} {\bibfnamefont {R.}~\bibnamefont {Sawant}}, \bibinfo {author} {\bibfnamefont {P.~D.}\ \bibnamefont {Gregory}}, \bibinfo {author} {\bibfnamefont {S.~L.}\ \bibnamefont {Bromley}}, \bibinfo {author} {\bibfnamefont {J.}~\bibnamefont {Aldegunde}}, \bibinfo {author} {\bibfnamefont {J.~M.}\ \bibnamefont {Hutson}},\ and\ \bibinfo {author} {\bibfnamefont {S.~L.}\ \bibnamefont {Cornish}},\ }\bibfield  {title} {\bibinfo {title} {Controlling the ac {Stark} effect of {RbCs} with dc electric and magnetic fields},\ }\href {https://doi.org/10.1103/PhysRevA.102.053316} {\bibfield  {journal} {\bibinfo  {journal} {Phys. Rev. A}\ }\textbf {\bibinfo {volume} {102}},\ \bibinfo {pages} {053316} (\bibinfo {year} {2020})}\BibitemShut {NoStop}%
\bibitem [{\citenamefont {Blackmore}\ \emph {et~al.}(2023)\citenamefont {Blackmore}, \citenamefont {Gregory}, \citenamefont {Hutson},\ and\ \citenamefont {Cornish}}]{blackmore2023}%
  \BibitemOpen
  \bibfield  {author} {\bibinfo {author} {\bibfnamefont {J.~A.}\ \bibnamefont {Blackmore}}, \bibinfo {author} {\bibfnamefont {P.~D.}\ \bibnamefont {Gregory}}, \bibinfo {author} {\bibfnamefont {J.~M.}\ \bibnamefont {Hutson}},\ and\ \bibinfo {author} {\bibfnamefont {S.~L.}\ \bibnamefont {Cornish}},\ }\bibfield  {title} {\bibinfo {title} {Diatomic-py: A {P}ython module for calculating the rotational and hyperfine structure of $^{1}{\Sigma}$ molecules},\ }\href {https://doi.org/https://doi.org/10.1016/j.cpc.2022.108512} {\bibfield  {journal} {\bibinfo  {journal} {Comput. Phys. Commun.}\ }\textbf {\bibinfo {volume} {282}},\ \bibinfo {pages} {108512} (\bibinfo {year} {2023})}\BibitemShut {NoStop}%
\bibitem [{\citenamefont {Vexiau}\ \emph {et~al.}(2017)\citenamefont {Vexiau}, \citenamefont {Borsalino}, \citenamefont {Lepers}, \citenamefont {Orbán}, \citenamefont {Aymar}, \citenamefont {Dulieu},\ and\ \citenamefont {Bouloufa-Maafa}}]{Vexiau2017}%
  \BibitemOpen
  \bibfield  {author} {\bibinfo {author} {\bibfnamefont {R.}~\bibnamefont {Vexiau}}, \bibinfo {author} {\bibfnamefont {D.}~\bibnamefont {Borsalino}}, \bibinfo {author} {\bibfnamefont {M.}~\bibnamefont {Lepers}}, \bibinfo {author} {\bibfnamefont {A.}~\bibnamefont {Orbán}}, \bibinfo {author} {\bibfnamefont {M.}~\bibnamefont {Aymar}}, \bibinfo {author} {\bibfnamefont {O.}~\bibnamefont {Dulieu}},\ and\ \bibinfo {author} {\bibfnamefont {N.}~\bibnamefont {Bouloufa-Maafa}},\ }\bibfield  {title} {\bibinfo {title} {Dynamic dipole polarizabilities of heteronuclear alkali dimers: optical response, trapping and control of ultracold molecules},\ }\href {https://doi.org/10.1080/0144235X.2017.1351821} {\bibfield  {journal} {\bibinfo  {journal} {Int. Rev. Phys. Chem.}\ }\textbf {\bibinfo {volume} {36}},\ \bibinfo {pages} {709} (\bibinfo {year} {2017})}\BibitemShut {NoStop}%
\bibitem [{\citenamefont {Stefanazzi}\ \emph {et~al.}(2022)\citenamefont {Stefanazzi}, \citenamefont {Treptow}, \citenamefont {Wilcer}, \citenamefont {Stoughton}, \citenamefont {Bradford}, \citenamefont {Uemura}, \citenamefont {Zorzetti}, \citenamefont {Montella}, \citenamefont {Cancelo}, \citenamefont {Sussman}, \citenamefont {Houck}, \citenamefont {Saxena}, \citenamefont {Arnaldi}, \citenamefont {Agrawal}, \citenamefont {Zhang}, \citenamefont {Ding},\ and\ \citenamefont {Schuster}}]{Stefanazzi2022}%
  \BibitemOpen
  \bibfield  {author} {\bibinfo {author} {\bibfnamefont {L.}~\bibnamefont {Stefanazzi}}, \bibinfo {author} {\bibfnamefont {K.}~\bibnamefont {Treptow}}, \bibinfo {author} {\bibfnamefont {N.}~\bibnamefont {Wilcer}}, \bibinfo {author} {\bibfnamefont {C.}~\bibnamefont {Stoughton}}, \bibinfo {author} {\bibfnamefont {C.}~\bibnamefont {Bradford}}, \bibinfo {author} {\bibfnamefont {S.}~\bibnamefont {Uemura}}, \bibinfo {author} {\bibfnamefont {S.}~\bibnamefont {Zorzetti}}, \bibinfo {author} {\bibfnamefont {S.}~\bibnamefont {Montella}}, \bibinfo {author} {\bibfnamefont {G.}~\bibnamefont {Cancelo}}, \bibinfo {author} {\bibfnamefont {S.}~\bibnamefont {Sussman}}, \bibinfo {author} {\bibfnamefont {A.}~\bibnamefont {Houck}}, \bibinfo {author} {\bibfnamefont {S.}~\bibnamefont {Saxena}}, \bibinfo {author} {\bibfnamefont {H.}~\bibnamefont {Arnaldi}}, \bibinfo {author} {\bibfnamefont {A.}~\bibnamefont {Agrawal}}, \bibinfo {author} {\bibfnamefont {H.}~\bibnamefont {Zhang}}, \bibinfo {author} {\bibfnamefont {C.}~\bibnamefont
  {Ding}},\ and\ \bibinfo {author} {\bibfnamefont {D.~I.}\ \bibnamefont {Schuster}},\ }\bibfield  {title} {\bibinfo {title} {The {QICK} ({Q}uantum {I}nstrumentation {C}ontrol {K}it): Readout and control for qubits and detectors},\ }\href {https://doi.org/10.1063/5.0076249} {\bibfield  {journal} {\bibinfo  {journal} {Rev. Sci. Instrum.}\ }\textbf {\bibinfo {volume} {93}},\ \bibinfo {pages} {044709} (\bibinfo {year} {2022})}\BibitemShut {NoStop}%
\bibitem [{\citenamefont {Weber}\ \emph {et~al.}(2017)\citenamefont {Weber}, \citenamefont {Tresp}, \citenamefont {Menke}, \citenamefont {Urvoy}, \citenamefont {Firstenberg}, \citenamefont {Büchler},\ and\ \citenamefont {Hofferberth}}]{Weber2017}%
  \BibitemOpen
  \bibfield  {author} {\bibinfo {author} {\bibfnamefont {S.}~\bibnamefont {Weber}}, \bibinfo {author} {\bibfnamefont {C.}~\bibnamefont {Tresp}}, \bibinfo {author} {\bibfnamefont {H.}~\bibnamefont {Menke}}, \bibinfo {author} {\bibfnamefont {A.}~\bibnamefont {Urvoy}}, \bibinfo {author} {\bibfnamefont {O.}~\bibnamefont {Firstenberg}}, \bibinfo {author} {\bibfnamefont {H.~P.}\ \bibnamefont {Büchler}},\ and\ \bibinfo {author} {\bibfnamefont {S.}~\bibnamefont {Hofferberth}},\ }\bibfield  {title} {\bibinfo {title} {Calculation of {R}ydberg interaction potentials},\ }\href {https://doi.org/10.1088/1361-6455/aa743a} {\bibfield  {journal} {\bibinfo  {journal} {J. Phys. B}\ }\textbf {\bibinfo {volume} {50}},\ \bibinfo {pages} {133001} (\bibinfo {year} {2017})}\BibitemShut {NoStop}%
\bibitem [{\citenamefont {Mögerle}\ \emph {et~al.}(2026)\citenamefont {Mögerle}, \citenamefont {Hummel}, \citenamefont {Keil}, \citenamefont {Legrand}, \citenamefont {Braun}, \citenamefont {Menke}, \citenamefont {King}, \citenamefont {Olmos}, \citenamefont {Hofferberth}, \citenamefont {Büchler},\ and\ \citenamefont {Weber}}]{Mogerle2026}%
  \BibitemOpen
  \bibfield  {author} {\bibinfo {author} {\bibfnamefont {J.}~\bibnamefont {Mögerle}}, \bibinfo {author} {\bibfnamefont {F.}~\bibnamefont {Hummel}}, \bibinfo {author} {\bibfnamefont {A.}~\bibnamefont {Keil}}, \bibinfo {author} {\bibfnamefont {T.}~\bibnamefont {Legrand}}, \bibinfo {author} {\bibfnamefont {E.~J.}\ \bibnamefont {Braun}}, \bibinfo {author} {\bibfnamefont {H.}~\bibnamefont {Menke}}, \bibinfo {author} {\bibfnamefont {J.}~\bibnamefont {King}}, \bibinfo {author} {\bibfnamefont {B.}~\bibnamefont {Olmos}}, \bibinfo {author} {\bibfnamefont {S.}~\bibnamefont {Hofferberth}}, \bibinfo {author} {\bibfnamefont {H.~P.}\ \bibnamefont {Büchler}},\ and\ \bibinfo {author} {\bibfnamefont {S.}~\bibnamefont {Weber}},\ }\href {https://arxiv.org/abs/2605.14993} {\bibinfo {title} {Accurate modeling of {R}ydberg atoms and their interactions: Theory and implementation in {PairInteraction}}} (\bibinfo {year} {2026}),\ \Eprint {https://arxiv.org/abs/2605.14993} {arXiv:2605.14993 [physics.atom-ph]} \BibitemShut {NoStop}%
\bibitem [{\citenamefont {Steck}(2025)}]{Steck_Rb}%
  \BibitemOpen
  \bibfield  {author} {\bibinfo {author} {\bibfnamefont {D.~A.}\ \bibnamefont {Steck}},\ }\href@noop {} {\bibinfo {title} {Rubidium 87 {D} line data}},\ \bibinfo {howpublished} {available online at \url{http://steck.us/alkalidata}} (\bibinfo {year} {revision 2.3.4, 2025})\BibitemShut {NoStop}%
\bibitem [{\citenamefont {Johansson}\ \emph {et~al.}(2012)\citenamefont {Johansson}, \citenamefont {Nation},\ and\ \citenamefont {Nori}}]{Johansson2012}%
  \BibitemOpen
  \bibfield  {author} {\bibinfo {author} {\bibfnamefont {J.}~\bibnamefont {Johansson}}, \bibinfo {author} {\bibfnamefont {P.}~\bibnamefont {Nation}},\ and\ \bibinfo {author} {\bibfnamefont {F.}~\bibnamefont {Nori}},\ }\bibfield  {title} {\bibinfo {title} {{QuTiP}: An open-source {Python} framework for the dynamics of open quantum systems},\ }\href {https://doi.org/10.1016/j.cpc.2012.02.021} {\bibfield  {journal} {\bibinfo  {journal} {Comput. Phys. Commun.}\ }\textbf {\bibinfo {volume} {183}},\ \bibinfo {pages} {1760–1772} (\bibinfo {year} {2012})}\BibitemShut {NoStop}%
\bibitem [{\citenamefont {Johansson}\ \emph {et~al.}(2013)\citenamefont {Johansson}, \citenamefont {Nation},\ and\ \citenamefont {Nori}}]{Johansson2013}%
  \BibitemOpen
  \bibfield  {author} {\bibinfo {author} {\bibfnamefont {J.}~\bibnamefont {Johansson}}, \bibinfo {author} {\bibfnamefont {P.}~\bibnamefont {Nation}},\ and\ \bibinfo {author} {\bibfnamefont {F.}~\bibnamefont {Nori}},\ }\bibfield  {title} {\bibinfo {title} {{QuTiP} 2: A python framework for the dynamics of open quantum systems},\ }\href {https://doi.org/https://doi.org/10.1016/j.cpc.2012.11.019} {\bibfield  {journal} {\bibinfo  {journal} {Comput. Phys. Commun.}\ }\textbf {\bibinfo {volume} {184}},\ \bibinfo {pages} {1234} (\bibinfo {year} {2013})}\BibitemShut {NoStop}%
\bibitem [{\citenamefont {Lambert}\ \emph {et~al.}(2026)\citenamefont {Lambert}, \citenamefont {Gigu\`ere}, \citenamefont {Menczel}, \citenamefont {Li}, \citenamefont {Hopf}, \citenamefont {Su\'arez}, \citenamefont {Gali}, \citenamefont {Lishman}, \citenamefont {Gadhvi}, \citenamefont {Agarwal}, \citenamefont {Galicia}, \citenamefont {Shammah}, \citenamefont {Nation}, \citenamefont {Johansson}, \citenamefont {Ahmed}, \citenamefont {Cross}, \citenamefont {Pitchford},\ and\ \citenamefont {Nori}}]{Lambert2026}%
  \BibitemOpen
  \bibfield  {author} {\bibinfo {author} {\bibfnamefont {N.}~\bibnamefont {Lambert}}, \bibinfo {author} {\bibfnamefont {E.}~\bibnamefont {Gigu\`ere}}, \bibinfo {author} {\bibfnamefont {P.}~\bibnamefont {Menczel}}, \bibinfo {author} {\bibfnamefont {B.}~\bibnamefont {Li}}, \bibinfo {author} {\bibfnamefont {P.}~\bibnamefont {Hopf}}, \bibinfo {author} {\bibfnamefont {G.}~\bibnamefont {Su\'arez}}, \bibinfo {author} {\bibfnamefont {M.}~\bibnamefont {Gali}}, \bibinfo {author} {\bibfnamefont {J.}~\bibnamefont {Lishman}}, \bibinfo {author} {\bibfnamefont {R.}~\bibnamefont {Gadhvi}}, \bibinfo {author} {\bibfnamefont {R.}~\bibnamefont {Agarwal}}, \bibinfo {author} {\bibfnamefont {A.}~\bibnamefont {Galicia}}, \bibinfo {author} {\bibfnamefont {N.}~\bibnamefont {Shammah}}, \bibinfo {author} {\bibfnamefont {P.}~\bibnamefont {Nation}}, \bibinfo {author} {\bibfnamefont {J.~R.}\ \bibnamefont {Johansson}}, \bibinfo {author} {\bibfnamefont {S.}~\bibnamefont {Ahmed}}, \bibinfo {author} {\bibfnamefont {S.}~\bibnamefont {Cross}},
  \bibinfo {author} {\bibfnamefont {A.}~\bibnamefont {Pitchford}},\ and\ \bibinfo {author} {\bibfnamefont {F.}~\bibnamefont {Nori}},\ }\bibfield  {title} {\bibinfo {title} {{QuTiP} 5: The quantum toolbox in {Python}},\ }\href {https://doi.org/10.1016/j.physrep.2025.10.001} {\bibfield  {journal} {\bibinfo  {journal} {Phys. Rep.}\ }\textbf {\bibinfo {volume} {1153}},\ \bibinfo {pages} {1} (\bibinfo {year} {2026})}\BibitemShut {NoStop}%
\bibitem [{\citenamefont {Holland}\ \emph {et~al.}(2023{\natexlab{b}})\citenamefont {Holland}, \citenamefont {Lu},\ and\ \citenamefont {Cheuk}}]{Holland2023Entanglement}%
  \BibitemOpen
  \bibfield  {author} {\bibinfo {author} {\bibfnamefont {C.~M.}\ \bibnamefont {Holland}}, \bibinfo {author} {\bibfnamefont {Y.}~\bibnamefont {Lu}},\ and\ \bibinfo {author} {\bibfnamefont {L.~W.}\ \bibnamefont {Cheuk}},\ }\bibfield  {title} {\bibinfo {title} {On-demand entanglement of molecules in a reconfigurable optical tweezer array},\ }\href {https://doi.org/10.1126/science.adf4272} {\bibfield  {journal} {\bibinfo  {journal} {Science}\ }\textbf {\bibinfo {volume} {382}},\ \bibinfo {pages} {1143} (\bibinfo {year} {2023}{\natexlab{b}})}\BibitemShut {NoStop}%
\bibitem [{\citenamefont {Bao}\ \emph {et~al.}(2023)\citenamefont {Bao}, \citenamefont {Yu}, \citenamefont {Anderegg}, \citenamefont {Chae}, \citenamefont {Ketterle}, \citenamefont {Ni},\ and\ \citenamefont {Doyle}}]{Bao2023Entanglement}%
  \BibitemOpen
  \bibfield  {author} {\bibinfo {author} {\bibfnamefont {Y.}~\bibnamefont {Bao}}, \bibinfo {author} {\bibfnamefont {S.~S.}\ \bibnamefont {Yu}}, \bibinfo {author} {\bibfnamefont {L.}~\bibnamefont {Anderegg}}, \bibinfo {author} {\bibfnamefont {E.}~\bibnamefont {Chae}}, \bibinfo {author} {\bibfnamefont {W.}~\bibnamefont {Ketterle}}, \bibinfo {author} {\bibfnamefont {K.-K.}\ \bibnamefont {Ni}},\ and\ \bibinfo {author} {\bibfnamefont {J.~M.}\ \bibnamefont {Doyle}},\ }\bibfield  {title} {\bibinfo {title} {Dipolar spin-exchange and entanglement between molecules in an optical tweezer array},\ }\href {https://doi.org/10.1126/science.adf8999} {\bibfield  {journal} {\bibinfo  {journal} {Science}\ }\textbf {\bibinfo {volume} {382}},\ \bibinfo {pages} {1138} (\bibinfo {year} {2023})}\BibitemShut {NoStop}%
\bibitem [{\citenamefont {Picard}\ \emph {et~al.}(2025)\citenamefont {Picard}, \citenamefont {Park}, \citenamefont {Patenotte}, \citenamefont {Gebretsadkan}, \citenamefont {Wellnitz}, \citenamefont {Rey},\ and\ \citenamefont {Ni}}]{Picard2025}%
  \BibitemOpen
  \bibfield  {author} {\bibinfo {author} {\bibfnamefont {L.~R.~B.}\ \bibnamefont {Picard}}, \bibinfo {author} {\bibfnamefont {A.~J.}\ \bibnamefont {Park}}, \bibinfo {author} {\bibfnamefont {G.~E.}\ \bibnamefont {Patenotte}}, \bibinfo {author} {\bibfnamefont {S.}~\bibnamefont {Gebretsadkan}}, \bibinfo {author} {\bibfnamefont {D.}~\bibnamefont {Wellnitz}}, \bibinfo {author} {\bibfnamefont {A.~M.}\ \bibnamefont {Rey}},\ and\ \bibinfo {author} {\bibfnamefont {K.-K.}\ \bibnamefont {Ni}},\ }\bibfield  {title} {\bibinfo {title} {Entanglement and {{iSWAP}} gate between molecular qubits},\ }\href {https://doi.org/10.1038/s41586-024-08177-3} {\bibfield  {journal} {\bibinfo  {journal} {Nature}\ }\textbf {\bibinfo {volume} {637}},\ \bibinfo {pages} {821} (\bibinfo {year} {2025})}\BibitemShut {NoStop}%
\end{thebibliography}
\end{document}